\documentstyle[fleqn,run2col,epsfig,epsf]{article}

\def\ifm{\ifmmode}
\def\lam{\ifm  \lambda    \else $ \{ \lambda \} $ \fi}

\title{PARTON DISTRIBUTIONS WORKING GROUP}

\author{ Co-Conveners: 
Lucy de Barbaro\address{Fermi National Accelerator
    Laboratory, Batavia, IL 60510},
Stephane A. Keller\address{Theory Division, CERN, CH 1211 Geneva 23,
    Switzerland}~\thanks{Supported by the European Commission under 
contract number ERB4001GT975210, TMR - Marie Curie Fellowship},
Steve Kuhlmann\address{Argonne National Lab, Argonne, IL, 60439},
Heidi Schellman\address{Northwestern University, Dept. of Physics, 
Evanston, IL, 60208},
and Wu-Ki Tung\address{Michigan State University}.}
\begin{document}
\begin{abstract}
This report summarizes the activities of the Parton Distributions
Working Group of the 'QCD and Weak Boson Physics workshop' held in
preparation for Run II at the Fermilab Tevatron.  The main focus of
this working group was to investigate the different issues associated
with the development of quantitative tools to estimate parton
distribution functions uncertainties.  In the conclusion, we introduce
a "Manifesto" that describes an optimal method for reporting data.
\end{abstract}
\maketitle
\section*{INTRODUCTION}

With Run II and its large increase in integrated luminosity, the
Tevatron will enter an era of high precision measurements.  In this
era, parton distribution function (PDF) uncertainties will play a major role.

The basic questions for PDFs at the Tevatron Run II are simple and
common to all other experiment:
\begin{itemize}
\item What limitations will the PDFs put on physics analysis?
\item What information can we gain about the PDFs?
\end{itemize}
There are some qualitative tools that exists and can be used to try to
answer these questions.  However, beside S.~Alekhin's pioneer
work~\cite{a96}, quantitative tools that attempt to include all
sources of uncertainties are not available yet.  The main focus of
this working group has therefore been to investigate the different
issues associated with the development of those tools, although obviously 
other topics have also been investigated.

We have divided this summary of activities into individual contributions:
\begin{itemize}
\item UNCERTAINTIES OF PARTON DISTRIBUTION FUNCTIONS AND THEIR IMPLICATION
ON PHYSICAL PREDICTIONS.  R.~Brok {\sl et al.} describe preliminary
results from an effort to quantify the uncertainties in PDFs and 
the resulting uncertainties in predicted
physical quantities. The production cross section of the W boson is
given as a first example.
\item PARTON DISTRIBUTION FUNCTION UNCERTAINTIES.
Giele {\sl et al.} review the status of their effort to extract PDFs
from data with a quantitative estimate of the
uncertainties.

\item EXPERIMENTAL UNCERTAINTIES AND THEIR DISTRIBUTIONS IN THE 
INCLUSIVE JET CROSS SECTION.  R.~Hirosky summarizes the current CDF and
D0 analysis for the inclusive jet cross sections.  So far the
uncertainties have been assumed to be Gaussian distributed.  He
investigates what information can be extracted about the shape of the
uncertainties with the goal of being able to provide a way to
calculate the Likelihood.

\item PARTON DENSITY UNCERTAINTIES AND SUSY PARTICLE PRODUCTION.
T.~Plehn and M.~Kr\"amer study the current status of PDF's
uncertainties on SUSY particle mass bounds or mass determinations.

\item SOFT-GLUON RESUMMATION AND PDF THEORY UNCERTAINTIES.
G.~Sterman and W.~Vogelsang discuss the interplay of higher order
corrections and PDF determinations, and the possible use of soft-gluon
resummation in global fits.

\item PARTON DISTRIBUTION FUNCTIONS: EXPERIMENTAL DATA AND THEIR 
INTERPRETATION.  L.~de~Barbaro review current issues in the
interpretation of experimental data and the outlook for future data.

\item HEAVY QUARK PRODUCTION.
Olness {\sl et al.} present a status report of a variety of projects
related to heavy quark production.

\item PARTON DENSITIES FOR HEAVY QUARKS.
J. Smith compares different PDFs for heavy quarks.

\item CONSTRAINTS ON THE GLUON DENSITY FROM LEPTON PAIR PRODUCTION.
E.~L.~Berger and M.~Klasen study the sensitivity of the
hadroproduction of lepton pairs to the gluon density.

\end{itemize}

Note that the individual references are at the end of the
corresponding contribution.  The references for the introduction and
the conclusion are at the end.

\newpage
\setcounter{section}{0}

\newcommand{\figWprod}  
{
\begin{figure}[tbh]
%
 \centerline{
\epsfxsize=8cm \epsfbox{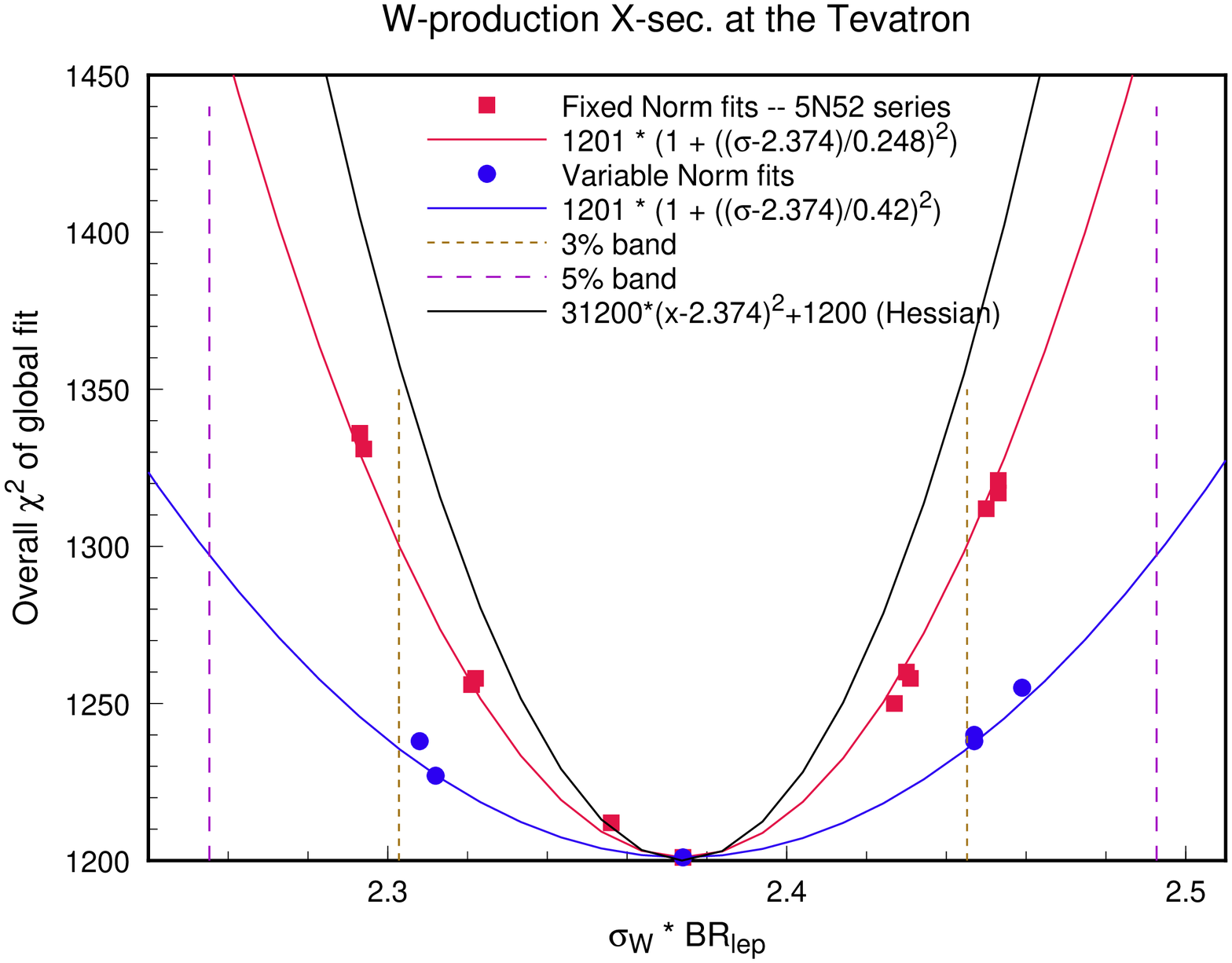} }
 \caption{$\chi^2$ of the base experimental data sets
{\it versus} $\sigma_{W}\cdot{BR}_{\rm lep}$, the $W$ production cross-section
at the Tevatron times lepton branching ratio, in nb.}
 \label{fig:Wprod}
\end{figure}
}

\newcommand{\figEigenVectLow} 
{
\begin{figure}[t]
 \epsfxsize=8cm
 \centerline{\epsfbox{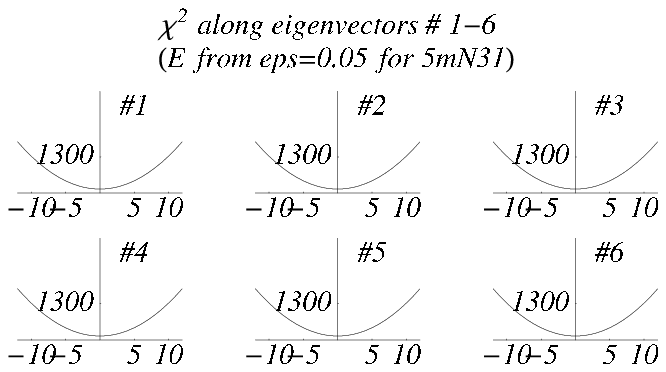}}
 \caption{Value of $\chi^2$ along the six eigenvectors with
     the largest eigenvalues.}
 \label{fig:EigenVect1to6}
\end{figure}
}
\newcommand{\figEigenVectHigh}  
{
\begin{figure}[ht]
 \epsfxsize=8cm
 \centerline{\epsfbox{EigenVect7-18.eps}}
 \caption{Value of $\chi^2$ along the 12 eigenvectors with
     the smallest eigenvalues.}
 \label{fig:EigenVect7to18}
\end{figure}
}
\newcommand{\figEigenValues}  
{
\begin{figure}[t]
 \epsfxsize=7cm
 \centerline{\epsfbox{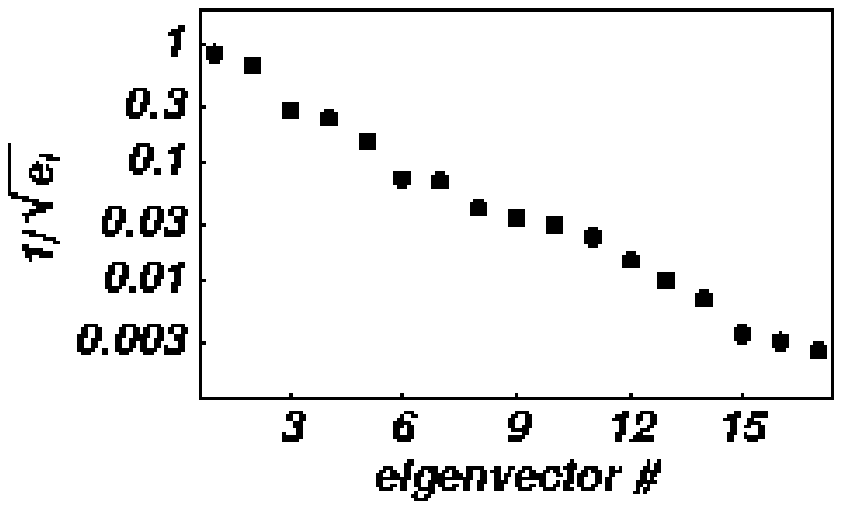}}
 \caption{Plot of the eigenvalues of the Hessian.
The vertical axis is $\ell_{i}=1/\sqrt{e_{i}}$.}
 \label{fig:EigenValues}
\end{figure}
}
\newcommand{\figWprodPDF}  
{
\begin{figure}[t]
 \epsfxsize=7.5cm
 \centerline{\epsfbox{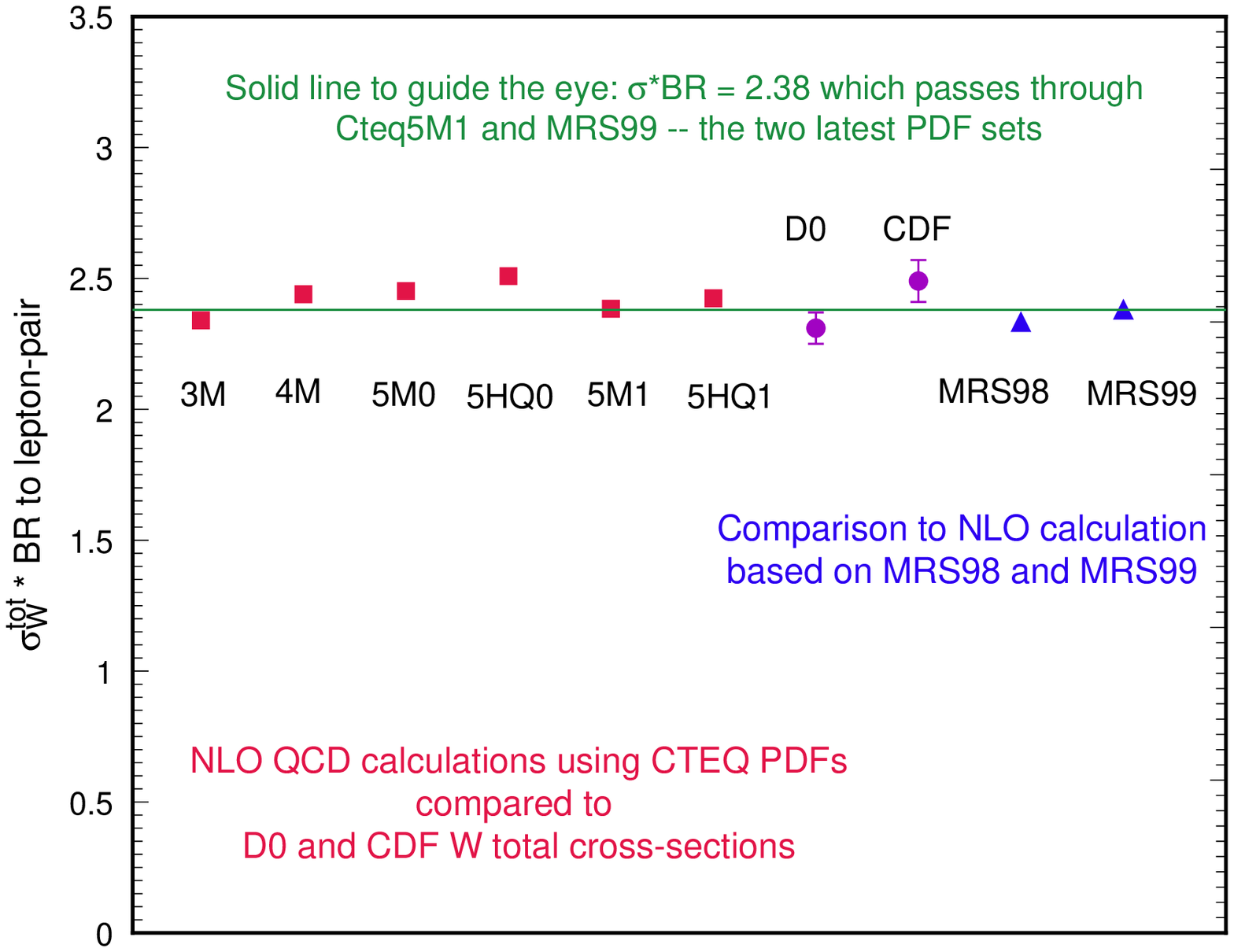}}
 \caption{Predicted cross section for $W$ boson production for various PDFs.}
 \label{fig:WprodPDF}
\end{figure}
}

\newcommand{\figPDFHist}  
{
\begin{figure}[tbh]
 \epsfxsize=8cm
 \centerline{\epsfbox{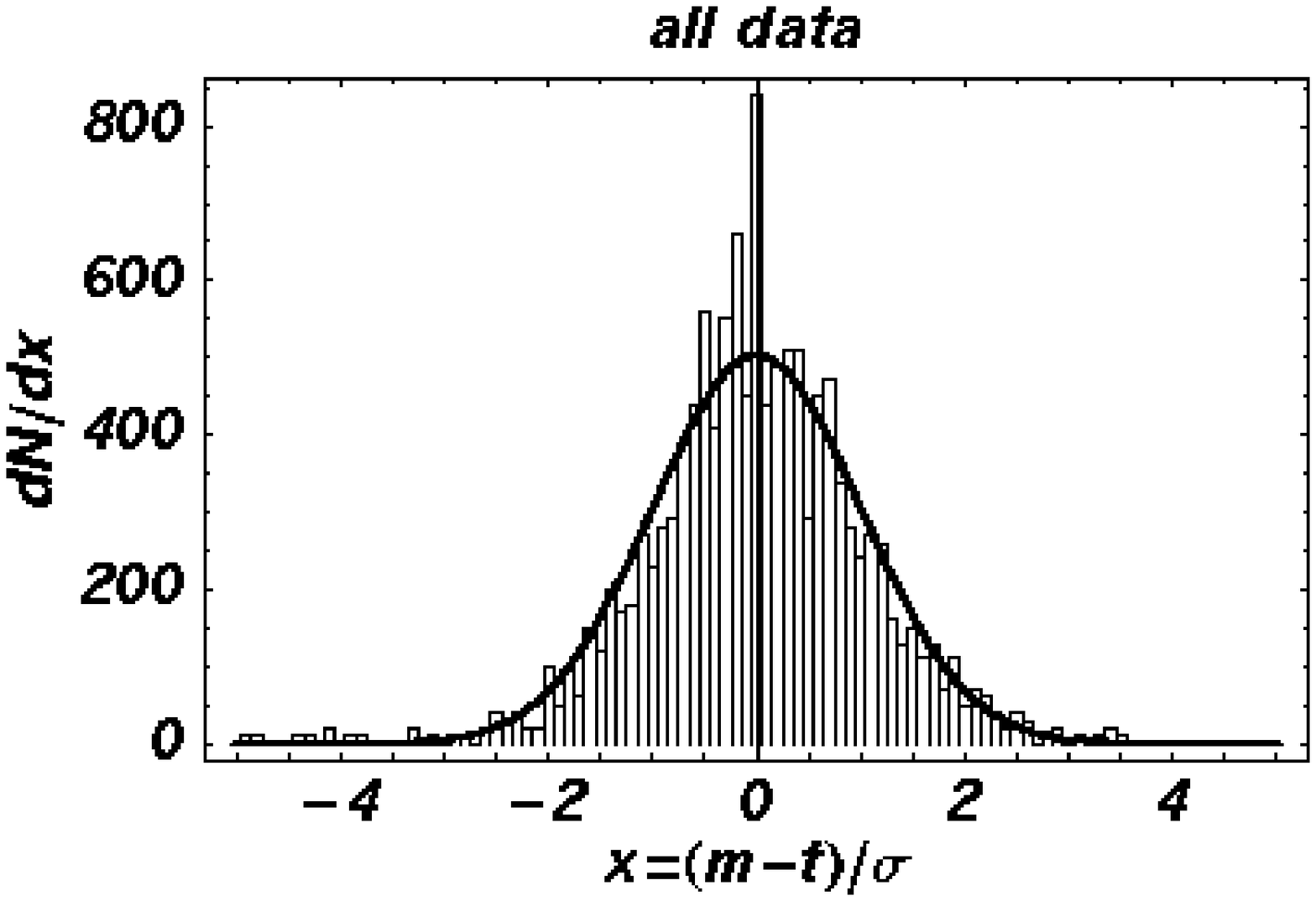}}
 \caption{Histogram of the $(measurement-theory)$ for all data points
 in the CTEQ5m fit.}
 \label{fig:PDFHist}
\end{figure}
}

\newcommand{\figHChisq}   
{
\begin{figure}[tbh]
 \epsfxsize=7cm
 \centerline{\epsfbox{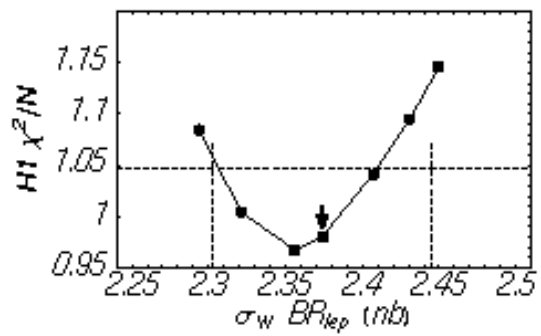}}
 \caption{$\chi^{2}/N$ of the H1 data, including
error correlations, compared to PDFs obtained by the Lagrange multiplier method
for constrained values of $\sigma_{W}$.}
 \label{fig:H1Chisq}
\end{figure}
}

\newcommand{\figName}
{
\begin{figure}[tbh]
 \epsfxsize=4in
 \centerline{\epsfbox{NAME.eps}}
 \caption{fig:NAME}
 \label{fig:NAME}
\end{figure}
}

\newcommand{\tblExptList}
{
\begin{table}
\begin{center}
\begin{tabular}{|c|c|c|c|}
\hline Process & Experiment & Measurable & $N_{data}$
\\ \hline\hline
DIS & BCDMS\cite{bcdms} & $F_{2\ H}^\mu, F_{2\ D}^\mu $ & 324   \\ \hline & NMC
\cite{NewNmc} & $F_{2\ H}^\mu, F_{2\ D}^\mu  $ & 240 \\ \hline & H1
\cite{NewH1}& $F_{2\ H}^e $ & 172   \\ \hline & ZEUS\cite{NewZeus} & $F_{2\
H}^e $ & 186   \\ \hline & CCFR  \cite{ccfr}& $F_{2\ Fe}^\nu, x\ F_{3\ Fe}^\nu
$ & 174  \\ \hline Drell-Yan  & E605\cite{E605} & $sd\sigma /d\sqrt{\tau }dy$ &
119  \\ \hline & E866 \cite{E866} & $\sigma(pd)/2\sigma(pp)$ & 11 \\ \hline &
NA-51\cite{NA51} & $A_{DY}$ & 1   \\ \hline W-prod. & CDF \cite{Wasym} & Lepton
asym. & 11  \\ \hline\hline Incl. Jet & CDF \cite{CdfJets} & $d\sigma /dE_t$ &
33  \\ \hline & D0\cite{D0Jets} & $d\sigma /dE_t$ & 24   \\ \hline
\end{tabular}
\end{center}
  \caption{List of processes and experiments used in the CTEQ5M Global
  analysis. The total number of data points is 1295.
\label{tbl:exptlis}}

\end{table}
}

\newcommand{\tblEigenVectors}
{
\begin{table}
\begin{center}
\begin{tabular}{|c|c|c|c|}

\end{tabular}
\end{center}
  \caption{Table of eigenvector mapping to PDF parameters.
  \label{tbl:EigenVectors}}

\end{table}
}

\newcommand{\tblChisqCorr}
{
\begin{table}
\begin{center}
\begin{tabular}{rrrr}
\\ \hline \hline
Lagrange & $\sigma_{W}\cdot{B}$ & $\chi^{2}/172$ & probability\\ multiplier &
in nb & & \\ \hline 3000 & 2.294 & 1.0847 & 0.212\\ 2000 & 2.321 & 1.0048 &
0.468\\ 1000 & 2.356 & 0.9676 & 0.605\\ 0 & 2.374 & 0.9805 & 0.558\\ -1000 &
2.407 & 1.0416 & 0.339\\ -2000 & 2.431 & 1.0949 & 0.187\\ -3000 & 2.450 &
1.1463 & 0.092\\ \hline \hline
\end{tabular}
\end{center}
  \caption{Comparison of H1 data to the PDF fits with
constrained values of $\sigma_{W}$.
  \label{tbl:ChisqCorr}}

\end{table}
}

\def\pt{$p_T$}                          

\hyphenation{author another created financial paper re-commend-ed}

\newenvironment{Simlis}
{\begin{list}{$\bullet$}
 {
  \settowidth{\labelwidth}{$\bullet$}
  \setlength{\labelsep}{0.5em}
  \setlength{\leftmargin}{.5em}
  \setlength{\rightmargin}{0em}
  \setlength{\itemsep}{0ex}
  \setlength{\topsep}{0ex}
 }
} {\end{list}}

\section*{UNCERTAINTIES OF PARTON DISTRIBUTION FUNCTIONS AND THEIR IMPLICATIONS
ON PHYSICAL PREDICTIONS}

R.\ Brock, D.\ Casey, J.\ Huston, J.\ Kalk,
                   J.\ Pumplin, D.\ Stump, W.K.\ Tung \\

Department of Physics and Astronomy,
Michigan State University, East Lansing, MI 48824\\

\begin{center}
		Abstract
\end{center}
We describe preliminary results from an effort to quantify the
uncertainties in parton distribution functions and the resulting
uncertainties in predicted physical quantities. The production cross
section of the $W$ boson is given as a first example. Constraints due
to the full data sets of the CTEQ global analysis are used in this
study.  Two complementary approaches, based on the Hessian and the
Lagrange multiplier method respectively, are outlined. We discuss
issues on obtaining meaningful uncertainty estimates that include the
effect of correlated experimental systematic uncertainties and
illustrate them with detailed calculations using one set of precision
DIS data.

\section{Introduction}
Many measurements at the Tevatron rely on parton distribution
functions (PDFs) for significant portions of their data analysis as
well as the interpretation of their results.  For example, in cross
section measurements the acceptance calculation often relies on Monte
Carlo (MC) estimates of the fraction of unobserved events.  As another
example, the measurement of the mass of the $W$ boson depends on PDFs
via the modeling of the production of the vector boson in MC.  In such
cases, uncertainties in the PDFs contribute, by necessity, to uncertainties
on the measured quantities.  Critical comparisons between experimental
data and the underlying theory are often even more dependent upon the
uncertainties in PDFs.  The uncertainties on the production cross
sections for $W$ and $Z$ bosons, currently limited by the uncertainty
on the measured luminosity, are approximately 4\%.  At this precision,
any comparison with the theoretical prediction inevitably raises the
question: How ``certain'' is the prediction itself?

A recent example of the importance of PDF uncertainty is the proper
interpretation of the
measurement of the high-$E_{T}$ jet cross-section at the Tevatron.
When the first CDF measurement was published \cite{CDFIa},
there was a great deal of controversy over whether the observed
excess, compared to theory, could be explained by deviations of the
PDFs, especially the gluon, from the conventionally assumed behavior, or
could it be the first signal for some new physics~\cite{jetrefs}.

With the unprecedented precision and reach of many of the Run I
measurements, understanding the implications of uncertainties in the
PDFs has become a burning issue.  During Run II (and later at LHC)
this issue may strongly affect the uncertainty estimates in precision
Standard Model studies, such as the all important $W$-mass
measurement, as well as the signal and background estimates in
searches for new physics.

In principle, it is the uncertainties on physical quantities due to
parton distributions, rather than on the PDFs themselves, that is of
primary concern. The latter are theoretical constructs which
depend on the renormalization and factorization
schemes; and there are strong correlations between PDFs of
different flavors and from
different values of $x$, which can compensate each other in the
convolution integrals that relate them to physical
cross-sections.  On the other hand, since PDFs are universal, if we
can obtain meaningful estimates of their uncertainties based on
analysis of existing data, then the results can be applied to all
processes that are of interest in the future. \cite{Alekhin,GieleEtal}

One can attempt to assess directly the uncertainty on a specific physical
prediction due to the full range of PDFs allowed by available
experimental constraints.  This approach will provide a more reliable
estimate for the range of possible predictions for the physical
variable under study, and may be the best course of action for
ultra-precise measurements such as the mass of the $W$ boson or the
$W$ production cross-section.  However, such results are
process-specific and therefore the analysis must be carried out for
each case individually.

Until recently, the attempts to quantify either the uncertainties on
the PDFs themselves (via uncertainties on their functional parameters,
for instance) or the uncertainty on derived quantities due to
variations in the PDFs have been rather unsatisfactory.  Two commonly
used methods are: (1) Comparing the predictions obtained with
different PDF sets, {\it e.g.,} various CTEQ \cite{cteq5t}, MRS \cite{MRSTt}
and GRV \cite{GRV98} sets; (2)
\figWprodPDF
Within a given global analysis effort, varying individual functional
parameters {\it ad hoc},
 within limits considered to be consistent
with the existing data, e.g.\ \cite{MRST2}.
Neither method provides a systematic,
quantitative measure of the uncertainties of the PDFs or their predictions.

As a case in point, Fig.~\ref{fig:WprodPDF} shows how the calculated
value of the cross section for $W$ boson production at the Tevatron
varies with a set of historical CTEQ PDFs as well as the most recent
CTEQ \cite{cteq5t} and MRST \cite{MRSTt} sets. Also shown are the most
recent measurements from D\O\ and CDF%
\footnote{It is
interesting to note that much of the difference between the D\O\ and
CDF $W$ cross sections is due to the different values of the total
$p\bar{p}$ cross sections used}.  While it is comforting to see that
the predictions have remained within a narrow range, the variation
observed cannot be characterized as a meaningful estimate of the
uncertainty: (i) the variation with time reflects mostly the changes
in experimental input to, or analysis procedure of, the global
analyses; and (ii) the perfect agreement between the values of the
most recent CTEQ5M1~\footnote{CTEQ5M1 is an updated version of
CTEQ5M differing only in a slight improvement in the QCD evolution
(cf.\ note added in proof of \cite{cteq5t}). 
The differences are
completely insignificant for our purposes. Henceforth, we shall refer
to them generically as CTEQ5M. Both sets can be obtained from the web
address http://cteq.org/.
}
and MRS99 sets must be fortuitous, since each group has also obtained
other satisfactory sets which give rise to much larger variations of the $W$
cross section. The MRST group, in particular has examined the range of this
variation by setting a variety of parameters to some extreme values
\cite{MRST2}. These studies are useful
but can not be considered quantitative or definitive. What is needed are
methods that explore thoroughly the possible variations of the parton
distribution functions.
%

It is important to recognize all potential \textbf{sources of
uncertainty} in the determination of PDFs. Focusing on some of these,
while neglecting significant others, may not yield practically useful
results. Sources of uncertainty are listed below:

\begin{Simlis}
\item
\textbf{Statistical uncertainties} of the experimental data used to determine
the PDFs. These vary over a wide range among the experiments used in a global
analysis, but are straightforward to treat.

\item
\textbf{Systematic uncertainties } within each data set. There are typically
many sources of experimental systematic uncertainty, some of which are
highly correlated.  These uncertainties can be treated by standard
methods of probability theory \emph{provided} they are precisely
known, which unfortunately is often not the case -- either because
they may not be randomly distributed and/or because their estimation
in practice involves subjective judgements.

\item
\textbf{Theoretical uncertainties} arising from higher-order PQCD corrections,
resummation corrections near the boundaries of phase space, power-law (higher
twist) and nuclear target corrections, etc.

\item
Uncertainties due to the \textbf{ parametrization of the
non-perturbative PDFs}, $f_{a}(x,Q_{0})$, at some low momentum scale
$Q_{0}$. The specific choice of the functional form used at $Q_{0}$
introduces implicit correlations between the various $x$-ranges, which
could be as important, if not more so, than the experimental
correlations in the determination of $f_{a}(x,Q)$ for all $Q$.

\end{Simlis}

Since strict quantitative statistical methods are based on idealized
assumptions, such as random measurement uncertainties, an important trade-off
must be faced in devising a \textbf{strategy} for the analysis of PDF
uncertainties. If emphasis is put on the ``rigor'' of the statistical
method, then many important experiments cannot be included in the
analysis, either because the published errors appear to fail strict
statistical tests or because data from different experiments appear
to be mutually exclusive in the parton distribution
parameter space \cite{GieleEtal}.
If priority is placed on using the maximal
experimental constraints from available data, then standard
statistical methods may not apply, but must be supplemented by
physical considerations, taking into account experimental and
theoretical limitations. We choose the latter tack, pursuing the
determination of the uncertainties in the context of the current CTEQ
global analysis. In particular, we include the same body of the
world's data as constraints in our uncertainty study as that used in
the CTEQ5 analysis; and adopt the ``best fit'' -- the CTEQ5M1 set
-- as the base set around which the uncertainty studies are performed. In
practice, there are unavoidable choices (and compromises) that must be made in
the analysis. (Similar subjective judgements often are also necessary in
estimating certain systematic errors in experimental analyses.)
The most important consideration is that quantitative results
must remain robust with respect to reasonable variations in these choices.

In this Report we describe preliminary results obtained by our group
using the two approaches
mentioned earlier.  In Section 3 we focus on the error matrix, which
characterizes the general uncertainties of the non-perturbative PDF
parameters.  In Sections 4 and 5 we study specifically the production
cross section $\sigma_{W}$ for $W^{\pm}$ bosons at the Tevatron, to
estimate the uncertainty of the prediction of $\sigma_{W}$ due to PDF
uncertainty.  We start in Section 2 with a review of some aspects of
the CTEQ global analysis on which this study is based.

\section{Elements of the Base Global Analysis}

Since our strategy is based on using the existing framework of the
CTEQ global analysis, it is useful to review some of its
features pertinent to the current study~\cite{cteq5t}.

\subparagraph{Data selection:}%
Table~\ref{tbl:exptlis} shows the experimental data sets included in
the CTEQ5 global analysis, and in the current study. For neutral
current DIS data only the most accurate proton and deuteron target
measurements are kept, since they are the ``cleanest'' and they are
already extremely extensive. For charged current (neutrino) DIS data,
the significant ones all involve a heavy (Fe) target.  Since these data
are crucial for the determination of the normalization of the gluon
distribution (indirectly via the momentum sum rule), and for quark
flavor differentiation (in conjunction with the neutral current data),
they play an important role in any comprehensive global analysis.  For this
purpose, a heavy-target correction is applied to the data, based on
measured ratios for heavy-to-light targets from NMC and other
experiments. Direct photon production data are not included because of
serious theoretical uncertainties, as well as possible inconsistencies
between existing experiments. Cf.\ \cite{cteq5t} and
\cite{kTissues}. The combination of neutral and charged DIS, lepton-pair
production, lepton charge asymmetry, and inclusive large-\pt\ jet
production processes provides a fairly tightly constrained system for
the global analysis of PDFs.
In total, there are $\sim$1300 data points which meet the minimum momentum
scale cuts which must be imposed to ensure that PQCD applies.
The fractional uncertainties on these points are distributed
roughly like $dF/F$ over the range $F=0.003-0.4$.
\tblExptList

\subparagraph{Parametrization:}%
The non-perturbative parton distribution functions $f_{a}(x,Q)$ at a low
momentum scale $Q=Q_0$ are parametrized by a set of functions of $x$,
corresponding to the various flavors $a$.  For this analysis, $Q_0$ is taken to
be 1 GeV.  The specific functional forms and the choice of $Q_0$ are not
important, as long as the parametrization is general enough to accommodate the
behavior of the true (but unknown) non-perturbative PDFs. The CTEQ analysis
adopts the functional form
\[ a_{0}x^{a_1}(1-x)^{a_2}(1+a_3x^{a_4}).\]
for most quark flavors as well as for the gluon.%
\footnote{ An exception is that recent data from E866 seem to require the
ratio $\bar{d}/\bar{u}$ to take a more unconventional functional
form.}
After momentum and quark number sum rules are enforced, there are 18 free
parameters left over, hereafter referred to as ``shape parameters'' $\{a_i\}$.
The PDFs at $Q>Q_{0}$ are determined from $f_{a}(x,Q_{0})$ by evolution
equations from the renormalization group.

\subparagraph{Fitting:}%
The values of $\{a_i\}$ are determined by fitting the global experimental data
to the theoretical expressions which depend on these parameters.
The fitting is done by minimizing a global ``chi-square'' function,
$\chi^{2}_{\rm global}$. The quotation mark indicates that this function serves
as a \emph{figure of merit} of the quality of the global fit; it does not
necessarily
have the full significance associated with rigorous statistical analysis, for
reasons to be discussed extensively throughout the rest of this report.
In practice, this function is defined as:
\begin{eqnarray}
\chi_{\rm global}^2 &=&
\sum_{n}\sum_{i} w_{n}\left[
\left(N_{n}d_{ni}-t_{ni}\right)/\sigma^d_{ni}\right]^{2}
\nonumber \\
  &+& \sum_{n}\left[\left(1-N_{n}\right)/\sigma^{N}_{n}\right]^{2}
\label{eq:chigbl}
\end{eqnarray}
where $d_{ni}$, $\sigma^d_{ni}$, and $t_{ni}$ denote the data,
measurement uncertainty, and theoretical value (dependent on
$\{a_i\}$) for the $i^{\rm th}$ data point in the $n^{\rm th}$
experiment.  The second term allows the absolute normalization
($N_{n}$) for each experiment to vary, constrained by the published
normalization uncertainty ($\sigma^N_{n}$).  The $w_{n}$ factors are
weights applied to some critical experiments with very few data
points, which are known (from physics considerations) to provide useful
constraints on certain unique features of PDFs not afforded by other
experiments.  Experience shows that without some judiciously chosen
weights, these experimental data points will have no influence in the
global fitting process. The use of these weighing factors, to enable
the relevant unique constraints, amounts to imposing certain prior
probability (based on physics knowledge) to the statistical analysis.

In the above form, $\chi^{2}_{\rm global}$ includes for each data point
the random statistical uncertainties and the combined systematic
uncertainties in uncorrelated form, as presented by most experiments
in the published papers.  These two uncertainties are \emph{combined
in quadrature} to form $\sigma^d_{ni}$ in Eq.~\ref{eq:chigbl}.
Detailed point to point correlated systematic uncertainties are not
available in the literature in general; however, in some cases, they
can be obtained from the experimental groups.  For global fitting,
uniformity in procedure with respect to all experiments favors the
usual practice of merging them into the uncorrelated uncertainties.
For the study of PDF uncertainties, we shall discuss this issue in
more detail in Section
\ref{sec:correlations}.

\subparagraph{Goodness-of-fit for CTEQ5M:}%
Without going into details, Fig.~\ref{fig:PDFHist} gives an overview
of how well CTEQ5m fits the total data set.  The graph is a histogram
of the variable $x \equiv (d-t)/\sigma$ where $d$ is a data value,
$\sigma$ the uncertainty of that measurement (statistical and
systematic combined), and $t$ the theoretical value for CTEQ5m.  The
curve in Fig.~\ref{fig:PDFHist} has no adjustable parameters; it is
the Gaussian with width 1 normalized to the total number of data
points (1295).  Over the entire data set, the theory fits the data
within the assigned uncertainties $\sigma^{d}_{ni}$, indicating that
those uncertainties are numerically consistent with the actual
measurement fluctuations.  Similar histograms for the individual
experiments reveal various deviations from the theory, but {\em
globally} the data have a reasonable Gaussian distribution around
CTEQ5M.

\figPDFHist

\section{Uncertainties on PDF parameters: The Error Matrix}
\label{sec:ErrorMatrix}

We now describe results from an investigation of the behavior of the
$\chi^{2}_{\rm global}$ function at its minimum, using the standard error
matrix approach~\cite{CSCTEQ}.  This allows us to determine which combinations
of parameters are contributing the most to the uncertainty.

At the minimum of $\chi^{2}_{\rm global}$, the first derivatives with respect to
the $\{a_{i}\}$ are zero; so near the minimum, $\chi^{2}_{\rm global}$ can be
approximated by
\begin{center}
\begin{equation}
\chi_{\rm global}^2 = \chi^{2}_{0} + \frac{1}{2}\sum_{i,j}F_{ij}y_{i}y_{j}
\end{equation}
\end{center}
where $y_{i}=a_{i}-a_{0i}$ is the displacement from the minimum, and
$F_{ij}$ is the \emph{Hessian}, the matrix of second derivatives.  It
is natural to define a new set of coordinates using the complete
orthonormal set of eigenvectors of the symmetric matrix $F_{ij}$ as
basis vectors.  These vectors can be ordered by their eigenvalues
$e_{i}$.  Each eigenvalue is a quantitative measure of the
uncertainties in the shape parameters $\{a_{i}\}$ for displacements in
parameter space in the direction of the corresponding eigenvector.
The quantity $\ell_{i}\equiv 1/\sqrt{e_{i}}$ is the distance in the 18
dimensional parameter space, in the direction of eigenvector $i$, that
makes a unit increase in $\chi^{2}_{\rm global}$.  If the only
measurement uncertainty were uncorrelated gaussian uncertainties, then
$\ell_{i}$ would be one standard deviation from the best fit in the
direction of the eigenvector. The inverse of the Hessian is the error
matrix.

Because the real uncertainties, for the wide variety of experiments included,
are far more complicated than assumed in the ideal situation, the
quantitative measure of a given increase in $\chi^2_{global}$ carries
\figEigenValues
little true statistical meaning.  However, qualitatively, the Hessian
gives an analytic picture of $\chi^{2}_{\rm global}$ near its minimum in
$\{a_{i}\}$ space, and hence allows us to identify the
particular degrees of freedom that need further experimental input in
future global analyses.

From calculations of the Hessian we find that the eigenvalues vary
over a wide range.  Figure~\ref{fig:EigenValues} shows a graph of the
eigenvalues of $F_{ij}$, on a logarithmic scale.  The vertical axis is
$\ell_{i}=1/\sqrt{e_{i}}$, the distance of a ``standard deviation''
along the $i^{\rm th}$ eigenvector.  These distances range over 3
orders of magnitude.%
Large eigenvalues of $F_{ij}$ correspond to ``steep directions'' of
$\chi^{2}_{\rm global}$.  The corresponding eigenvectors are
combinations of shape parameters that are well determined by current
data.  For example, parameters that govern the valence $u$ and $d$
quarks at moderate $x$ are sharply constrained by DIS data.  Small
eigenvalues of $F_{ij}$ correspond to ``flat directions'' of
$\chi^{2}_{\rm global}$.  In the directions of these eigenvectors,
$\chi^{2}_{\rm global}$ changes little over large distances in
$\{a_{i}\}$ space.  For example, parameters that govern the large-$x$
behavior of the gluon distribution, or differences between sea quarks,
properties of the nucleon that are not accurately determined by
current data, contribute to the flat directions.  The existence of
flat directions is inevitable in global fitting, because as the data
improve it only makes sense to maintain enough flexibility for
$f_{a}(x,Q_{0})$ to fit the available experimental constraints.

Because the eigenvalues of the Hessian have a large range of values,
efficient calculation of $F_{ij}$ requires an adaptive algorithm.  In
principle $F_{ij}$ is the matrix of second derivatives at the minimum
of $\chi^{2}_{\rm global}$, which could be calculated from very small
finite differences.  In practice, small computational errors in the
evaluation of $\chi^{2}_{\rm global}$ preclude the use of a very small
step size.  Coarse grained finite differences yield a more accurate
calculation of the second derivatives.  But because the variation of
$\chi^{2}_{\rm global}$ varies markedly in different directions, it is
important to use a grid in $\{a_{i}\}$ space with small steps in steep
directions and large steps in flat directions.  This grid is generated
by an iterative procedure, in which $F_{ij}$ converges to a good
estimate of the second derivatives.

From calculations of $F_{ij}$ we find that the minimum of $\chi_{\rm
global}^2$ is fairly quadratic over large distances in the parameter
space.  Figures \ref{fig:EigenVect1to6}
and \ref{fig:EigenVect7to18}
\figEigenVectLow
\figEigenVectHigh
show the behavior of $\chi^{2}_{\rm global}$ near the minimum along each
of the 18 eigenvectors.  $\chi^{2}_{\rm global}$ is plotted on the
vertical axis, and the variable on the horizontal axis is the distance
in $\{a_{i}\}$ space in the direction of the eigenvector, in units of
$\ell_{i}=1/\sqrt{e_{i}}$.  There is some nonlinearity, but it is
small enough that the Hessian can be used as an analytic model of the
functional dependence of $\chi^{2}_{\rm global}$ on the shape
parameters.

In a future paper we will provide details on the uncertainties of the
original shape parameters $\{a_{i}\}$.  But it should be remembered
that these parameters specify the PDFs at the low $Q$ scale, and
applications of PDFs to Tevatron experiments use PDFs at a high $Q$
scale.  The evolution equations determine $f(x,Q)$ from $f(x,Q_{0})$,
so the functional form at $Q$ depends on the $\{a_{i}\}$ in a
complicated way.


\section{Uncertainty on $\sigma_W$: the Lagrange Multiplier Method }
\label{sec:Lagrange}

In this Section, we determine the variation of $\chi^{2}_{\rm global}$
as a function of a single measurable quantity. We use the production
cross section for $W$ bosons ($\sigma_W$) as an archetype example.
The same method can be applied to any other physical observable of
interest, for instance the Higgs production cross section, or to certain
measured differential distributions.  The aim is to quantify the
uncertainty on that physical observable due to uncertainties of the
PDFs integrated over the entire PDF parameter space.

Again, we use the standard CTEQ5 analysis tools and results
\cite{cteq5t} as the starting point. The ``best fit'' is the CTEQ5M1
set.  A natural way to find the limits of a physical quantity $X$,
such as $\sigma_{W}$ at $\sqrt{s}=1.8$\,TeV, is to take $X$ as one of
the search parameters in the global fit and study the dependence of
$\chi^{2}_{\rm global}$ for the 15 base experimental data sets on $X$.

Conceptually, we can think of the function $\chi^{2}_{\rm global}$
that is minimized in the fit as a function of $a_{1}, \dots , a_{17},
X$ instead of $a_{1},\dots, a_{18}$. This idea could be implemented
directly in principle, but a more convenient way to do the same thing
in practice is through Lagrange's method of undetermined
multipliers. One minimizes, with respect to the $\{a_i\}$, the
quantity
\begin{center}
\begin{equation}
F(\lambda)=\chi_{\rm global}^2+\lambda X(a_{1},\dots,a_{18})
\end{equation}
\end{center}
for a fixed value of $\lambda$, the Lagrange multiplier. By minimizing
$F(\lambda)$ for many values of $\lambda$, we map out $\chi^{2}_{\rm global}$
as a function of $X$. The minimum of $F$ for a given value of $\lambda$ is the
best fit to the data for the corresponding value of $X$, {\it i.e.,} evaluated
at the minimum.

Figure~\ref{fig:Wprod} shows $\chi^{2}_{\rm global}$ for the 15 base
experimental data sets as a function of $\sigma_{W}$ at the Tevatron.
The horizontal axis is $\sigma_{W}$ times the branching ratio for
$W\rightarrow$ leptons, in nb.  The CTEQ5m prediction is
$\sigma_{W}\cdot{BR}_{\rm lep}=2.374$\,nb.
\figWprod
The vertical dashed lines
are $\pm{3}$\% and $\pm{5}$\% deviations from the CTEQ5m prediction.

The two parabolas associated with points in Fig.~\ref{fig:Wprod}
correspond to different treatments of the normalization factor $N_{n}$
in Eq.\ \ref{eq:chigbl}.  The dots ($\bullet$) are variable norm fits,
in which $N_{n}$ is allowed to float, taking into account the
experimental normalization uncertainties, and $F(\lambda)$ is
minimized with respect to $N_{n}$.  The justification for this
procedure is that overall normalization is a common systematic
uncertainty.  The boxes ($\Box$) are fixed norm fits, in which all
$N_{n}$ are held fixed at their values for the global minimum
(CTEQ5m).  These two procedures represent extremes in the treatment of
normalization uncertainty.  The parabolas associated with $\bullet$'s
and $\Box$'s are just least-square fits to the points.

The other curve in Fig.~\ref{fig:Wprod} was calculated using the
Hessian method.  The Hessian $F_{ij}$ is the matrix of second
derivatives of $\chi^{2}_{\rm global}$ with respect to the shape
parameters $\{a_{i}\}$.  The derivatives (first and second) of
$\sigma_{W}$ may also be calculated by finite differences.  Using the
resultant quadratic approximations for $\chi_{\rm global}^{2}(a)$ and
$\sigma_{W}(a)$, one may minimize $\chi^{2}_{\rm global}$ with
$\sigma_{W}$ fixed.  Since this calculation keeps the normalization
factors constant, it should be compared with the fixed norm fits from
the Lagrange multiplier method.  The fact that the Hessian and
Lagrange multiplier methods yield similar results lends support to
both approaches; the small difference between them indicates that the
quadratic functional approximations for $\chi^{2}_{\rm global}$ and
$\sigma_{W}$ are only approximations.

For the quantitative analysis of uncertainties, the important question
is: How large an increase in $\chi^{2}_{\rm global}$ should be taken
to define the likely range of uncertainty in $X$?  There is an
elementary statistical theorem that states that $\Delta\chi^{2}=1$ in
a constrained fit corresponds to 1 standard deviation of the
constrained quantity $X$.  However, the theorem relies on the
assumption that the uncertainties are gaussian, uncorrelated, and
correctly estimated in magnitude.  Because these conditions do not
hold for the full data set (of $\sim$ 1300 points from 15 different
experiments), this theorem cannot be naively applied quantitatively.%
\footnote{It has been shown by Giele {\it et.al.} \cite{GieleEtal},
that, taken literally, only one or two selected experiments satisfy
the standard statistical tests.}  Indeed, it can be shown that, if
the measurement uncertainties are correlated, and the correlation is
not properly taken into account in the definition of $\chi_{\rm
global}^{2}$, then a standard deviation may vary over the entire
range from $\Delta\chi^{2}= 1$ to $\Delta\chi^{2} = N$ (the total
number of data points -- $\sim$ 1300 in our case). 
\section{Statistical Analysis with Systematic Uncertainties}
\label{sec:correlations}

Fig.~\ref{fig:Wprod} shows how the fitting function $\chi_{\rm
global}^{2}$ increases from its minimum value, at the best global fit,
as the cross-section $\sigma_{W}$ for $W$ production is forced away
from the prediction of the global fit.  The next step in our analysis
of PDF uncertainty is to use that information, or some other analysis,
to estimate the uncertainty in $\sigma_{W}$.  In ideal circumstances
we could say that a certain increase of $\chi_{\rm global}^{2}$ from
the minimum value, call it $\Delta\chi^{2}$, would correspond to a
standard deviation of the global measurement uncertainty.  Then a
horizontal line on Fig.\ \ref{fig:Wprod} at $\chi^{2}_{\rm
min}+\Delta\chi^{2}$ would indicate the probable range of
$\sigma_{W}$, by the intersection with the parabola of $\chi_{\rm
global}^{2}$ versus $\sigma_{W}$.

However, such a simple estimate of the uncertainty of $\sigma_{W}$ is
not possible, because the fitting function $\chi^{2}_{\rm global}$
does not include the {\em correlations} between systematic
uncertainties.  The uncertainty $\sigma_{ni}^{d}$ in the definition
(\ref{eq:chigbl}) of $\chi^{2}_{\rm global}$ combines {\em in
quadrature} the statistical and systematic uncertainties for each data point;
that is, it treats the systematic uncertainties as uncorrelated.  The
standard theorems of statistics for Gaussian probability distributions
of random uncertainties do not apply to $\chi^{2}_{\rm global}$.

Instead of using $\chi^{2}_{\rm global}$ to estimate confidence levels
on $\sigma_{W}$, we believe the best approach is to carry out a
thorough statistical analysis, including the correlations of
systematic uncertainties, on individual experiments used in the global fit
for which detailed information is available.  We will describe
here such an analysis for the measurements of $F_{2}(x,Q)$ by the H1
experiment \cite{NewH1}
at HERA, as a case study.  In a future paper, we will
present similar calculations for other experiments.

The H1 experiment has provided a detailed table of measurement
uncertainties -- statistical and systematic -- for their measurements
of $F_{2}(x,Q)$. \cite{NewH1} The CTEQ program uses 172 data points
from H1 (requiring the cut $Q^{2}>5$\,GeV$^{2}$).  For each
measurement $d_{j}$ (where $j=1 \dots 172$) there is a statistical
uncertainty $\sigma_{0j}$, an uncorrelated systematic uncertainty
$\sigma_{1j}$, and a set of 4 correlated systematic uncertainties
$a_{jk}$ where $k=1\dots 4$.  (In fact there are 8 correlated
uncertainties listed in the H1 table.  These correspond to 4 pairs.
Each pair consists of one standard deviation in the positive sense,
and one standard deviation in the negative sense, of some experimental
parameter.  For this first analysis, we have approximated each pair of
uncertainties by a
single, symmetric combination, equal in magnitude to the average
magnitude of the pair.)

To judge the uncertainty of $\sigma_{W}$, as constrained by the H1
data, we will compare the H1 data to the global fits in
Fig.~\ref{fig:Wprod}.  The comparison is based on the true,
statistical $\chi^{2}$, including the correlated uncertainties, which is
given by
\begin{equation}\label{chisqcorr}
\chi^{2}= \sum_{j}\frac{\left(d_{j}-t_{j}\right)^{2}}
{\sigma_{j}^{2}}
- \sum_{kk'}B_{k}\left(A^{-1}\right)_{kk'}B_{k'}.
\end{equation}
The index $j$ labels the data points and runs from 1 to 172.  The
indices $k$ and $k'$ label the source of systematic uncertainty and run from
1 to 4.  The combined uncorrelated uncertainty $\sigma_{j}$ is
$\sqrt{\sigma_{0j}^{2}+\sigma_{1j}^{2}}$.  The second term in
(\ref{chisqcorr}) comes from the correlated uncertainties.  $B_{k}$ is the
vector
\begin{equation}
B_{k}=\sum_{j}\frac{\left(d_{j}-t_{j}\right)a_{jk}}
{\sigma_{j}^{2}},
\end{equation}
and $A_{kk'}$ is the matrix
\begin{equation}
A_{kk'}=\delta_{kk'}+\sum_{j}\frac{a_{jk}a_{jk'}}
{\sigma_{j}^{2}}.
\end{equation}

Assuming the published uncertainties $\sigma_{0j}$, $\sigma_{1j}$ and
$a_{jk}$ accurately reflect the measurement fluctuations, $\chi^{2}$
would obey a chi-square distribution if the measurements were repeated
many times.  Therefore the chi-square distribution with 172 degrees of
freedom provides a basis for calculating {\em confidence levels} for
the global fits in Fig.~\ref{fig:Wprod}.

\tblChisqCorr

Table \ref{tbl:ChisqCorr} shows $\chi^{2}$ for the H1 data compared to seven of
the PDF fits in Fig.~\ref{fig:Wprod}.  The center row of the Table is the
global best fit -- CTEQ5m.  The other rows are fits obtained by the Lagrange
multiplier method for different values of the Lagrange multiplier.  The best
fit to the H1 data, {\it i.e.}, the smallest $\chi^{2}$, is not CTEQ5m (the
best global fit) but rather the fit with Lagrange multiplier 1000 for which
$\sigma_{W}$ is 0.8\% smaller than the prediction of CTEQ5m. Forcing the $W$
cross section values away from the prediction of CTEQ5m causes an increase in
$\chi^2$ for the DIS data. At $\sqrt{s}=1.8$\,TeV, $W$ production is mainly
from $q\bar{q}\rightarrow W^{+}W^{-}$ with moderate values of $x$ for $q$ and
$\bar{q}$, {\em i.e.,} values in the range of DIS experiments. Forcing
$\sigma_{W}$ higher (or lower) requires a higher (or lower) valence quark
density in the proton, in conflict with the DIS data, so $\chi^2$ increases.

The final column in Table \ref{tbl:ChisqCorr}, labeled
``probability'', is computed from the chi-square distribution with 172
degrees of freedom.  This quantity is the probability for $\chi^{2}$
to be greater than the value calculated from the existing data, if the
H1 measurements were to be repeated.  So, for example, the fit with
Lagrange multiplier $-3000$, which corresponds to $\sigma_{W}$ being
3.2\% larger than the CTEQ5m prediction, has probability 0.092.  In
other words, if the H1 measurements could be repeated many times, in
only 9.2\% of trials would $\chi^{2}$ be greater than or equal to the
value that has been obtained with the existing data.  This probability
represents a confidence level for the value of $\sigma_{W}$ that was
forced on the PDF by setting the Lagrange multiplier equal to -3000.
At the 9.2\% confidence level we can say that
$\sigma_{W}\cdot{BR}_{\rm lep}$ is less than 2.450\,nb, based on the
H1 data.  Similarly, at the 21.2\% confidence level we can say
that $\sigma_{W}\cdot{BR}_{\rm lep}$ is greater than 2.294\,nb.

\figHChisq

Fig.~\ref{fig:H1Chisq} is a graph of $\chi^{2}/N$ for the H1 data
compared to the PDF fits in Table
\ref{tbl:ChisqCorr}.
This figure may be compared to Fig.~\ref{fig:Wprod}.  The CTEQ5
prediction of the $W$ production cross-section is shown as an arrow,
and the vertical dashed lines are $\pm{3}$\% away from the CTEQ5m
prediction.  The horizontal dashed line is the 68\% confidence level
on $\chi^{2}/N$ for $N=172$ degrees of freedom.  The comparison with
H1 data alone indicates that the uncertainty on $\sigma_{W}$ is $\sim
3$\%.

There is much more to say about $\chi^{2}$ and confidence levels.  In
a future paper we will discuss statistical calculations for other
experiments in the global data set.  The H1 experiment is a good case,
because for H1 we have detailed information about the correlated
uncertainties.  But it may be somewhat fortuitous that the $\chi^{2}$
per data point for CTEQ5m is so close to 1 for the H1 data set.  In
cases where $\chi^{2}/N$ is not close to 1, which can easily happen if
the estimated systematic uncertainties are not textbook-like, we must
supply further arguments about confidence levels.  For experiments
with many data points, like 172 for H1, the chi-square distribution is
very narrow, so a small inaccuracy in the estimate of $\sigma_j$ may
translate to a large uncertainty in the calculation of confidence levels
based on the absolute value of $\chi^{2}$.  Because the estimation of
experimental uncertainties introduces some uncertainty in the value of
$\chi^{2}$, it is not really the {\em absolute} value of $\chi^{2}$
that is important, but rather the {\em relative} value compared to the
value at the global minimum.  Therefore, we might study {\em ratios}
of $\chi^{2}$'s to interpret the variation of $\chi^{2}$ with
$\sigma_{W}$.

\section{Conclusions}

It has been widely recognized by the HEP community, and it has been
emphasized at this workshop, that PDF phenomenology must progress from
the past practice of periodic updating of \emph{representative} PDF
sets to a systematic effort to map out the uncertainties, both on the
PDFs themselves and on physical observables derived from them.  For
the analysis of PDF uncertainties, we have only addressed the issues
related to the treatment of experimental uncertainties. Equally important for
the ultimate goal, one must come to grips with uncertainties
associated with theoretical approximations and phenomenological
parametrizations. Both of these sources of uncertainties induce highly
correlated uncertainties, and they can be numerically more important than
experimental uncertainties in some cases.  Only a balanced approach is likely
to produce truly useful results. Thus, great deal of work lies ahead.

This report described first results from two methods for quantifying
the uncertainty of parton distribution functions associated with
experimental uncertainties. The specific work is carried out as extensions of
the CTEQ5 global analysis.  The same methods can be applied using
other parton distributions as the starting point, or using a different
parametrization of the non-perturbative PDFs.  We have indeed tried a
variety of such alternatives.  The results are all similar to those
presented above. The robustness of these results lends confidence to
the general conclusions.

The Hessian, or error matrix method reveals the uncertainties
of the shape parameters used in the functional parametrization.
The behavior of $\chi^{2}_{\rm global}$ in the neighborhood
of the minimum is well described by the Hessian if the
minimum is quadratic.

The Lagrange multiplier method produces constrained fits,
{\it i.e.,} the best fits to the global data set for specified
values of some observable.
The increase of $\chi^{2}_{\rm global}$, as the observable is
forced away from the predicted value, indicates how well the
current data on PDFs determines the observable.

The constrained fits generated by the Lagrange multiplier
method may be compared to data from individual experiments,
taking into account the uncertainties in the data, to estimate
confidence levels for the constrained variable.
For example, we estimate that the uncertainty of
$\sigma_{W}$ attributable to PDFs is $\pm{3}$\%.

Further work is needed to apply these methods to other
measurements, such as the $W$ mass or the forward-backward
asymmetry of $W$ production in $p\bar{p}$ collisions.
Such work will be important in the era of high precision
experiments.


\newpage
\setcounter{section}{0}
\section*{PARTON DISTRIBUTION FUNCTION UNCERTAINTIES}

Walter T. Giele$^a$, Stephane A. Keller$^b$
and David A. Kosower$^c$.\\

a) Fermi National Accelerator
    Laboratory, Batavia, IL 60510; 
b) Theory Division, CERN, CH 1211 Geneva 23,
    Switzerland;
c) CEA--Saclay, F--91191 Gif-sur-Yvette cedex,  France.

\begin{center}
		Abstract
\end{center}
  We review the status of our effort to
  extract parton distribution functions from data with a quantitative
  estimate of the uncertainties.

\section{Introduction}

The goal of our work is to extract parton distribution functions (PDF)
from data with a quantitative estimation of the uncertainties.  There
are some qualitative tools that exist to estimate the uncertainties,
see e.g. Ref.~\cite{mrs98}.  These tools are clearly not adequate when
the PDF uncertainties become important.  One crucial example of a
measurement that will need a quantitative assessment of the PDF
uncertainty is the planned high precision measurement of the mass of
the $W$-vector boson at the Tevatron.  Clearly, quantitative tools
along the line of S.~Alekhin's pionner work~\cite{a96_k} are needed.

The method we have developed in Ref.~\cite{gk98} is flexible and can
accommodate non-Gaussian distributions for the uncertainties associated
with the data and the fitted parameters as well as all their
correlations.  New data can be added in the fit without having to redo
the whole fit.  Experimenters can therefore include their own data
into the fit during the analysis phase, as long as correlation with
older data can be neglected.  Within this method it is trivial to
propagate the PDF uncertainties to new observables, there is for
example no need to calculate the derivative of the observable with
respect to the different PDF parameters.  The method also provides
tools to assess the goodness of the fit and the compatibility of new
data with current fit.  The computer code has to be fast as there is a
large number of choices in the inputs that need to be tested.

It is clear that some of the uncertainties are difficult to quantify
and It might not be possible to quantify all of them.  All the plots
presented here are for illustration of the method only, our results
are {\sl preliminary}.  At the moment we are not including all the
sources of uncertainties and our results should therefore be
considered as lower limits on the PDF uncertainties.  Note that all
the techniques we use can be
found in books and papers on statistics~\cite{a95} and/or in Numerical
Recipes~\cite{nr}.

\section{Outline of the Method}

We only give a brief overview of the method in this section.  More
details are available in Ref.\cite{gk98}.  Our method follows the
Bayesian methodology~\footnote{we also plan to present results within
  the ``classical frequentist'' framework~\cite{cf98}}.  Once a set of
core experiments is selected, a large number of uniformly distributed
sets of parameters $\lambda \equiv\lambda_1, \lambda_2, \ldots,
\lambda_{N_{par}}$ (each set corresponds to one PDF) can be generated
and the probability density of the set $P(\lambda)$ calculated from
the likelihood (the probability) that the predictions based on
$\lambda$ describe the data, see Ref.~\cite{a95} and next section.
  
Knowing $P(\lambda)$, then for any observable $x$ 
(or any quantity that depends on $\lambda$) the probability density, $P(x)$ 
can be evaluated, and using a Monte Carlo integration, the average
value and the standard deviation of $x$ can be calculated with the standard 
expressions:

\begin{eqnarray}
\mu_{x} & = & \int \left(\prod_{i=1}^{N_{par}}  d\lambda_i \right) \, x(\lambda) P(\lambda)  
\nonumber \\ 
\sigma_{x}^2 & = & \int \left(\prod_{i=1}^{N_{par}} d\lambda_i \right) \, (x(\lambda)- \mu_{x})^2 P(\lambda).
\label{eq:musigma}
\end{eqnarray}

If $P(x)$ is Gaussian distributed, then the standard deviation is a
sufficient measure of the PDF uncertainties.  If $P(x)$ is not Gaussian
distributed, then one should refer to the distribution itself and not
try to ``summarize'' it by a single number, all the information is in
the distribution itself.  The uncertainties due to the Monte Carlo can
also be calculated with standard technique.  

The above is correct but computationally inefficient, instead we use a
Metropolis algorithm, see Ref.~\cite{nr}, to generate $N_{pdf}$
unit-weighted PDFs distributed according to $P(\lambda)$.  With this set
of PDFs, the expressions in Eq.~\ref{eq:musigma} become:

\begin{eqnarray}
\mu_{x} & \approx & \frac{1}{N_{pdf}}\sum_{j=1}^{N_{pdf}} 
x\left(\lambda_j\right)
\nonumber \\ 
\sigma_{x}^2 & \approx & \frac{1}{N_{pdf}}\sum_{j=1}^{N_{pdf}} 
\left(x\left(\lambda_j\right)-\mu_{x}\right)^2\ .
\end{eqnarray}

This is equivalent to importance sampling in Monte Carlo integration
techniques.  It is very efficient because the number of PDFs needed to
reach a given level of accuracy in the evaluation of the integrals is
much smaller than when using a set of PDFs uniformly distributed.
Given the unit-weighted set of PDFs, a new experiment can be added to the
fit by assigning a weight (a new probability) to each of the PDFs,
using Bayes' theorem.  The above summations become weighted.  There is
no need to redo the whole fit {\sl if} there is no correlation between
the old and new data.  If we know how to calculate $P(\lambda)$
properly, the only uncertainty in the method comes from the
Monte-Carlo integrations.

\section{Calculation of $P(\lambda)$} 

Given a set of experimental points $\{ x^e
\}=x^e_1,x^e_2,\ldots,x^e_{N_{obs}}$ the probability of a set of PDF
is in fact the conditional probability of \lam given that $\{ x^e \}$
has been measured, this conditional probability can be calculated using 
Bayes theorem:

\begin{equation}
\label{eq:plam}
P(\lam)=P(\lam|x^e)=\frac{P(x^e|\lam)}{P(x^e)} P_{init} (\lam),
\end{equation}

where, as already mentioned, the prior distribution of the parameters,
$P_{init} (\lam)$, has been assumed to be uniform.  A prior
sensitivity should be performed.  ${P(x^e|\lam)}$ is the likelihood,
the probability to observe the data given that the theory is fixed by
the set of \lam.  ${P(x^e)}$ is the probability density of the data 
(integrated over the PDFs) and act as a normalization coefficient in
Eq.~\ref{eq:plam}.

If all the uncertainties are Gaussian distributed, then it is well known that:

\begin{equation}
P({x^e}|\lambda) \approx e^{-\frac{\chi^2(\lam)}{2}},
\label{eq:pxegaus}
\end{equation}

where $\chi^2$ is the {\it usual} chi-square:

\begin{equation}
\chi^2(\lam) = \sum_{k,l}^{N_{obs}} 
                      \left(x^e_k-x_k^{t}(\lambda)\right)
           M^{tot}_{kl}\left(x^e_l-x_l^{t}(\lambda)\right)\ ,
\label{eq:chi2}
\end{equation}

$x_k^t(\lam)$ are the theory prediction for the experimental
observables calculated with the parameters \lam.  The matrix $M^{tot}$ is the
inverse of the total covariance matrix.

When the uncertainties are not Gaussian distributed, the result is not
as well known.  We first present two simple examples to illustrate how
the likelihood should be calculate and then give a generalization.

\subsection{The simplest example}

We first consider the simplest example to setup the notation,
one experimental point with a statistical uncertainty:

\begin{equation}
x^t(\lam)=x^e+u \Delta,
\label{eq:xt}
\end{equation}

where $u$ is a random variable that has it own distribution, f(u) (assumed
to be Gaussian in this case).  By convention, we take the average of
$u$ equal to 0 and its standard deviation equal to 1.  $\Delta$ gives
the size of the statistical uncertainty.  For each experimental
measurement there is a different value of $u$ and $x^e$.  The
probability to find $x^e$ in an element of length $dx^e$ given that
the theory is fixed by \lam is equal to the probability to find $u$ in
a corresponding element of length $du$\footnote{the repetition of the
  experiment will only be distributed according to $u$ around the true
  nature value of $x^t$.  However we are trying to calculate the
  likelihood, the conditional probability of the data given that the
  true nature value of $x^t$ is given by the value of the \lam under
  study}:

\begin{equation}
P({x^e}|\lam) dx^e = f(u) du. 
\end{equation}

The variable $u$ and the Jacobian for the change of variable from $u$
to $x^e$ can be extracted from Eq.~\ref{eq:xt}:

\begin{equation}
u = \frac{x^t(\lam)-x^e}{\Delta}; \left|\frac{du}{dx^e}\right| = \frac{1}{\Delta}
\end{equation}

such that:

\begin{eqnarray}
P({x^e}|\lam) &=& \frac{f(\frac{x^t(\lam)-x^e}{\Delta})}{\Delta} \nonumber \\
&=&\frac{1}{\sqrt{2 \pi} \Delta} e^{{-\frac{(x^t-x^e)^2}{2 \Delta^2}}}.
\end{eqnarray}

This is the expected result.

\subsection{ A simple example}

We now consider the case of one experimental point with a statistical 
and a systematic uncertainty:

\begin{equation}
x^t(\lam)=x^e+u_1 \Delta_1 + u_2 \Delta_2
\label{eq:xt2}
\end{equation}

$\Delta_1$ and $\Delta_2$ give the size of the uncertainties.  $u_1$
and $u_2$ have their own distribution $f^1(u_1)$ and $f^2(u_2)$ and we
use the same convention for their average and standard deviation as
for $u$ in the first example.  This time for each experimental
measurement, there is an infinite number of sets of $u_1,u_2$ that
correspond to it, because there is only one equation that relate $x^t$,
$x^e$ and $u_1$ and $u_2$.  The probability to find $x^e$ in an
element of length $dx^e$ given that the theory is fixed by \lam is
here equal to the probability to find $u_1$ and $u_2$ in a
corresponding element of area $du_1\, du_2$, with an integration over
one of the two variables:

\begin{equation}
P(x^e|\lam) dx^e = du^1 \int du_2 f^1(u_1) f^2(u_2).
\end{equation}

We choose to integrate over $u_2$. $u_1$ and the Jacobian for the
change of variable from $u_1$ to $x^e$ are given by Eq.~\ref{eq:xt2}:

\begin{equation}
u_1=\frac{x^t-x^e - u_2 \Delta_2}{\Delta_1}; \left|\frac{du_1}{dx^e}\right| 
=\frac{1}{\Delta_1}
\end{equation}

such that:

\begin{equation}
P(x^e|\lam) =  \int du_2  f^2(u_2) 
\frac{f^1(\frac{x^t-x^e - u_2 \Delta_2}{\Delta_1})}{\Delta_1}
\label{eq:pxe}
\end{equation}

If both $f^1(u_1)$ and $f^2(u_2)$ are Gaussian distribution then we
recover the expected result, as in Eq~\ref{eq:pxegaus}.  Note that
this expected result is recovered if the uncertainties are Gaussian
distributed \underline{and} the relationship between the theory, the
data and the uncertainties are given by Eq.~\ref{eq:xt2}.  If that
relationship is more complex there is no guarantee to recover
Eq.~\ref{eq:pxegaus}.  In the general case, the integral in
Eq.~\ref{eq:pxe} has to be done numerically.

\subsection{Generalization:}

We are now ready to give a generalization of the calculation of the
likelihood.  We are considering $N_{obs}$ observables, and $N_{unc}$
uncertainties (statistical and systematic) parametrised by $N_{unc}$
random variables $\{u\}=u_1, u_2, \ldots, u_{N_{unc}}$ with their own
distributions, $f^i(u_i)$.

There are $N_{obs}$ relations between $\{x^t\}$, $\{x^e\}$ and $\{u\}$, 
one for each observable:

\begin{equation}
F_i(x_i^e,\{x^t (\lambda)\},\{u\})=0.
\label{eq:F}
\end{equation}

This gives $N_{unc}-N_{obs}$ independent $u_i$ that we
choose by convenience to be the $u_i's$ corresponding to the
systematic uncertainties.  Without loosing generality we assume that
there is one statistical uncertainty for each observable, and we
organize the corresponding $u_i$ with the same index as $x^e_i$, such
that the last $N_{sys}$($=N_{unc}-N_{obs}$) $u_i$ are the random variables
for the systematic uncertainties.  For each set of measured $\{x^e\}$
there is an infinite number of $\{u\}$ sets that correspond to it.

The probability to find $\{x^e\}$ in an element of volume
$\prod_{i=1}^{N_{obs}} dx^e_i$ given that the theory is fixed by \lam
is equal to the probability to find the $\{u\}$ in a corresponding
element of volume $\prod_{i=1}^{N_{unc}} dx^u$ , with an integration
over the independent $u_i$ \footnote{if there are correlations between
  the $u_i$ replace $ \prod_{j=1}^{N_{unc}} f^j(u_j)$ by $ f(u_1, u_2,
  ..., u_{N_{obs}})$ the global probability distribution of the
  $\{u\}$} :
\begin{eqnarray}
P(\{ x^e\}| \lam )\prod_{i=1}^{N_{obs}} dx^e_i &=& 
(\prod_{k=1}^{N_{obs}}du_k) 
\int (\prod_{i=N_{obs}+1}^{N_{unc}}du_i) \nonumber \\ 
&&* \ \ \prod_{j=1}^{N_{unc}} f^j(u_j)
\end{eqnarray}

The values of the $\{u_i,i=1,N_{obs}\}$ (corresponding to the
statistical uncertainties) and the Jacobian, $J(u \rightarrow x^e)$,
for the change of variable from those $u_i$ to the $x^e_i$ can be
extracted from the $N_{obs}$ relations in Eq.~\ref{eq:F}.  The
likelihood is then given by:

\begin{eqnarray}
P(\{ x^e\}| \lam ) = 
\int (\prod_{i=N_{obs}+1}^{N_{unc}}du_i) \ \ \prod_{j=1}^{N_{unc}} f^j(u_j)
J(u \rightarrow x^e) 
\end{eqnarray}

Often, the $F_i$ relationship in Eq~\ref{eq:F} have a simple dependence on 
$\{x^e\}$ and the $u's$ corresponding to the statistical uncertainties:

\begin{equation}
F_i(x_i^e,\{x^t (\lambda)\},\{u\})=x_i^e+u_i \Delta_i +\cdots,
\end{equation}

where the $\Delta_i$ are the size of the statistical uncertainties.
In that case, the Jacobian is simply given by:

\begin{equation}
J(u \rightarrow x^e) = \prod_{i=1}^{N_{obs}}\frac{1}{\Delta_i} 
\end{equation}

In most cases, the likelihood will not
be analytically calculable, and has to be calculated
numerically again with Monte Carlo technique.

In order to be able to calculate the likelihood we therefore need:

\begin{itemize}
\item the relations between $\{x^t\}$, $\{x^e\}$ and $\{u\}$ as in
  Eq.~\ref{eq:F}.
\item the probability distribution of the random variable associated
with the uncertainties: $f^i(u_i)$.
\end{itemize}

Unfortunately most of the time that information is not reported by the
experimenters, and/or is not available and certainly difficult to extract
from papers.  It is only in the case that all the uncertainties are
Gaussian distributed~\footnote{or can be considered as Gaussian
  distributed, see later} that it is sufficient to report the size of
the uncertainties and their correlation~\footnote{with an explicit
  statement that the uncertainties can be assumed to be Gaussian
  distributed}.  This is a very important issue, simply put,
experiments should always provide a way to calculate the likelihood,
$P(\{x^e\}|\lambda)$.  This last fact was also the unanimous
conclusion of a recent workshop on confidence limits held at 
CERN~\cite{cl00}.  This is particularly crucial when combining
different experiments together: the pull of each experiment will
depend on it and, as a result, so will the central values of the
deduced PDFs.

\subsection{The central limit theorem}

Assuming that the uncertainties are Gaussian distributed when they are
not can lead to some serious problems.  For example, 
minimizing the $\chi^2$ constructed assuming Gaussian distribution 
will not even maximize the likelihood.  Indeed in the general case, the
usually defined $\chi^2$ will not appear in the likelihood.

It is often assumed that the central limit theorem can be used
to justify the assumption of Gaussian distribution for the uncertainties.
It is therefore useful to revisit this theorem. 
$Y$ is a linear combination of $n$ independent $X_i$:
\begin{eqnarray}
Y=\sum_i c_i X_i \\
\sigma_Y^2=\sum_i c_i^2 \sigma_{X_i}^2
\nonumber
\end{eqnarray}
where the $c_i$ are constants and the $\sigma$ are the standard deviations. 
The theorem states that in the limit of large $n$ the distribution 
of $Y$ will be approximately Gaussian if $\sigma_Y^2$ is much larger than any
component $c_i^2 \sigma_{X_i}^2$ from a non-Gaussian distributed $X_i$.
For some examples of how large $n$ has to be, see Ref.~\cite{a95}.

Here is one way the theorem could be used:
\underline{If} the $F_i$ relations are given by:
\begin{eqnarray}
x^t_i(\lam)=x^e_i+\sum_{k=1}^{N_{unc}} u_k \Delta_{ik}
\nonumber
\end{eqnarray}
and \underline{if} there is a large number of uncertainties, the $u_k$
are independent and none of the $\Delta_{ik}$ for a non-Gaussian-like
$u_k$ dominate then we know that the sum will be approximately
Gaussian distributed.  One way to express this fact is simply to assume 
that all the uncertainties are Gaussian distributed.
In this case, we recover the usual expression for the
likelihood.

A direct consequence is that if there are a few uncertainties 
that dominate a measurement, then we certainly need to know their 
distribution.  See Ref.~\cite{h}, for an example of a non-Gaussian 
dominant uncertainty in a real life experiment.

\subsection{Luminosity Uncertainty}

We now turn to the calculation of the likelihood when there is a
normalization uncertainty, like the Luminosity uncertainty.  The $F$
relation of Eq.~\ref{eq:F} is given by:

\begin{equation}
{\cal L} \lam=x^e+u_1 \Delta_1,
\label{eq:Lxt}
\end{equation}
where we have assumed that we are measuring 
the parameter directly, $x^t=\lam$.  
The Luminosity, ${\cal L}$, has also an uncertainty:

\begin{equation}
{\cal L}={\cal L}_0+u_2 \Delta_2.
\label{eq:L}
\end{equation}

We assume that both $u_1$ and $u_2$ are Gaussian distributed.
Replacing Eq.~\ref{eq:L} in Eq.~\ref{eq:Lxt}, we obtain:
\begin{equation}
{\cal L}_0 \lam-x^e=u_1 \Delta_1-u_2 \Delta_2 x^t.
\end{equation}
This expression shows that ${\cal L}_0 \lam-x^e$ is the sum of two
Gaussian, such that the likelihood is a Gaussian
distribution with the standard deviation given by:
\begin{eqnarray}
\sigma^2=\Delta_1^2 + (\Delta_2 x^t)^2.
\end{eqnarray}
The systematic uncertainty due to the Luminosity uncertainty is proportional
to the theory.  Explicitly:
\begin{equation}
P(x^e|\lam)=\frac{1} {\sqrt{2 \pi} \sqrt{\Delta_1^2+(\Delta_2 \lam)^2}}    
e^{-\frac{(L_0 \lam-x^e)^2}{2 (\Delta_1^2+(\Delta_2 \lam)^2)}}
\end{equation}
This result can also be derived from the general expression of the
likelihood, after doing the appropriate integral analytically.  

A few remarks are in order.  In this case, eventhough all the
uncertainties are Gaussian distributed, the minimization of the
$\chi^2$ would not maximize the likelihood because the theory appears
in the normalization of the likelihood.  Another mistake that leads to
problems in this case is to replace $\lambda$ by $x^e/{\cal L}_0$ in
the uncertainty.  This mistake leads to a downwards bias.  If $x^e$
has a downward statistical fluctuation, a smaller systematic
uncertainty is assigned to it, such that when it is combined with
other measurements, it is given a larger weight than it should.

This example shows clearly that we have to know if the uncertainties
are proportional to the theory or to the experimental value.  Assuming
one when the other is correct can lead to problems.  It is
clear that many other systematic uncertainties depend on the theory
and that should also be taken into account.

%
%
\section{Sources of uncertainties} 

There are many sources of uncertainties beside the experimental
uncertainties.  They either have to be shown to be small enough to be
neglected or they need to be included in the PDF uncertainties.  For
examples: variation of the renormalization and factorization scales;
non-perturbative and nuclear binding effects; the choice of functional
form of the input PDF at the initial scale; 
accuracy of the evolution; Monte-Carlo uncertainties; and dependence on 
theory cut-off. 

\begin{figure*}[t]
   \epsfig{file=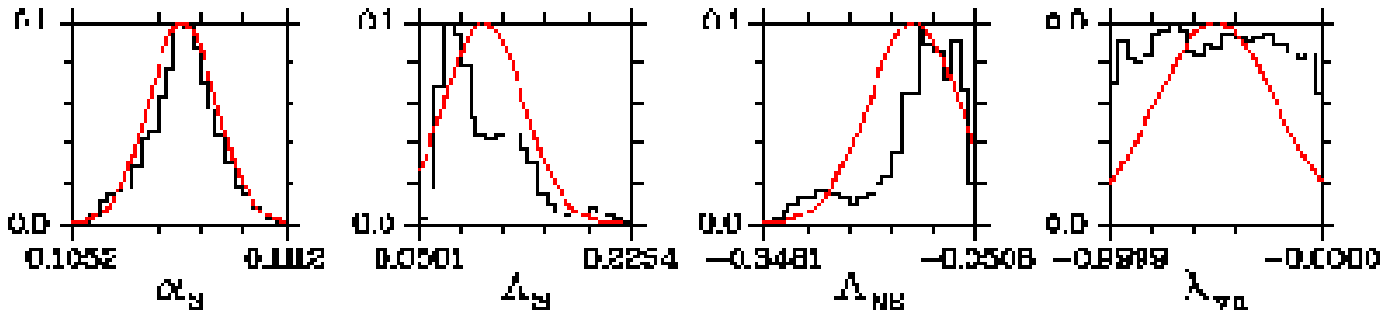,bbllx=91,bblly=342,bburx=522pt,bbury=449pt}
\caption{Plot of the distribution (black histograms) of four of the parameters.
  The first one is $\alpha_s$, the strong coupling constant at the
  mass of the $Z$-boson.  The line is a Gaussian distribution with
  same average and standard deviation as the histogram}
\label{fig:parameters}
\end{figure*}

\section{Current fit}

Draconian measures were needed to restart from scratch and re-evaluate
each issue.  We fixed the renormalization and factorisation scales,
avoided data affected by nuclear binding and non-perturbative effects,
and use a MRS-style parametrization for the input PDFs.  The evolution
of the PDFs is done by Mellin transform method, see Ref.~\cite{k97}.  All the
quarks are considered massless.  We imposed a positivity constraint on
F2.  A positivity constraint on other ``observables'' could also be
imposed.

At the moment we are using H1 and BCDMS (proton data) measurement of
$F_2^p$ for our core set.  In order to be able to use these data we
have to assume that all the uncertainties are Gaussian
distributed~\footnote{no information being given about the
  distribution of the uncertainties}.  We then can calculate the
$\chi^2 (\lambda)$ and $P(\lambda)$ ($ \approx \exp{-\chi^2/2}$) with all
the correlations taken into account~\footnote{here we assumed that
  none of the systematic uncertainties depend on the theory}.  We
generated 50000 unit-weighted PDFs according to the probability function.
For 532 data points, we obtained a minimum $\chi^2$ of 530 for 24
parameters.  We have plotted in Fig.~\ref{fig:parameters}, the
probability
distribution of some of the parameters.  Note that the first parameter
is $\alpha_s$.  The value is smaller than the current world average.
However, it is known that the experiments we are using prefer a lower
value of this parameter, see Ref.~\cite{f21}, and as already pointed
out, our current uncertainties are lower limits.  Note that the
distribution of the parameter is not Gaussian, indicating that the
asymptotic region is not reached yet.  In this case, the blind use of
the so-called chi-squared fitting method might be misleading.

\begin{figure}[t]
 \begin{center}
   \epsfig{file=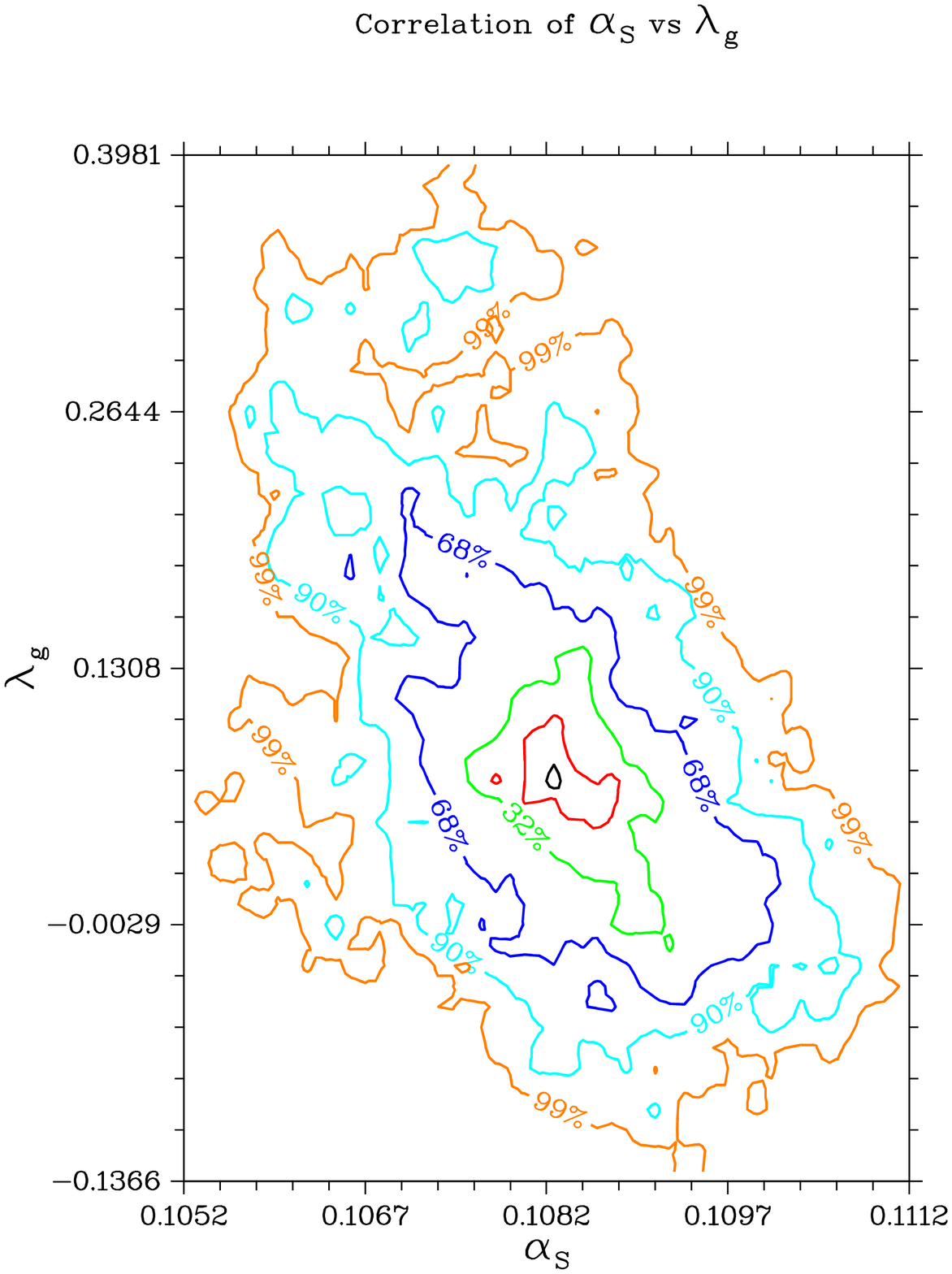,height=7cm,width=6cm,bbllx=65,bblly=92,bburx=546pt,bbury=701pt}
 \end{center}
\vspace*{-1cm}
\caption{Correlation between two of the parameters: $\alpha_s$ and 
  $\lambda_g$, see the text for their definition.  Constant
  probability density levels are plotted.}
\label{fig:contour}
\end{figure}

From this large set of PDFs, it is straightforward to plot, for
example, the correlation between different parameters and to propagate
the uncertainties to other observables.  In Fig.~\ref{fig:contour},
the correlation between $\alpha_s$ and $\lambda_g$ is presented.
$\lambda_g$ parametrizes the small Bjorken-$x$ behavior of the gluon
distribution function at the initial scale: $x g(x) \sim
x^{-\lambda_g}$.  The lines are constant probability density levels
that are characterized by a percentage, $\alpha$, wich is defined such
that $1-\alpha$ is the ratio of the probability density corresponding to
the level to the maximum probability density. 

\begin{figure}[t]
 \begin{center}
  \epsfig{file=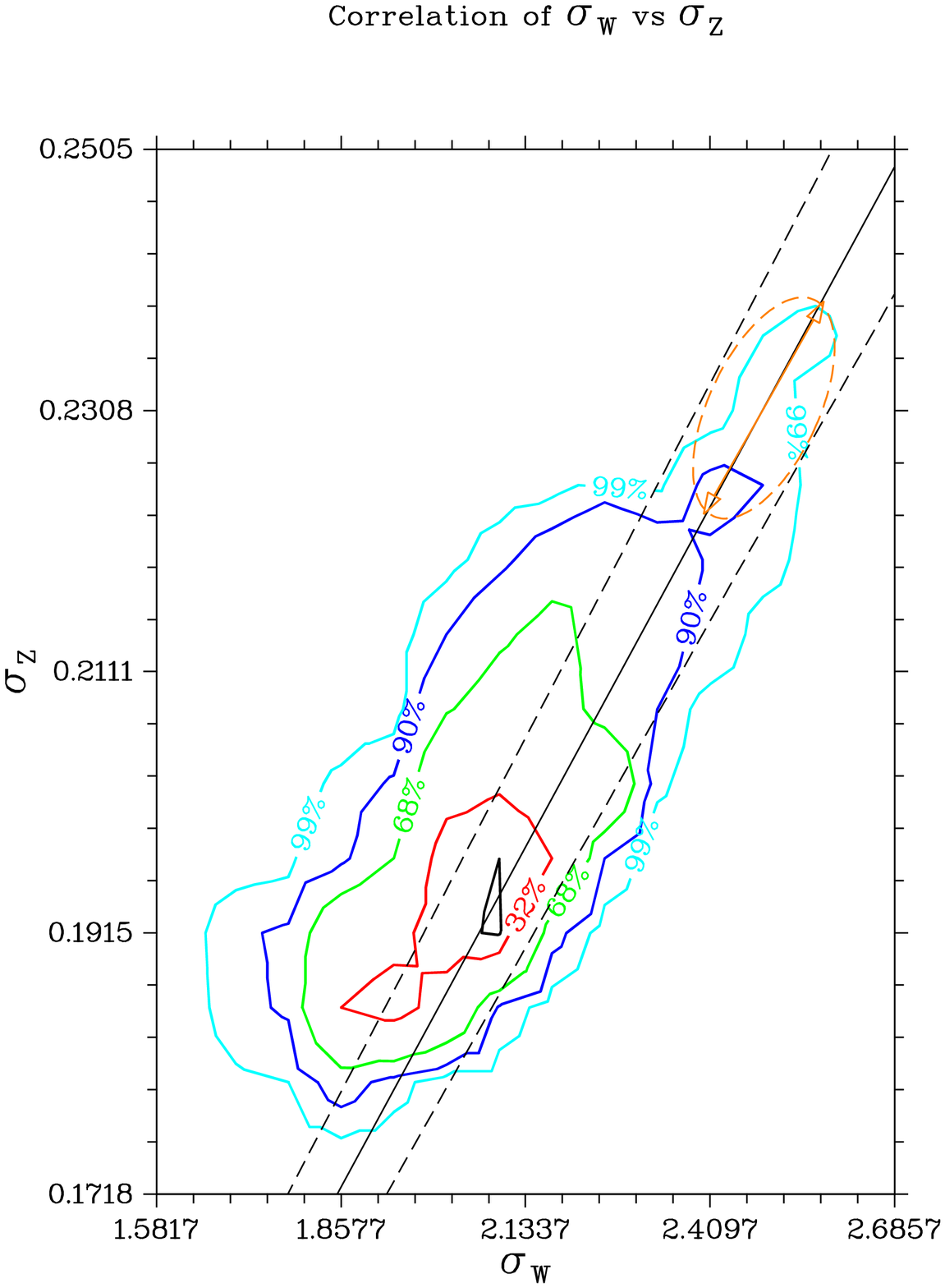,height=7cm,width=6cm,bbllx=48,bblly=45,bburx=510pt,bbury=659pt}
 \end{center}
\vspace*{-1cm} 
\caption{Correlation between the production cross sections for the $W$ 
  and $Z$ vector bosons at the Tevatron, $\sigma_W$ and $\sigma_Z$ (in
  nbarns, includes leptonic branching fraction).  The solid and dashed
  lines show the constraint due to the CDF measurement of the cross
  section ratio.}
\label{fig:sigmas}
\end{figure}

In Fig.~\ref{fig:sigmas}, we show the correlation between two
observables, the production cross sections for the $W$ and $Z$ vector
bosons at the Tevatron along with the experimental result from CDF.
The constant probability density levels are shown.  The agreement
between the theory and the data is qualitatively good.  

\begin{figure}[t]
 \begin{center}
   \epsfig{file=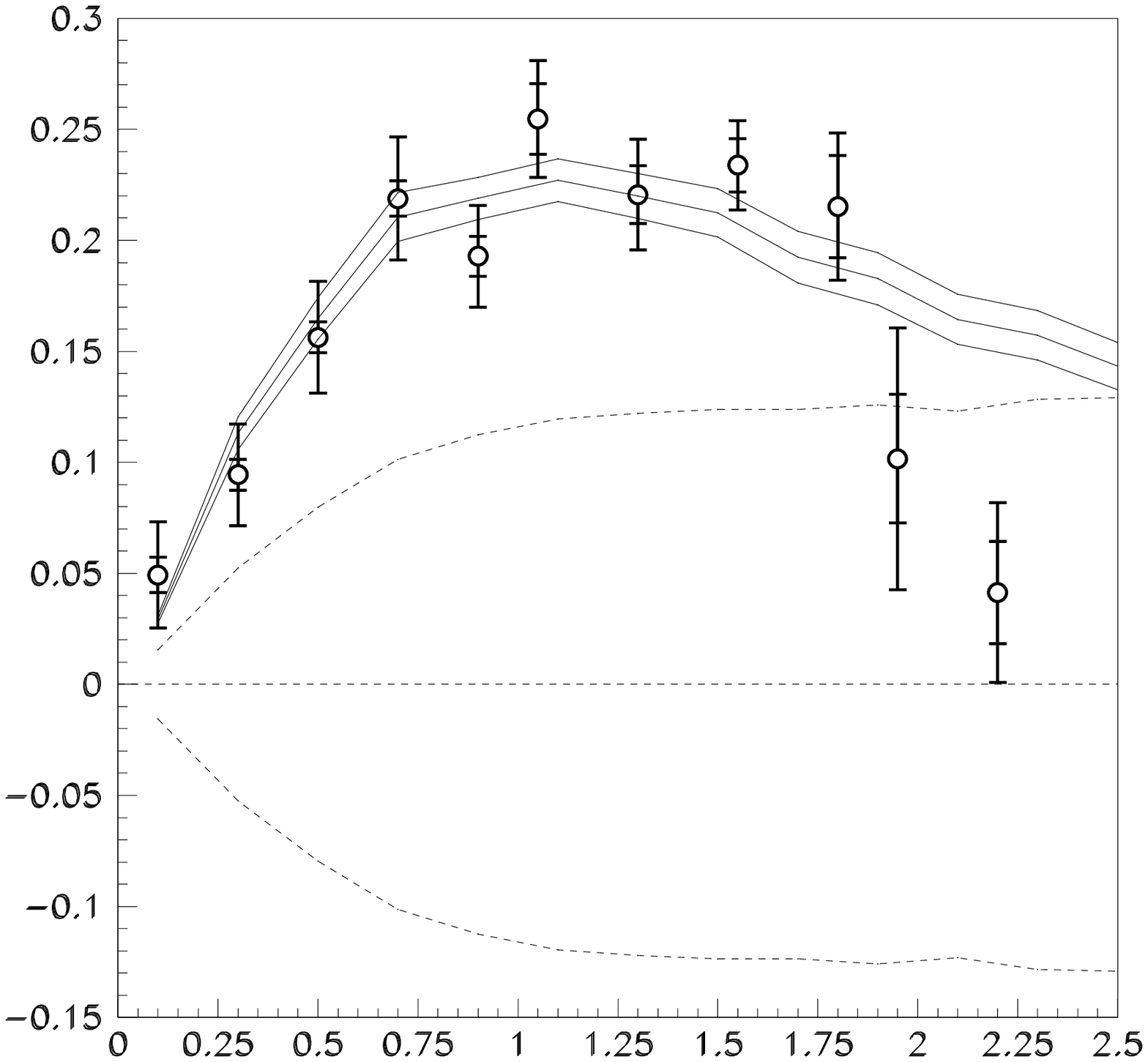,height=7cm,width=6cm,bbllx=48,bblly=45,
     bburx=510pt,bbury=659pt}
 \end{center}
\vspace*{-1cm}
\caption{Data-theory for the lepton charge asymmetry in $W$ decay at 
the Tevatron.}
\label{fig:wasym}
\end{figure}

In Fig.~\ref{fig:wasym}, we present data-theory for the lepton charge
asymmetry in $W$ decay at the Tevatron.  The data are the CDF
result~\cite{cdf98} and the theory correspond to the average value
over the PDF sets for each data point, as defined in
Eq.~\ref{eq:musigma}.  The dashed line are the theory plots
corresponding to the one standard deviation over the PDF sets, also
defined in Eq.~\ref{eq:musigma}.  The inner error bars are the
statistical and systematic uncertainties added in
quadrature\footnote{The distribution of the uncertainties and the
  point to point correlation of the systematic uncertainties were not
  published such that we had to assume Gaussian uncertainties and no
  correlation}.  The outer error bar correspond to the experiment and
theory uncertainties added in quadrature.  The theory uncertainty is
the uncertainty associated with the Monte-Carlo integration, the
factorization and renormalization scale dependence are small and can
be neglected.  5000 PDFs were used to generate this plot.  It is well
known that the data we have included so far in our fit mainly
constraint the sum of the quark parton distribution weighted by the
square of the charges.  The lepton charge asymmetry is sensitive to
the ratio of up-type to down-type quark and is therefore not well
constraint.  We can add this data set by simply weighting each PDF
from our set with the likelihood of the new data.  The resulting new
range of the theory (calculated with weighted sums) is given by the
band of solid curves in Fig~\ref{fig:wasym}.

\begin{figure}[t]
 \begin{center}
   \epsfig{file=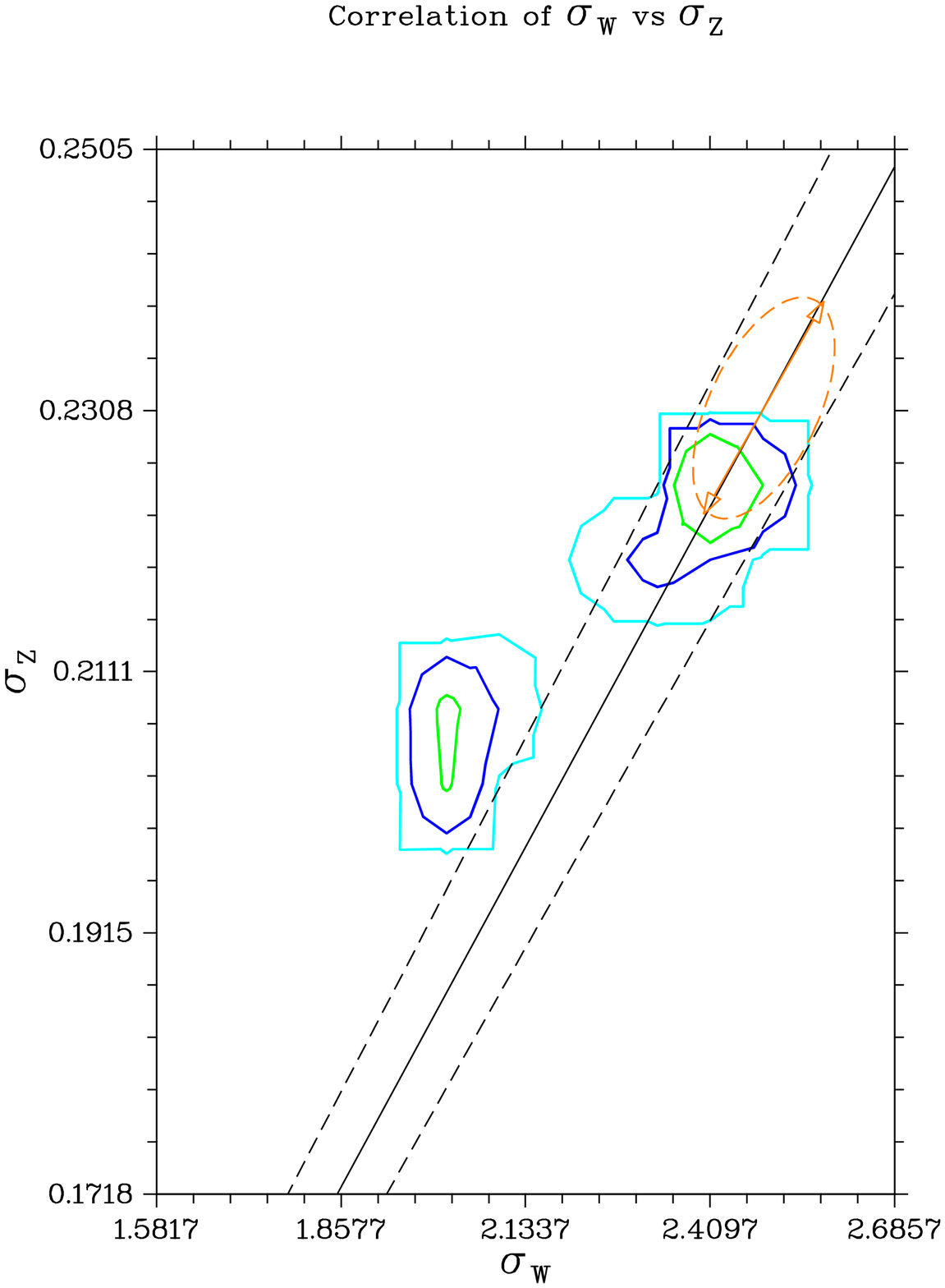,height=7cm,width=6cm,bbllx=48,bblly=45,bburx=510pt,bbury=659pt}
 \end{center}
\vspace*{-1cm}
\caption{Same as in Fig.~\ref{fig:sigmas} for the weighted PDFs.}
\label{fig:sigmas2}
\end{figure}

The effect of the inclusion of the lepton charge asymmetry
can be seen in Fig.~\ref{fig:sigmas2}, where the correlation between
the $W$ and the $Z$ cross section is shown again but for the weighted
PDFs.  The agreement with the data is better than before, but the
probability density has now two maxima.

\begin{figure}[t]
 \begin{center}
   \epsfig{file=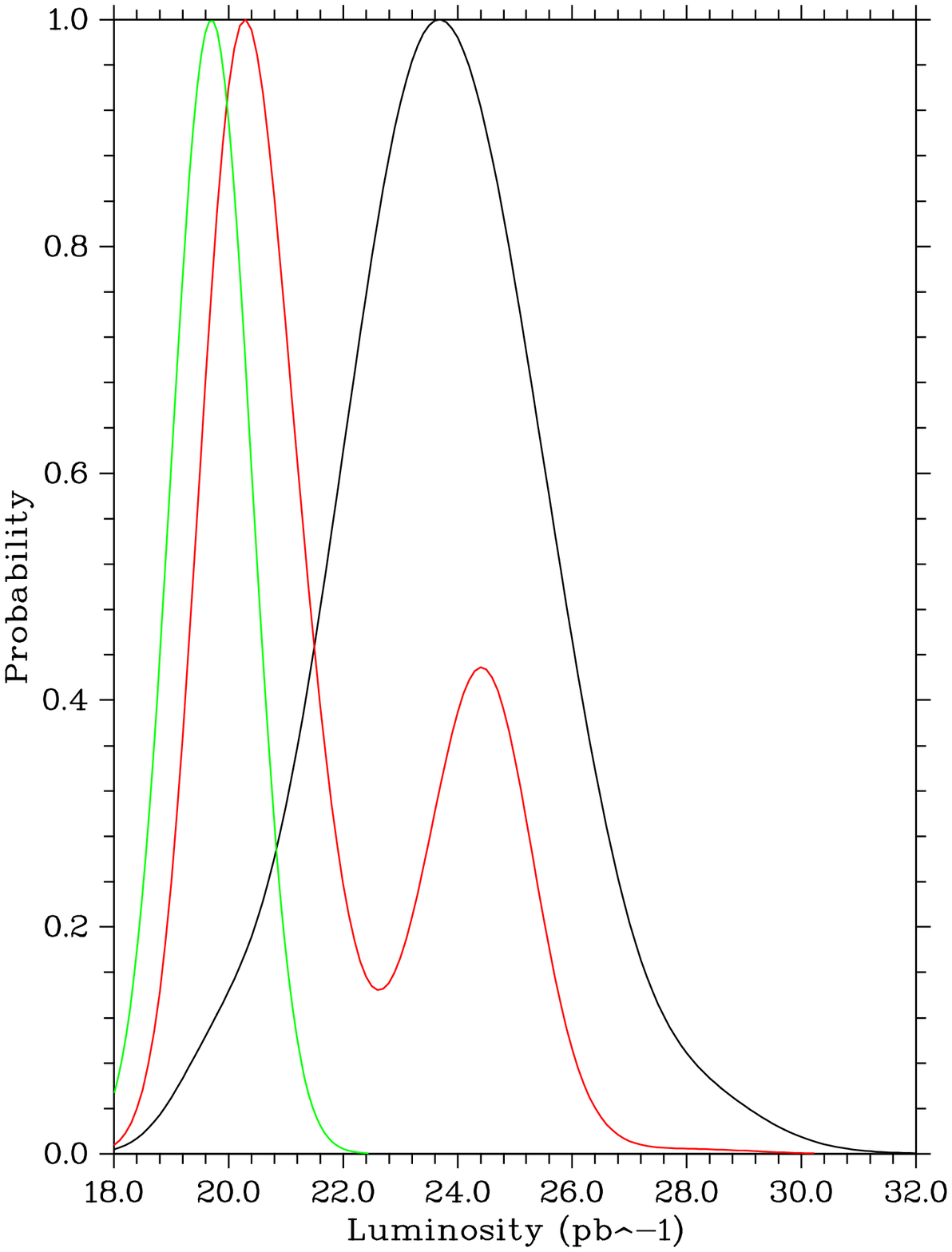,height=7cm,width=6cm,bbllx=48,bblly=45,bburx=510pt,bbury=659pt}
 \end{center}
\vspace*{-1cm}
\caption{Probability distribution of the luminosity (run1a in $pb^{-1}$) 
for the unit-weighted (right plot) and weighted (middle plot) PDFs, 
compared to the value used by CDF (left plot).}
\label{fig:lumi}
\end{figure}

It has been argued that for Run II at the Tevatron, the
measurement of the number of $W$ and $Z$ produced could be used as a
measurement of the Luminosity.  That of course requires the knowledge
of the cross section with a small enough uncertainties.  In
Fig.~\ref{fig:lumi}, the luminosity probability
distribution is presented for the unit-weighted and weighted PDF sets
along with the the luminosity used by CDF.  The plot for the weighted
set has also two maxima, has in Fig.~\ref{fig:sigmas2}.

\subsection{Conclusions}

In conclusion, we remind the reader again that all the results should 
be taken as illustration of the method and that not all the uncertainties
have been included in the fitting.

\newpage
\newpage
\setcounter{section}{0}
\hyphenation{author another created financial paper re-commend-ed 
	mea-sure-ment CDF sta-tis-ti-cal pre-dic-tions next fluctu-a-tions}


\begin{center}\section*{EXPERIMENTAL UNCERTAINTIES AND THEIR DISTRIBUTIONS IN THE 
INCLUSIVE JET CROSS SECTION.}
\end{center}

R.~Hirosky\\
University of Illinois, Chicago, IL 60607\\

\section{Introduction}

This workshop has been an important channel of communication between
those performing global parton distribution function (pdf) fits
and the experimental groups who provide the data at the Tevatron.
In the particular case of jets analyses we have initiated a detailed dialog
on the sources and distributions of experimental uncertainties.  As part of my
participation in the workshop, I have used the D\O\ inclusive jet
cross section as an example of a jet measurement with a complex ensemble of 
uncertainties and have provided descriptions of each component uncertainty.
Such dialogs will prove crucial in obtaining the best constraints on allowable
pdf models from the data.

\section{Uncertainties on the CDF and D\O\ inclusive jet cross sections}

In the first meeting we summarized the jet inclusive cross section measurements
from the D\O\ \cite{d0csprl} and CDF~\cite{cdfcsprl} experiments.  
In particular, we illustrated the major corrections applied to the data, namely
jet $E_T$ scale and $E_T$ resolution corrections, as well as the
derivation methods for these corrections employed by each experiment.  To
review these methods see~\cite{CDF_D0_JETS}-\cite{d0jesnim} and references 
therein.

\begin{figure}[htb]
\begin{center}
\epsfig{figure=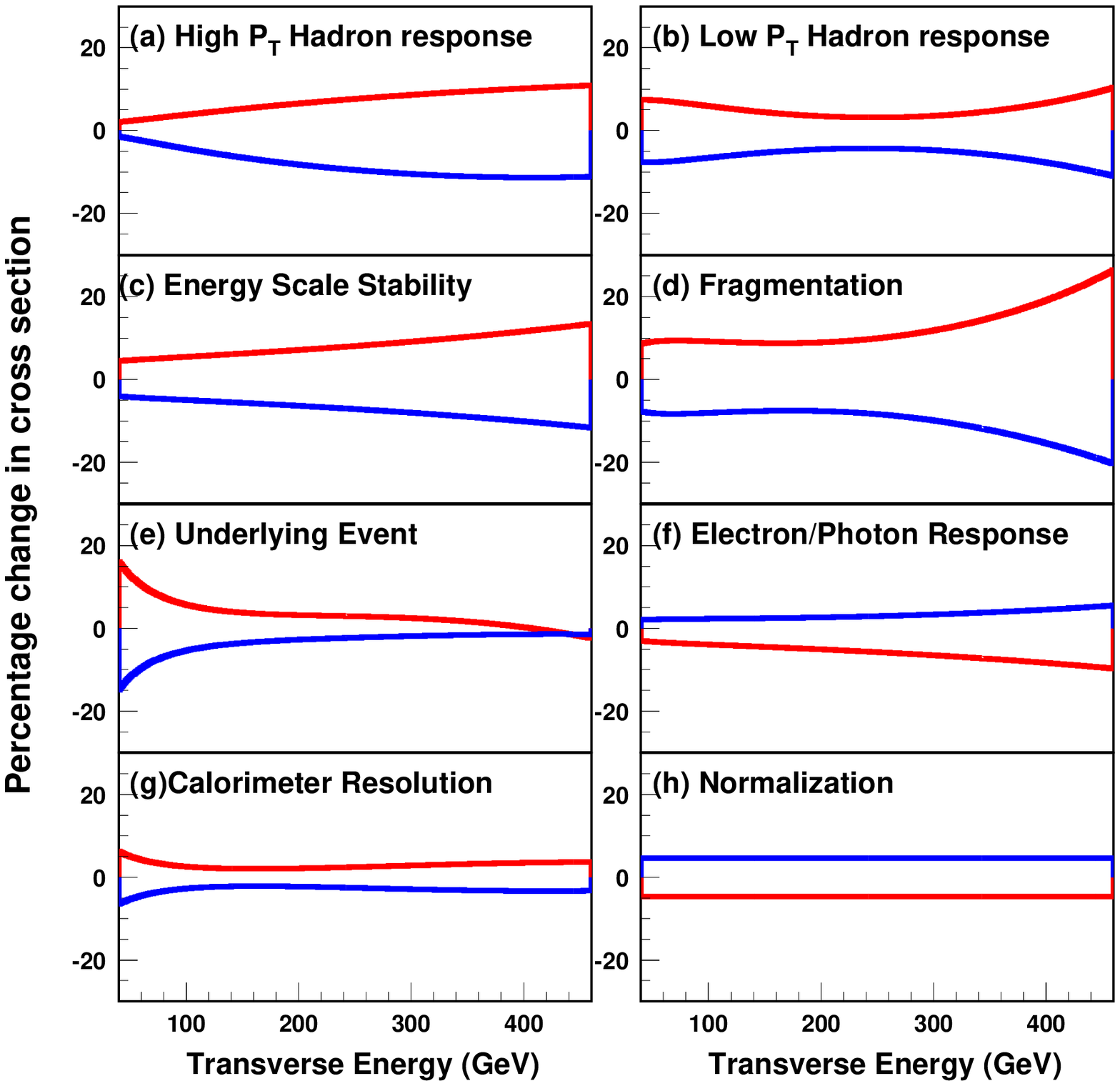,width = 75mm}
\end{center}
\vspace{-10mm}
\caption{\mbox{Uncertainties by component in the CDF }
\mbox{inclusive jet cross section, }
$1/(\Delta \eta\Delta E_T) \int\int d^2\sigma/(dE_Td\eta)dE_Td\eta$, 
$0.1<|\eta|<0.7$ }
\label{cdf_xc_errors}
\end{figure}

\begin{figure}[htb]
\begin{center}
\epsfig{figure=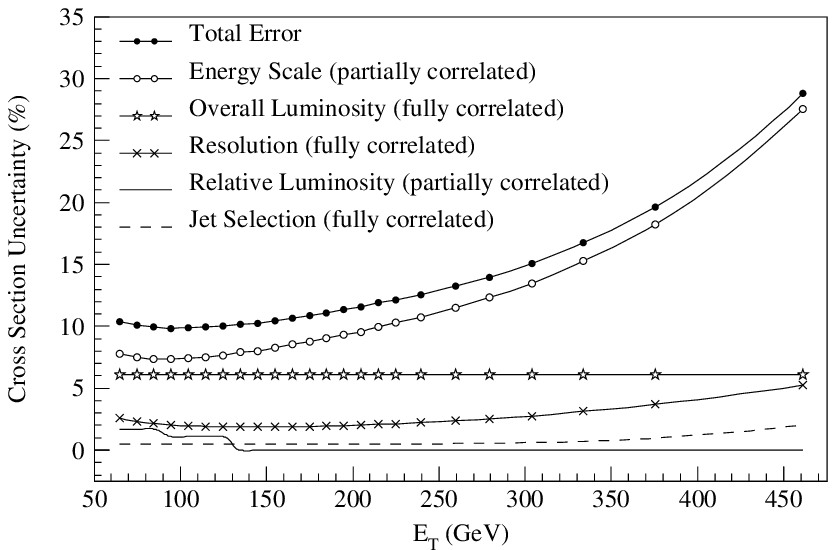,width = 75mm}
\end{center}
\vspace{-10mm}
\caption{\mbox{Uncertainties by component in the D\O\ } 
\mbox{inclusive jet cross section, }
$1/(\Delta \eta\Delta E_T) \int\int d^2\sigma/(dE_Td\eta)dE_Td\eta$, 
$|\eta|<0.5$ }
\label{d0_xc_errors}
\end{figure}

The uncertainties by component in the CDF and D\O\ inclusive 
jet cross sections are shown in Figs.\ref{cdf_xc_errors}-\ref{d0_xc_errors}.
Each component of the uncertainty reported for the CDF cross section
is taken to be completely correlated across jet $E_T$, while individual 
components are independent of one another.  The D\O\ uncertainties (shown
here symmetrized) are also independent of one another, however each 
component may be either fully or partially correlated across jet $E_T$.
In the case of the energy scale uncertainty the band shown is constructed from 
eight subcomponents.

\subsection{Comparisons with theory}

The two experiments have used various means to compare their measurements
to theoretical predictions.  CDF has published a comparison of their
cross section to a next-to-leading order (NLO) QCD calculation using a 
variety of pdf models by means of various normalization-insensitive,
shape-dependent statistical measures~\cite{cdfcsprl} (Kolmogorov-Smirnov,
Cram\`{e}r-VonMises, Anderson-Darling).  D\O\ has formulated a 
covariance matrix using each uncertainty component in the cross section 
and its $E_T$ correlation information and employed a $\chi^2$ test to 
compare to NLO QCD~\cite{d0csprl}.
  It is difficult to generalize the various shape statistics to 
include non-trivial correlations in the systematic uncertainties and although
correlations may be easily added to a covariant error matrix $\chi^2$ tests
can show biases when faced with correlated scale errors.  
Reference~\cite{dag} illustrates how correlated scale errors may lead to biases
in parameter estimation by noting that systematic errors reported as a fraction
of the observed data can be evaluated as artificially small when applied
to a point that fluctuates low.  This bias may be mitigated by
parameterizing the systematic scale errors as percentages of a smooth
model of the data or by placing them on the smooth theory 
directly (see contributions to these proceedings by W. Giele, S. Keller, and
D. Kosower).

Other difficulties arise in interpretation of $\chi^2$ probabilities when
uncertainties show large correlations.
The probability that a prediction agrees with 
the data for a given $\chi^2$ is calculated assuming that 
the $\chi^2$ follows the distribution: 

\begin{eqnarray}
f(x;n) = \frac{(x)^{(n/2-1)}exp(x/2)}{2^{(n/2)}\Gamma(n/2)} 
\label{chidist}
\end{eqnarray}

\noindent
where $n$ is the number of degrees of freedom of the data set.  The probability
of getting a worse value of $\chi^2$ than the one obtained for the comparison 
is given by:

\begin{eqnarray}
P(\chi^2;n) = \int_{\chi^2}^{\infty}f(x;n)dx
\label{chiprob}
\end{eqnarray}

\begin{figure}[htb]
\begin{center}
\epsfig{figure=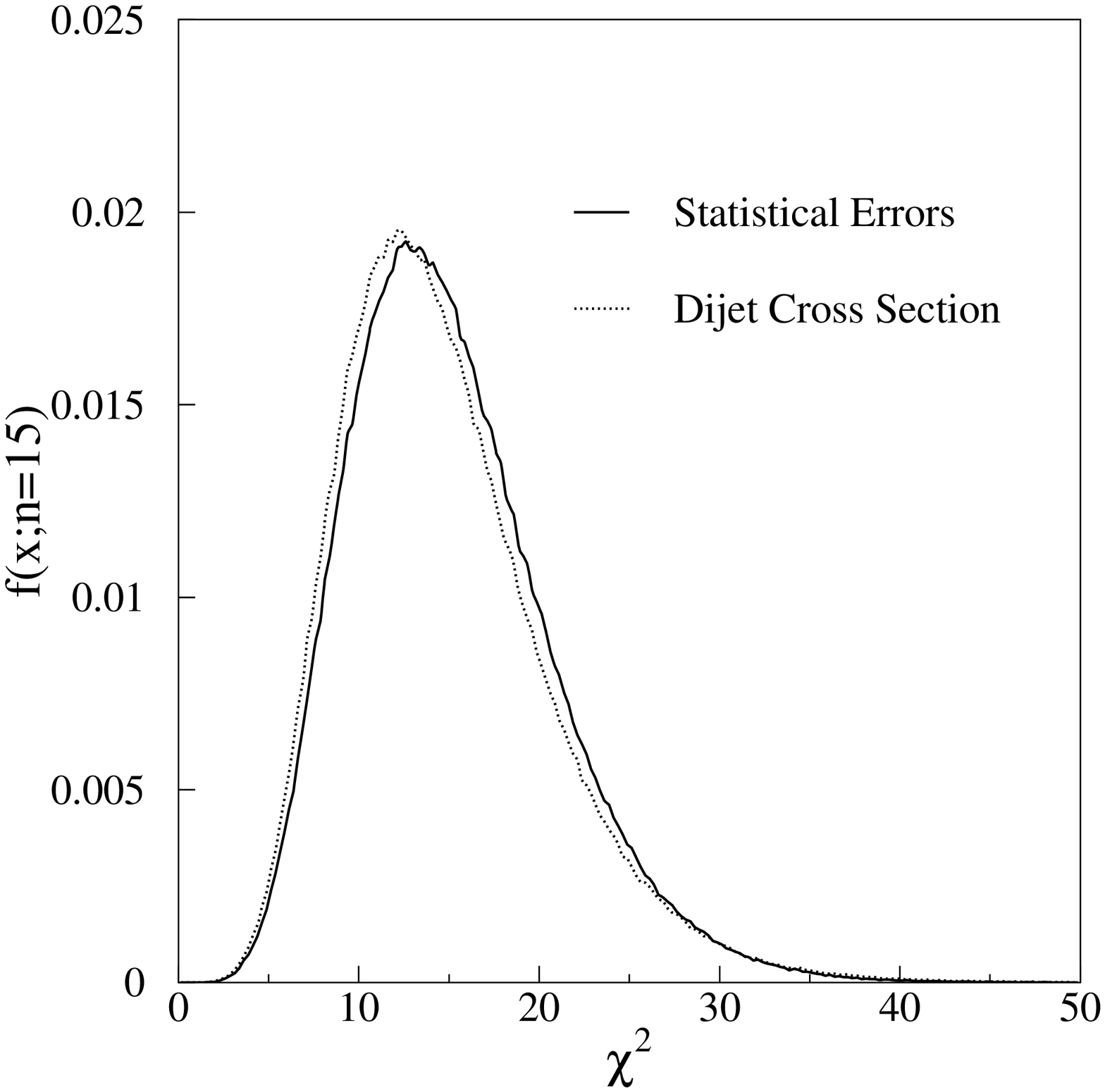,width = 75mm}
\end{center}
\vspace{-10mm}
\caption{$\chi^2$ distribution for random fluctuations around the 
nominal D\O\ Dijet Mass cross section.  (Solid) Errors are fluctuated
as uncorrelated.  (Dashed) $E_T$ correlations are included.}
\label{dijet_mass_chi}
\end{figure}

Hence, to verify the accuracy of the probabilities quoted in the recent D\O\
cross section papers (inclusive jet cross section~\cite{d0csprl} and dijet 
mass spectrum~\cite{d0djmprl}), 
the $\chi^2$ distribution may be compared to Equation~\ref{chidist} with the 
appropriate number of degrees of freedom.   The $\chi^2$ distribution 
for the D\O\ dijet mass spectrum was tested by developing a Monte Carlo 
program~\cite{IAIN} that generates many trial experiments based an
ansatz cross section determined from the best smooth fit to the data
(with a total of 15 bins, or 15 degrees of freedom).  
The first step generated trials based on statistical fluctuations taking
the true number of events per bin as given by the ansatz cross section. 
The trial spectra were then generated for each bin according to Poisson 
statistics.  The $\chi^2$ for each of these trials was calculated using the 
difference between the true and the generated values.  
Figure~\ref{dijet_mass_chi} (solid curve) shows the 
$\chi^2$ distribution for all of the generated trials. The distribution 
agrees well with Equation~\ref{chidist} for 15 degrees of freedom.
The next step assumes that the uncertainties correlated as in 
the measurement of the dijet mass cross 
section.  Trial spectra are generated using these 
uncertainties to generate a $\chi^2$ distribution (see the dotted 
curve in Fig.~\ref{dijet_mass_chi}). It is clear that 
$\chi^2$ distribution very  similar to the curve predicted by 
Equation~\ref{chidist}.   Hence, any probability generated 
using Equation~\ref{chiprob} will be approximately correct. 
The resulting $\chi^2$ distribution was fitted by Equation~\ref{chidist}
and the resulting fit is consistent with the distribution 
if 14.6 degrees of freedom are assumed. 

\begin{figure}[htb]
\begin{center}
\epsfig{figure=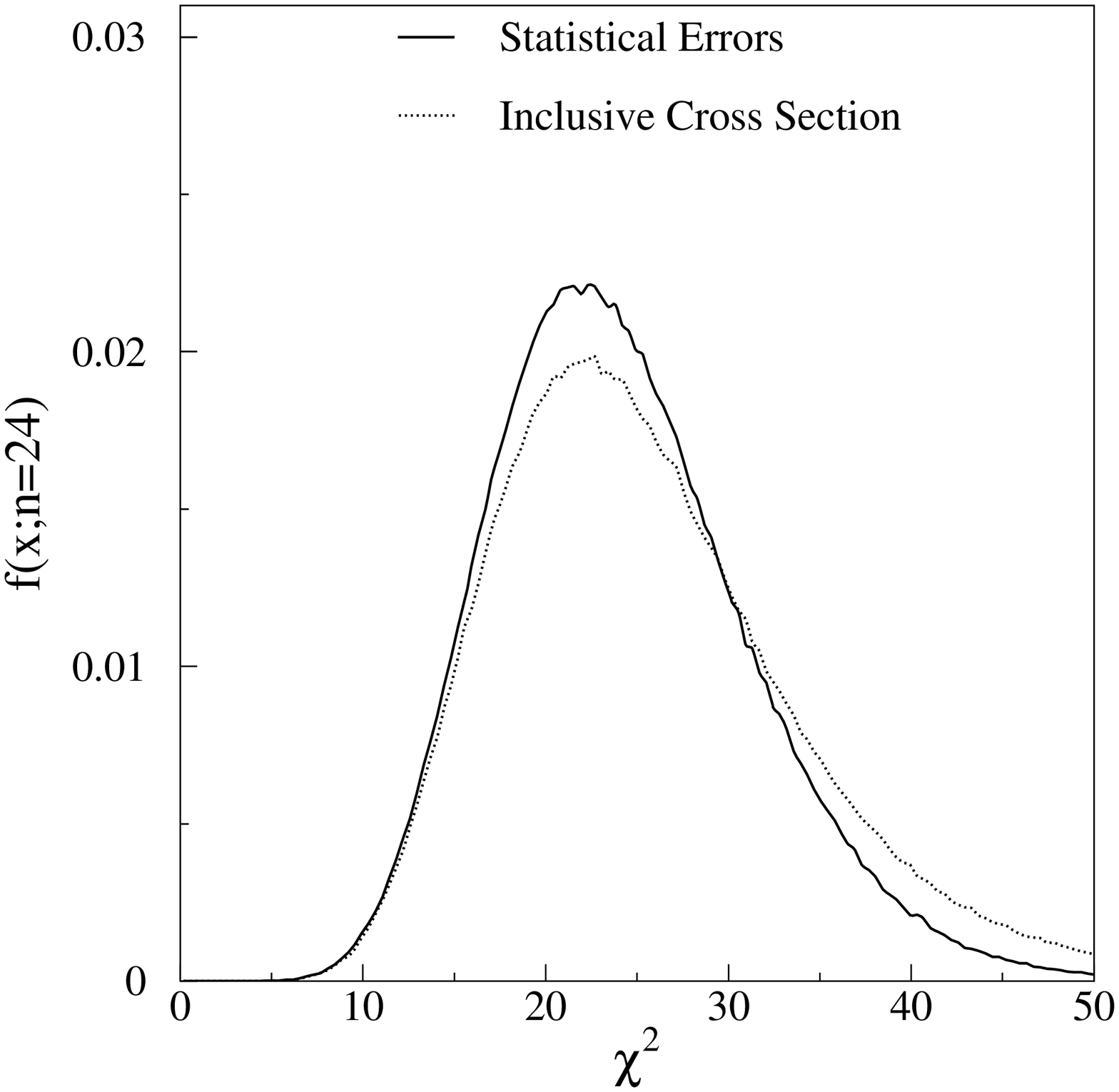,width = 75mm}
\end{center}
\vspace{-10mm}
\caption{$\chi^2$ distribution for random fluctuations around the 
nominal D\O\ inclusive jet cross section.  (Solid) errors are fluctuated
as uncorrelated.  (Dashed) $E_T$ correlations are included.}
\label{inc_cs_chi}
\end{figure}

A similar test using the D\O\ inclusive jet cross section finds the 
distributions shown in Fig.~\ref{inc_cs_chi}.  The two distributions agree 
well for $\chi^2$ values below approximately 15 and then begin to diverge 
slowly.  The distribution based on the cross section uncertainties
includes a larger tail than the $\chi^2$ distribution generated with the
wholly uncorrelated uncertainties,
implying that probabilities based on a $\chi^2$ analysis will be slightly 
underestimated.  See also the talks by B. Flaugher in this workshop
for additional observations and comments on $\chi^2$ analyses.

\section{Beyond the {\em Normal} assumption}

\begin{figure}[htb]
\begin{center}
\epsfig{figure=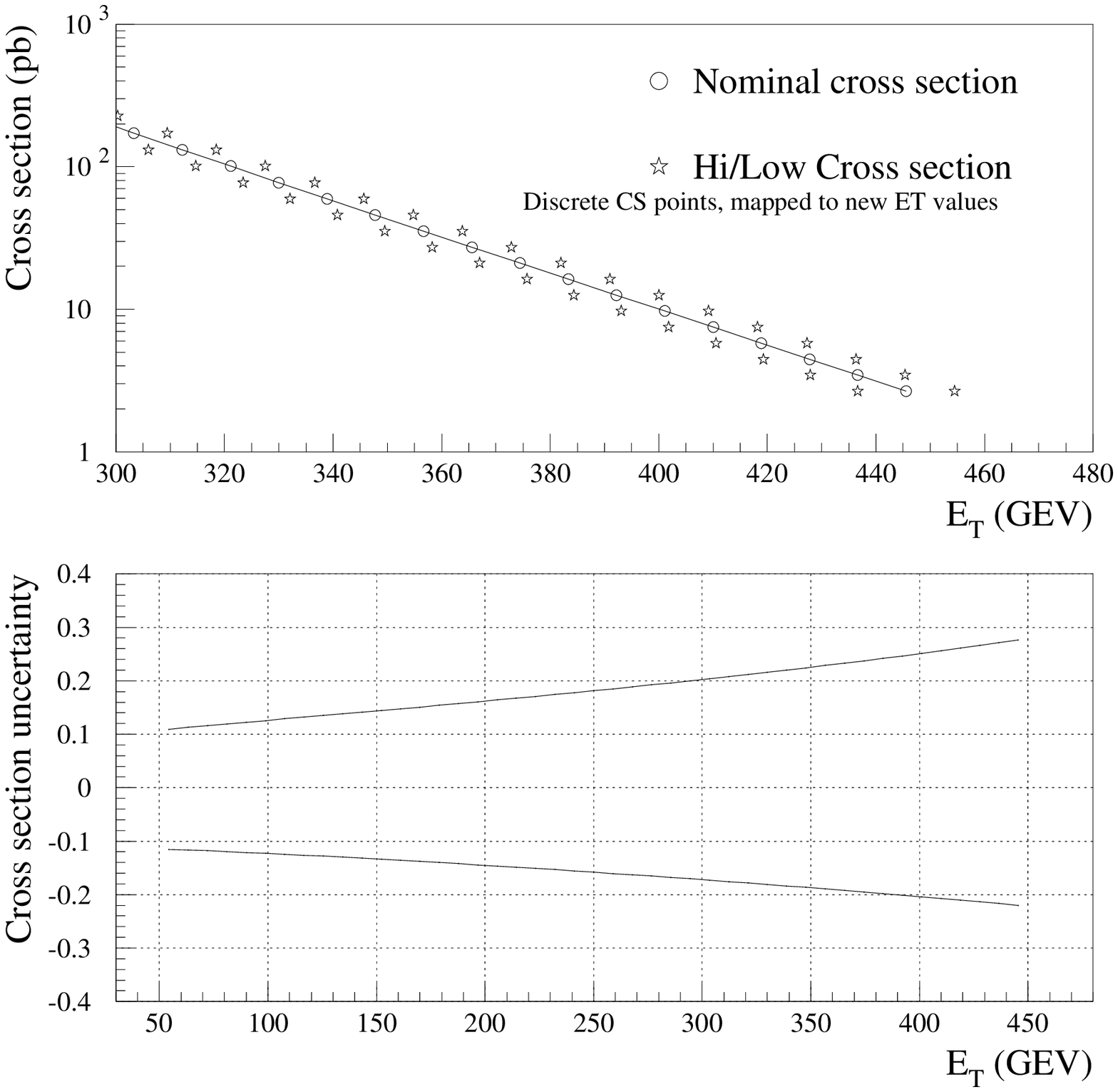,width = 75mm}
\end{center}
\vspace{-10mm}
\caption{Example of a 2\% $E_T$ scale error propagated through an
inclusive jet cross section measurement.}
\label{2percent}
\end{figure}

Independent of any difficulties due to correlated uncertainties, a 
$\chi^2$ test necessarily relies on the assumption that the
uncertainties follow a normal distribution.  This may be a reasonable 
approximation in some cases.  Upon close inspection we expect this 
assumption to be generally false for most rapidly varying 
observables (i.e. steeply falling cross section measurements).
Perhaps, as in the most obvious case, some experimental 
uncertainties will simply be non-Gaussian in their distribution 
and furthermore symmetric uncertainties in the abscissa variable
will develop into asymmetric uncertainties when propagated through to the
measured distribution.  The latter case is illustrated as follows.  
Consider an $E_T$-independent jet $E_T$ scale error of 2\%.  
What is it's effect on an inclusive jet cross section versus $E_T$?  
Jets are shifted bin-to-bin by fluctuating their $E_T$ values 
within the 2\% range and as a result of the steeply falling cross 
section, more jets from low $E_T$ values are shifted
into higher $E_T$ bins by one extreme of this scale uncertainty 
than the in reverse shift for higher $E_T$ jets.  Figure.~\ref{2percent}
shows how a flat 2\% $E_T$ scale uncertainty alters the measured cross section
using a smooth fit to the D\O\ data as the nominal cross section model.  
In general the degree of this asymmetry will depend on the steepness 
of the measured distribution.  In order to define a covariance matrix, 
such errors are typically symmetrized.

\begin{figure}[htb]
\begin{center}
\epsfig{figure=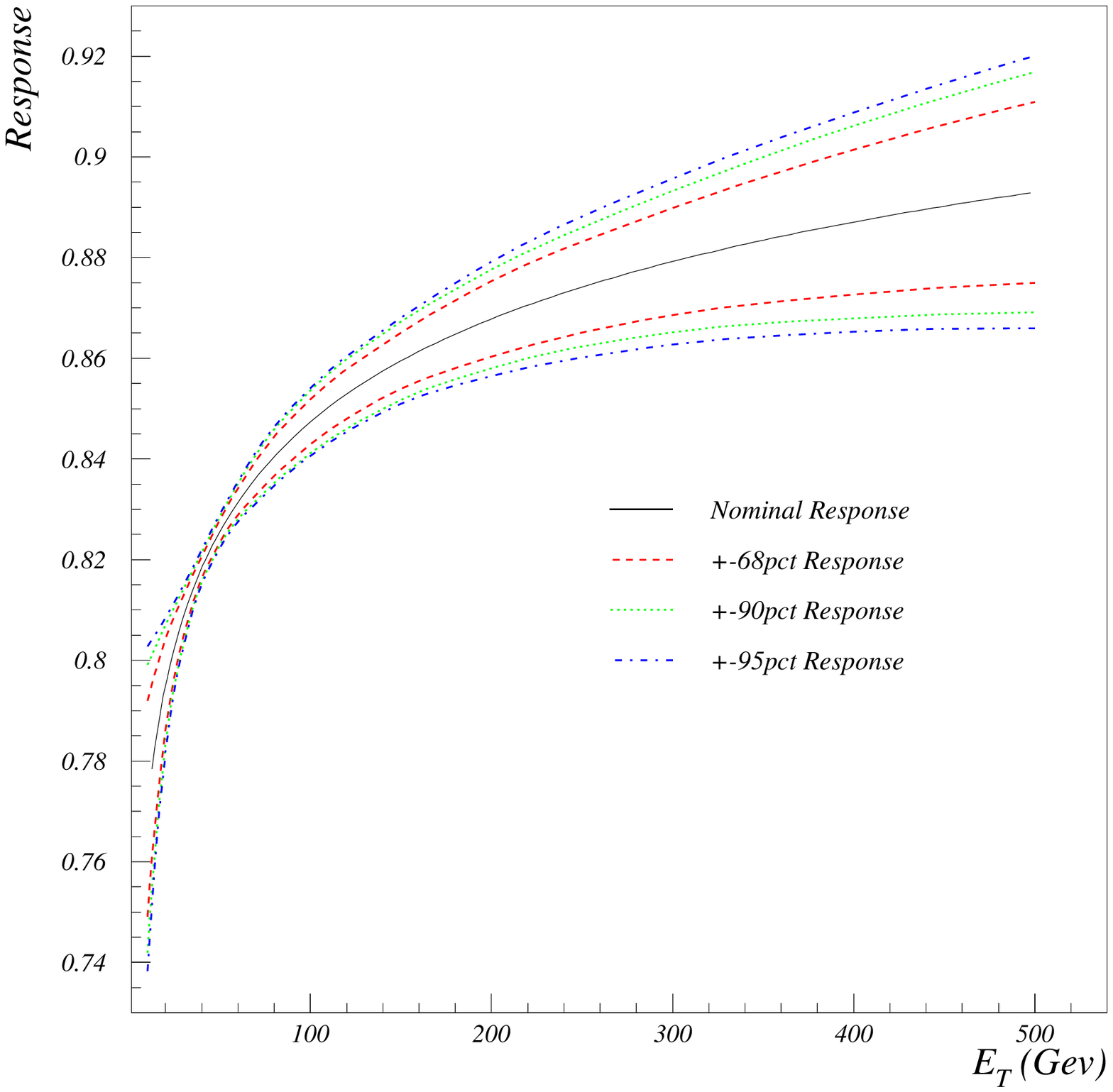,width=75mm}
\end{center}
\vspace{-10mm}
\caption{D\O\ Jet response correction versus Energy.  The outer bands
	show the extreme deviation in response at a given confidence level
	as a function of jet energy.}
\label{d0scale}
\end{figure}

\begin{figure}[htb]
\begin{center}
\epsfig{figure=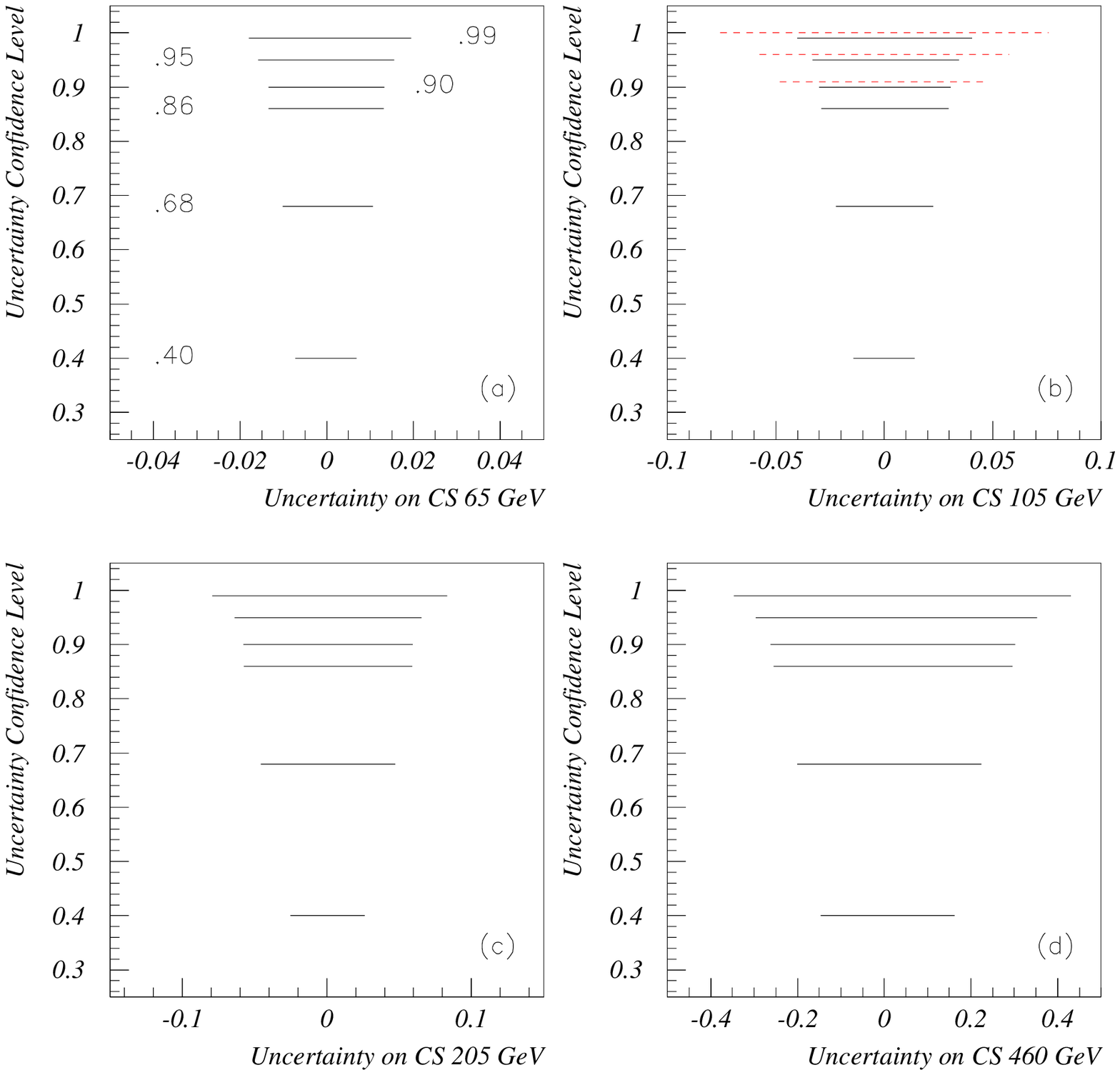,width = 75mm}
\end{center}
\vspace{-10mm}
\caption{D\O\ response uncertainty propagated through the
inclusive jet cross section measurement at various $E_T$ values.
The solid bands represent extreme variations at various confidence levels.
The dashed bands illustrate the overestimation of these variations by using a 
Gaussian approximation.}
\label{scaleprop}
\end{figure}

The use of an approximate covariance matrix will also result in a loss
of sensitivity when errors are shown to follow distributions with tails 
smaller than in a normal distribution.  As an example we show a correction 
factor with uncertainties of this type from the D\O\ jet cross section 
analysis in Fig.~\ref{d0scale}.  This figure shows the hadronic response 
correction for jets as a function of jet energy.  The correction is derived
from an analysis of $\gamma + jet$ data~\cite{d0jesnim}.  The bands
delimit regions that contain ensembles of deviations from the nominal response
within certain confidence limits.  It is evident that in this case
assuming the uncertainty follows a normal distribution with variance
equal to the 68\% limits shown
will tend of underestimate the sensitivity of the data for excluding
certain classes of theories.  Figure~\ref{scaleprop} shows the range 
of cross section uncertainty due to the response component only 
as a function of confidence level for several $E_T$ values of 
the D\O\ cross section.

\section{Application to pdf constraints}

In this workshop W. Giele, S. Keller, and D. Kosower have reported on
a method for extracting pdf distributions with quantitative estimates
of pdf uncertainties.
In effect their method~\cite{GKK2} uses a Bayesian approach that
integrates sets of pdf parameterizations over properly weighted samples 
of experimental uncertainties to produce a set of pdf models consistent
with the data within a given confidence level.  The basic method
may be extended to use data with arbitrary error distributions and
correlations.  For such methods to 
function reliably the experiments must be able to provide detailed 
descriptions of their error distributions.  
Giele et al. make a distinction between `errors on the data' and `errors 
on the theory' for estimation of the most likely pdf models.  In this context
we take only uncertainties depending directly on the number of events in a bin 
as `errors on the data'.  Other typical sources of uncertainty, luminosity,
energy scale, resolution, etc., may be treated as `errors on the theory'
in that they are in some sense independent of the statistical precision of
the data and represent how an underlying, {\em true}, distribution may be
distorted by observation in the experiment.

As a result of these dialogs, we have revisited the D\O\ response 
uncertainty (our largest uncertainty in the inclusive jet 
cross section measurement) from Fig.~\ref{d0scale} and generated a 
sampling of the probability density function for distributions in 
it parameters.  This probability density
function contains all the relevant information on both the shape
of the uncertainty distribution and point-to-point correlations.  It is 
clear that providing such information is a significant
enhancement from traditional methods of summarizing experimental 
uncertainties.   Optimum utilization of the data demands a
detailed understanding and reporting of its associated uncertainties.  
Through our fruitful discussions in this workshop, 
we look forward to setting an example for the reporting of experimental 
uncertainties and to fully exploiting our cross section data in pdf analyses 
in the near future.

\newpage
\setcounter{section}{0}
\def\simgt{\rlap{\lower 3.5 pt \hbox{$\mathchar \sim$}} \raise 1pt \hbox {$>$}}
\def\simlt{\rlap{\lower 3.5 pt \hbox{$\mathchar \sim$}} \raise 1pt \hbox {$<$}}

 \newcommand{\st}{\tilde{t}}
 \newcommand{\stb}{\bar{\tilde{t}}}
 \newcommand{\sq}{\tilde{q}}
 \newcommand{\sqb}{\bar{\tilde{q}}}
 \newcommand{\gl}{\tilde{g}}
 \newcommand{\gau}{\tilde{\chi}}
 \newcommand{\MS}{\mbox{$\overline{\rm MS}$}}
 \newcommand{\MSSM}{\mbox{$\MSSM}$}
 \newcommand{\SUSY}{\mbox{${\cal SUSY}$}}
 \newcommand{\gev}{\; \mbox{Ge$\!$V}}
 \newcommand{\tgb}{\mbox{$\tan\beta$}}

 \newcommand{\zp}[3]{{Z.\ Phys.} {\bf #1} (19#2) #3}
 \newcommand{\np}[3]{{Nucl.\ Phys.} {\bf #1} (19#2)~#3}
 \newcommand{\pl}[3]{{Phys.\ Lett.} {\bf #1} (19#2) #3}
 \newcommand{\pr}[3]{{Phys.\ Rev.} {\bf #1} (19#2) #3}
 \newcommand{\prl}[3]{{Phys.\ Rev. Lett.} {\bf #1} (19#2) #3}
 \newcommand{\prep}[3]{{\sl Phys. Rep.} {\bf #1} (19#2) #3}
 \newcommand{\fp}[3]{{\sl Fortschr. Phys.} {\bf #1} (19#2) #3}
 \newcommand{\nc}[3]{{\sl Nuovo Cimento} {\bf #1} (19#2) #3}
 \newcommand{\ijmp}[3]{{\sl Int. J. Mod. Phys.} {\bf #1} (19#2) #3}
 \newcommand{\ptp}[3]{{\sl Prog. Theo. Phys.} {\bf #1} (19#2) #3}
 \newcommand{\sjnp}[3]{{\sl Sov. J. Nucl. Phys.} {\bf #1} (19#2) #3}
 \newcommand{\cpc}[3]{{\sl Comp. Phys. Commun.} {\bf #1} (19#2) #3}
 \newcommand{\mpl}[3]{{\sl Mod. Phys. Lett.} {\bf #1} (19#2) #3}
 \newcommand{\cmp}[3]{{\sl Commun. Math. Phys.} {\bf #1} (19#2) #3}
 \newcommand{\jmp}[3]{{\sl J. Math. Phys.} {\bf #1} (19#2) #3}
 \newcommand{\nim}[3]{{\sl Nucl. Instr. Meth.} {\bf #1} (19#2) #3}
 \newcommand{\el}[3]{{\sl Europhysics Letters} {\bf #1} (19#2) #3}
 \newcommand{\epj}[3]{{\sl Eur. Phys. J.} {\bf #1} (19#2) #3}
 \newcommand{\ap}[3]{{\sl Ann. of Phys.} {\bf #1} (19#2) #3}
 \newcommand{\jetp}[3]{{\sl JETP} {\bf #1} (19#2) #3}
 \newcommand{\jetpl}[3]{{\sl JETP Lett.} {\bf #1} (19#2) #3}
 \newcommand{\acpp}[3]{{\sl Acta Physica Polonica} {\bf #1} (19#2) #3}
 \newcommand{\vj}[4]{{\sl #1~}{\bf #2} (19#3) #4}
 \newcommand{\ej}[3]{{\bf #1} (19#2) #3}
 \newcommand{\vjs}[2]{{\sl #1~}{\bf #2}}
 \newcommand{\hep}[1]{{hep--ph/}{#1}}
 \newcommand{\desy}[1]{{DESY-Report~}{#1}}

\begin{center}
\section*{PARTON DENSITY UNCERTAINTIES AND SUSY PARTICLE PRODUCTION}
\end{center}

T. Plehn$^{a,}$\footnote{Supported in
part by DOE grant DE-FG02-95ER-40896 and in part by the University of
Wisconsin Research Committee with funds granted by the Wisconsin
Alumni Research Foundation}
and M. Kr\"amer$^{b,}$\footnote{Supported in part by the EU Fourth Framework Programme
`Training and Mobility of Researchers', Network `Quantum
Chromodynamics and the Deep Structure of Elementary Particles',
contract FMRX-CT98-0194 (DG 12 - MIHT)}
\\ 

a) Department of Physics, University of Wisconsin, Madison WI 53706, USA; 
b) Department of Physics and Astronomy,
University of Edinburgh, Edinburgh EH9 3JZ, Scotland.\\
       
\begin{center}
                Abstract
\end{center}
  Parton densities are important input parameters for SUSY particle
  cross section predictions at the Tevatron. Accurate theoretical
  estimates are needed to translate experimental limits, or measured
  cross sections, into SUSY particle mass bounds or mass
  determinations.
 We study the PDF dependence of next-to-leading order cross section predictions, with
 emphasis on a new set of parton densities~\cite{gkk}. We compare the
 resulting error to the remaining theoretical uncertainty due to
 renormalization and factorization scale variation in next-to-leading
 order SUSY-QCD.

\section{Introduction}

The search for supersymmetric particles is among the most important
endeavors of present and future high energy physics. At the upgraded
$p\bar p$ collider Tevatron, the searches for squarks and gluinos (and
especially the lighter stops and sbottoms), as well as for the weakly
interacting charginos and neutralinos, will cover a wide range of the
MSSM parameter space~\cite{CCRFM,stop_cdf}.\smallskip

The hadronic cross sections for the production of SUSY particles
generally suffer from unknown theoretical errors at the Born
level~\cite{LO}. For strongly interacting particles the dependence on
the renormalization and factorization scale has been used as a measure
for this uncertainty, leading to numerical ambiguities of the order of
100$\%$. For Drell-Yan type weak production processes the dependence
on the factorization scale is mild. However, a comparison of leading
and next-to-leading order predictions~\cite{gaunlo} reveals that the
impact of higher-order corrections is much larger than the estimate
through scale variation would have suggested. The use of
next-to-leading order calculations~\cite{gaunlo,sqgl,stops} is thus
mandatory to reduce theoretical uncertainties to a level at which one
can reliably extract mass limits from the experimental data.\smallskip

In addition to the scale ambiguity and the impact of perturbative
corrections beyond next-to-leading order, hadron collider cross
section are subject to uncertainties coming from the parton densities
and the associated value of the strong coupling.  Previously, the only
way to estimate the PDF errors was to compare the best-fit results
from various global PDF analyses. Clearly, this is not a reliable
measure of the true uncertainty. As a first step towards a more
accurate error estimate, the widely used sets CTEQ~\cite{cteq} and
MRST~\cite{mrst} now offer different variants of PDF sets, {\sl e.g.}
using different values of the strong coupling constant. In this letter
we compare their predictions to the preliminary GKK parton
densities~\cite{gkk}, which provide a systematic way of propagating
the uncertainties in the PDF determination to new observables.

\section{Stop Pair Production}

For third generation squarks the off-diagonal left-right mass matrix
elements do not vanish, but lead to mixing stop (and sbottom) states.
The lighter mass eigenstate, denoted as $\st_1$, is expected to be the
lightest strongly interacting supersymmetric particle. Moreover, its
pair production cross section, to a very good approximation, only
depends on the stop mass, in contrast to the light flavor squark
production. Nevertheless, considering the different decay channels
complicates the analyses~\cite{stop_cdf,stop_dec}. At the Tevatron the
fraction of stops produced in quark-antiquark annihilation and in
gluon fusion varies strongly with the stop mass. Close to threshold
the valence quark luminosity is dominant, but for lower masses a third
of the hadronic cross section can be due to incoming
gluons~\cite{stops}.

\begin{figure}[hb]
\epsfig{file=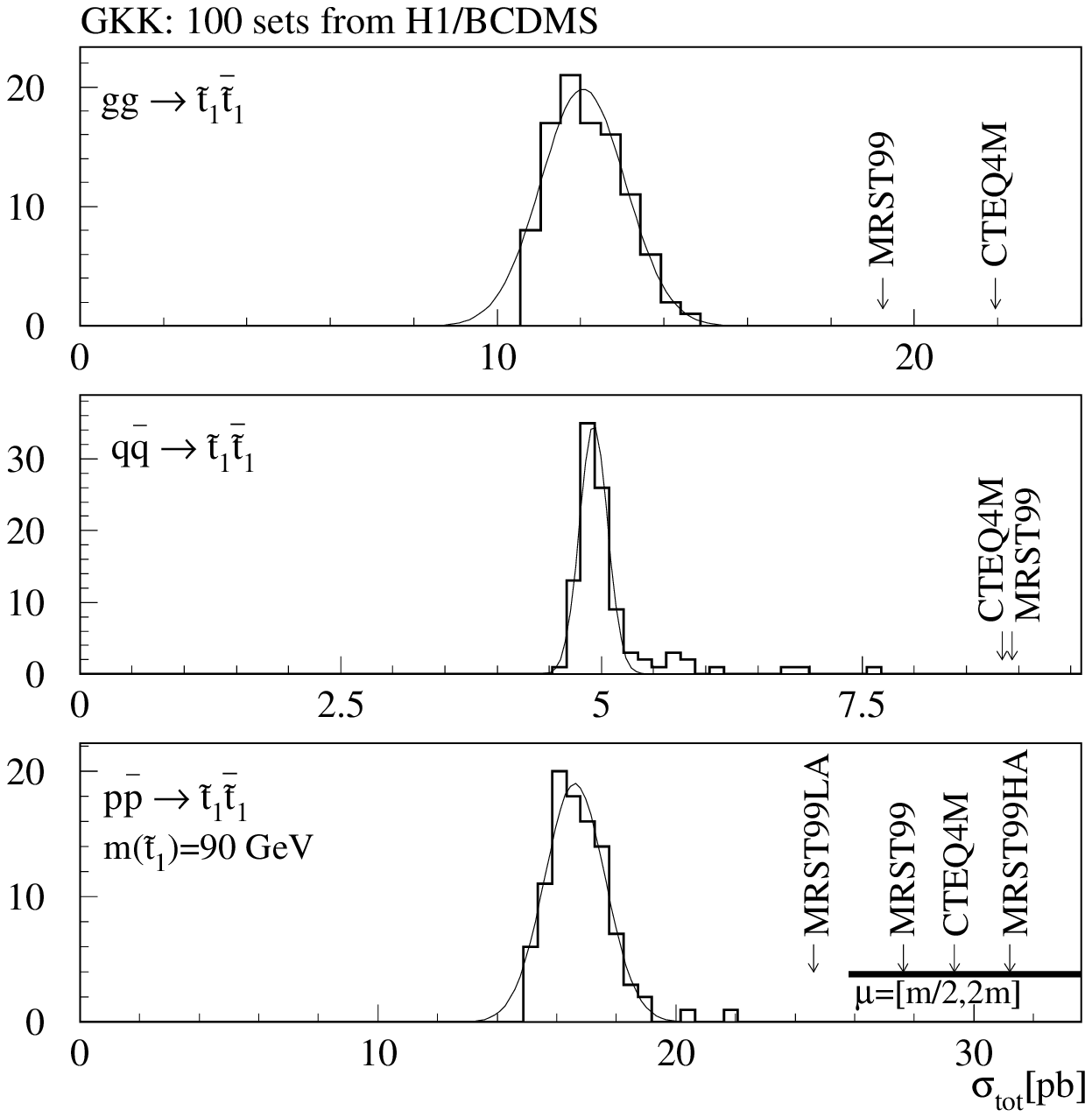,width=8.2cm} \vspace*{-10mm}
\caption[]{NLO production cross section for a light stop. The 
  Gaussian fits the preliminary GKK parton densities. The
  renormalization/factorization scale is varied around the average 
  final state mass.}
\label{fg:stop90}
\end{figure}

In Figure~\ref{fg:stop90} we compare the total $\st_1$-pair production
cross sections for three sets of parton densities: only for incoming
quarks do the CTEQ4 and MRST99 results lie on top of each other. For
gluon fusion the corresponding cross sections differ by $\sim 10
\%$. The GKK set centers around a significantly smaller value. This is
in part due to the low average value $\langle \alpha_s({\rm GKK})
\rangle =0.108$, which is expected to increase after including more
experimental information in the GKK analysis.  But even the normalized
cross section $\sigma/\alpha_s^2$ is still smaller by $35\%$ compared
to CTEQ4 and MRST99 because of the entangled fit of the strong
coupling constant and the parton densities. However, the width of the
Gaussian fit to the GKK results gives an uncertainty of $2\%$ and
$8\%$ for the quark-antiquark and gluon fusion channel, similar to the
difference between CTEQ4 and MRST99.

\begin{figure}[hb]
\vspace*{-8mm}
\epsfig{file=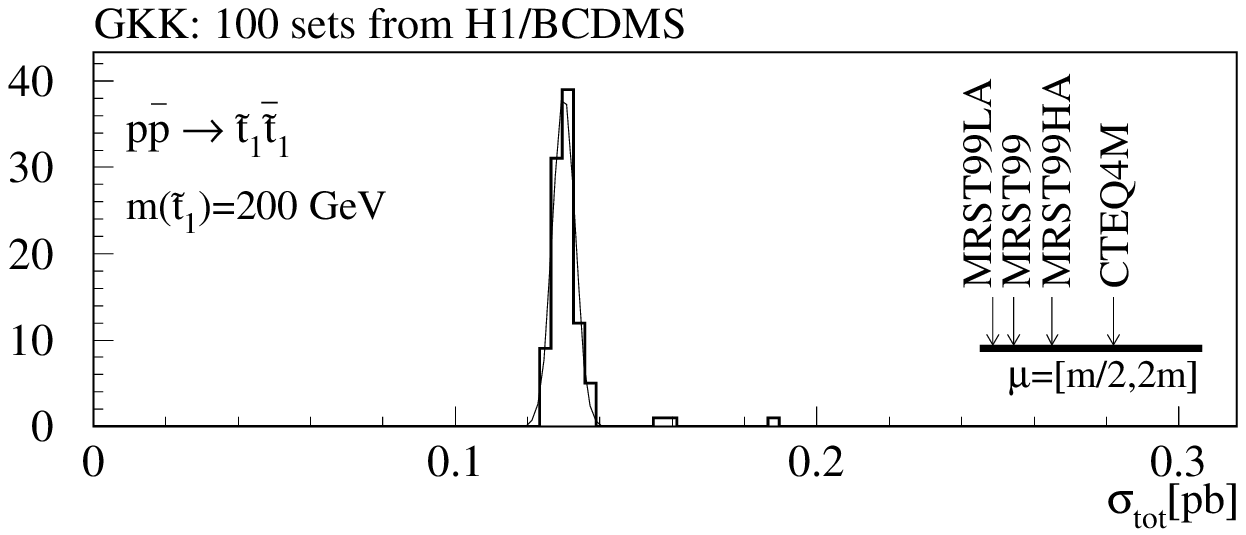,width=8.2cm} \vspace*{-10mm}
\caption[]{NLO production cross section for a heavier stop, 
  dominated by incoming valence quarks. The 
  Gaussian fits the preliminary GKK parton densities. The
  renormalization/factorization scale is varied around the average
  final state mass.}
\label{fg:stop200}
\end{figure}

For heavier stop particles, Figure~\ref{fg:stop200}, the gluon
luminosity is strongly suppressed due to the large final state mass,
and mainly valence quarks induced processes contribute to the cross
section. The Gaussian distribution of the GKK results has a width of
$\sim 2\%$. The comparably large difference between CTEQ4 and MRST99
is caused by the small fraction of gluon induced processes, since the
gluon flux at large values of $x$ differs for CTEQ4 and MRST99 by
approximately $40 \%$.

\section{Chargino/Neutralino Production}

The production of charginos and neutralinos at the Tevatron is
particularly interesting in the trilepton $\gau_2^0
\gau_1^\pm$ and the light chargino $\gau_1^+ \gau_1^-$ channels~\cite{trilepton}.
The next-to-leading order
corrections to the cross sections~\cite{gaunlo} reduce the
factorization scale dependence, but at the same time introduce a small
renormalization scale dependence. A reliable estimate of the
theoretical error from the scale ambiguity will thus only be possible
beyond next-to-leading order.
\smallskip

\begin{figure}[t]
\epsfig{file=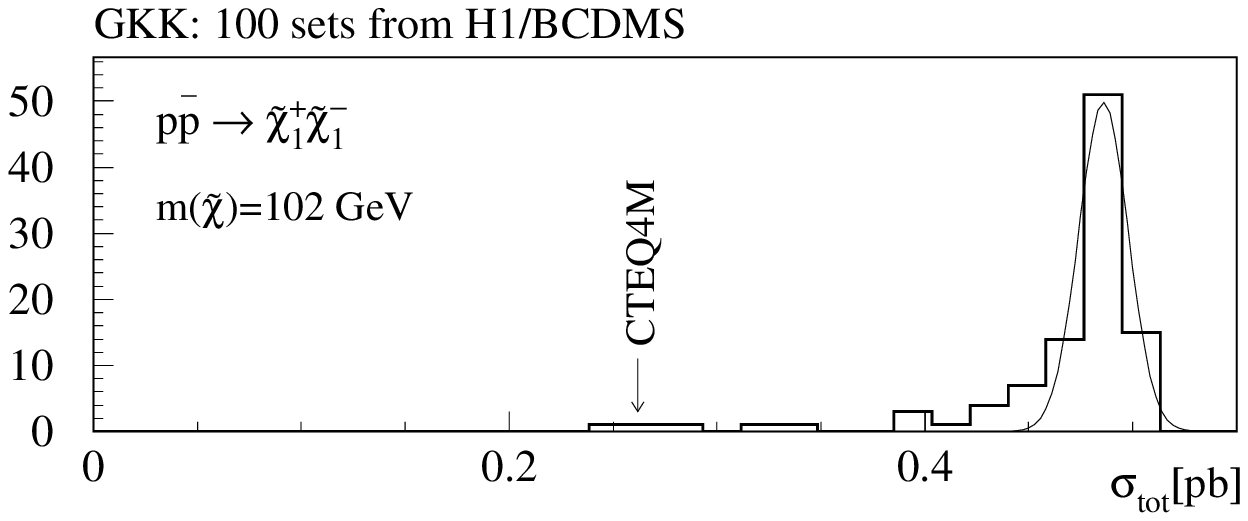,width=8.2cm} \vspace*{-10mm}
\caption[]{NLO production cross section for the chargino 
  channel $\gau_1^+ \gau_1^-$. The 
  Gaussian fits the preliminary GKK parton densities.}
\label{fg:trilepton}
\end{figure}

The Gaussian distribution of the GKK parton densities for 
light chargino pairs is shown in
Figure~\ref{fg:trilepton}. For the chosen mSUGRA parameters ($m_0=100
\gev, A_0=300 \gev, m_{1/2}=150 \gev$) the width is $\sim 2 \%$,
as one would expect from the quark-antiquark channel of the stop
production. 
But in contrast to the stop production, where all quark
luminosities add up, the chargino/neutralino channels can be extremely sensitive to
systematic errors in different parton densities due to destructive
interference between $s$ and $t$ channel diagrams. The total trilepton
cross section for example will therefore be a particular challenge for
a reliable error estimate.

\section{Outlook} 

We have briefly reviewed the status of the theoretical error analysis
of SUSY cross sections at the Tevatron. For strongly interacting final
state particles, the inclusion of next-to-leading order corrections
reduces the renormalization and factorization scale ambiguity to a
level $\simlt\; 10\%$ where the size of the PDF errors becomes
phenomenologically relevant. We have compared different recent PDF
sets provided by the CTEQ~\cite{cteq} and MRST~\cite{mrst}
collaborations to the preliminary GKK parton densities~\cite{gkk}.
The large spread in the cross section predictions can mainly be
attributed to the low average value of the strong coupling associated
with the GKK sets. We expect this spread to be reduced once more data
have been included in the GKK analysis and the corresponding average
value of the strong coupling becomes closer to the world average.  For
weak supersymmetric Drell-Yan type processes~\cite{gaunlo} the scale
dependence at NLO cannot serve as a measure for the theoretical error
since the renormalization scale dependence is only introduced at
NLO. The PDF induced errors for {\sl e.g.} the case of
$\gau_1^+ \gau_1^-$ production are small; however, interference
effects between the different partonic contributions must be taken
into account.

The recently available variants of PDF sets provided by CTEQ and MRST
and, in particular, the GKK parton densities allow for the first time
a systematic exploration of PDF uncertainties for the prediction of
SUSY particle cross sections. The preliminary GKK results do not yet
allow a conclusive answer, but they point the way towards a complete
and reliable error analysis in the near future.

\vspace*{2mm}

{\bf Acknowledgments} 

The authors want to thank W.\ Beenakker, R.\ H\"opker, M.\ Spira,
P.M.\ Zerwas, and M.\ Klasen for the collaboration during different
stages of this work. Furthermore we are grateful to S.\ Keller who
triggered this analysis by making his preliminary set of parton
densities available to us.

\newpage
\setcounter{section}{0}
\def\beq{\begin{equation}}
\def\eeq{\end{equation}}
\def\beqa{\begin{eqnarray}}
\def\eeqa{\end{eqnarray}}

\def\a {{\rm f}}
\def\e{r}
\def\g{\xi}
\def\w{\rho}
\def\y{\eta}
\def\Qg{Q_{\rm gap}}

\section*{SOFT-GLUON RESUMMATION AND PDF THEORY UNCERTAINTIES}

George Sterman and Werner Vogelsang\footnote{This work was supported
in part by the National Science Foundation, grant PHY9722101.}
\\ 

C.N.\ Yang Institute for Theoretical Physics SUNY at Stony Brook,
Stony Brook, NY 11794-3840, USA.\\

\begin{center}
	Abstract
\end{center}

Parton distribution functions are determined by the comparison of
finite-order calculations with data.  We briefly discuss the interplay
of higher order corrections and PDF determinations, and the use of
soft-gluon resummation in global fits.

\section{Factorization \& the nlo model}

A generic inclusive
cross section for the process $A+B\rightarrow F+X$
with observed final-state system $F$, of total mass $Q$, can be expressed as
\beqa
Q^4\, {d\sigma_{AB\rightarrow FX}\over dQ^2}
&=&
\phi_{a/A}(x_a,\mu^2)\; \otimes\;  \phi_{b/B}(x_b,\mu^2)
\nonumber\\
&\ &  \quad\quad  \otimes\;
{\hat \sigma}_{ab\rightarrow FX}\left (z,Q,\mu\right)\, ,
\label{basicfact}
\eeqa
with $z=Q^2/x_ax_bS$. The
$\hat{\sigma}_{ab}$ are partonic hard-scattering functions,
$
{\hat\sigma}=\sigma_{\rm Born}+(\alpha_s(\mu^2)/\pi){\hat\sigma}^{(1)}
+\dots\, .
$
They are known to NLO for most processes in the
standard model and its popular extensions.  Corrections
begin with higher, uncalculated orders in the hard scattering,
which respect  the
form of Eq.\ (\ref{basicfact}).
The discussion is simplified in terms of moments
with respect to $\tau=Q^2/S$,
\beqa
\tilde\sigma_{AB\rightarrow FX} &=& \int_0^1 d\tau\;
\tau^{N-1}\ Q^4\, {d\sigma_{AB\rightarrow FX} / dQ^2}
\nonumber\\
&& \hspace{-23mm}
= \sum_{a,b}\, \tilde \phi_{a/A}(N,\mu^2)\; \tilde\sigma_{ab\rightarrow 
FX}(N,Q,\mu)\;
\tilde\phi_{b/B}(N,\mu^2)\, ,
\label{momentact}
\eeqa
where the moments of the $\phi$'s and 
$\hat\sigma_{ab\rightarrow FX}$ are defined similarly.

Eqs.\ (\ref{basicfact}) and (\ref{momentact}) are starting-points for
both the determination and the application of
parton distribution functions (PDFs), $\phi_{i/H}$,
using 1-loop $\hat\sigma$'s \cite{mrst99,grv98,cteq5}
We may think of this collective enterprise as an ``NLO model" for
the PDFs, and for hadronic hard scattering in general.
For precision applications we ask how well we
really know the PDFs \cite{disPDF,uncerglue,bayseanpdfs}.
Partly this is a question of how
well data constrain them, and partly it is a question of
how well we {\it could} know them, given finite-order
calculations in Eqs.\ (\ref{basicfact}) and (\ref{momentact}).
We will not attempt here to assign error estimates
to theory.  We hope, however, to give a sense of how
to distinguish ambiguity from uncertainty, and how
our partial knowledge of higher orders can reduce
the latter.

\section{Uncertainties, schemes \& scales}

It is not obvious how to quantify
a ``theoretical uncertainty",
since the idea seems to require us to estimate corrections that
we haven't yet calculated.  We do not think an
unequivocal definition is possible, but we can try at least to
clarify the concept, by
considering a hypothetical set of nucleon PDFs
determined from DIS data alone \cite{disPDF}.  To make such a
determination, we would invoke isospin symmetry
to reduce the set of PDF's to those of the proton, $\phi_{a/P}$,
and then measure a set of singlet and
nonsinglet structure functions, which we denote $F^{(i)}$.
Each factorized structure function may be written
in moment space as
\beq
\tilde F^{(i)}(N,Q) = \sum_a \tilde C^{(i)}_a(N,Q,\mu)\, \tilde 
\phi_{a/P}(N,\mu^2)\, ,
\label{disfact}
\eeq
in terms of which we may solve for the parton
distributions by inverting the matrix $\tilde{C}$,
\beq
\tilde \phi_{a/P}(N,\mu^2) = \sum_i \, \tilde C^{-1}{}^{(i)}_a(N,Q,\mu)\
\tilde F^{(i)}(N,Q)\, .
\label{solvephi}
\eeq
With ``perfect" $\tilde F$'s at fixed $Q$,
and with a specific approximation for the coefficient functions,
we could solve for the moment-space distributions numerically,
without the need of a parameterization.  In a world of
perfect data, but of incompletely known coefficient functions, uncertainties
in the parton distributions would be entirely due to
the ``theoretical" uncertainties of the $C$'s:
\beq
\delta \tilde \phi_{a/P}(N,\mu) = \sum_i \, \delta \tilde 
C^{-1}{}^{(i)}_a(N,Q,\mu)\
\tilde F^{(i)}(N,Q)\, .
\label{delphidelC}
\eeq
Our question now becomes, how well do we know the $C$'s?
In fact this is a subtle question, because
the coefficient functions depend on choices of scheme
and scale.  

Factorization schemes are procedures
for defining coefficient functions perturbatively.
For example, choosing for $F_2$ the LO (quark) coefficient function
in Eq.\ (\ref{solvephi}) defines a DIS scheme
(with $\tilde{C}$ independent of $\mu$, which is then to
be taken as $Q$ in $\tilde{\phi}$).  Computing
the $C$'s from partonic cross sections by minimal subtraction
to NLO defines an NLO $\overline{\rm MS}$ scheme, and so on.
Once the choices of $C$'s and $\mu$
are made, the PDF's are defined
uniquely.  

Evolution in an $\overline{\rm MS}$
or related scheme, enters through
\beqa
\mu{d\over d\mu} \tilde \phi_{a/H}(N,\mu^2) \hspace*{-2mm} 
&=&\hspace*{-2mm}  -  \Gamma_{ab}(N,
\alpha_s(\mu^2))\, 
\tilde \phi_{b/H}(N,\mu^2) \nonumber \\
\mu{d\over d\mu}\tilde C^{(i)}_c(N,Q,\mu)\hspace*{-2mm}  &=&
\hspace*{-2mm}   \tilde C^{(i)}_d(N,Q,\mu)\, 
\Gamma_{dc}(N,\alpha_s(\mu^2))\, . 
\label{evol}
\eeqa
In principle, by Eq.\ (\ref{evol}), the scale-dependence of the $C_a^{(i)}$
exactly cancels that of the PDFs in Eq.\ (\ref{disfact}) and, by
extension, in Eq.\ (\ref{basicfact}).
This cancelation, however, requires that each $C$
and the anomalous dimensions $\Gamma$ be known
to all orders in perturbation theory.

To eliminate $\mu$-dependence
up to order $\alpha_s^{n+1}$, we need $\hat\sigma$ to order
$\alpha_s^n$ and the $\Gamma_{ab}$ to $\alpha_s^{n+1}$.
One-loop (NLO) QCD corrections to hard scattering require
two-loop splitting functions, which are known.  The complete
form of the  NNLO splitting functions, is still somewhere over
the horizon \cite{nnlo}.  Even when these are
known, it will take some time before more than a few hadronic
hard scattering functions are known at NNLO.

We can clarify the role of higher orders by relating  structure functions
at two scales,$Q_0$ and $Q$. Once we have measured
$F(N,Q_0)$, we may predict $F(N,Q)$ in terms of the relevant
anomalous dimensions and coefficient functions by
\beqa
F(N,Q) &=& F(N,Q_0)\ {\rm e}^{\int_{Q_0}^{Q} {d\mu'\over\mu'}\,
\Gamma(N,\alpha_s(\mu'{}^2))}
\nonumber\\
&\ & \quad \times \left [ {\tilde{C}(N,Q,Q)\over 
\tilde{C}(N,Q_0,Q_0)}\right]\, .
\label{predict}
\eeqa
This prediction, formally independent of PDFs
{\it and} independent of the factorization scale,
has corrections from the next, still uncalculated
order in the anomalous dimension and in the ratio of
coefficient functions.  The asymptotic freedom of QCD
gives a special role to LO: only the one-loop contribution
to $\Gamma$ diverges with $Q$ in the exponent, and contributes
to the leading, logarithmic scale breaking.  NLO
corrections already decrease as the inverse of the logarithm
of $Q$, NNLO as two powers of the log.  Thus, the theory
is self-regulating towards high energy,
where dependence on uncalculated pieces in
the coefficients and anomalous dimensions becomes less and
less important.

The general successes of the NLO model strongly
suggest that relations like (\ref{predict})
are well-satisfied for a wide range of observables
and values of $N$ (or $x$) in DIS and other processes.
This does not mean, however, that we have no knowledge of,
or use for, information from higher orders. In particular, 
near $x=1$ PDFs are rather poorly known \cite{cteqlargex}.
At the same time, the 
ratio of $C$'s depends on $N$, and if $\alpha_s\ln N$ is large, it becomes
important to control higher-order dependence on $\ln N$. This is a task 
usually referred to as resummation, to which we now turn.

\section{Resummation}

\begin{figure}[t]
\epsfig{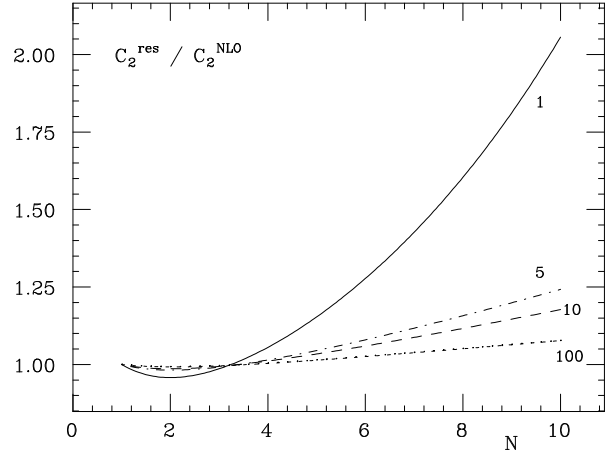}
\vspace{-0.6cm}
\caption{Ratio of Mellin-$N$ moments of resummed and NLO 
$\overline{\rm MS}$-scheme quark coefficient functions for $F_2$.
The numbers denote the value of $Q^2$ in GeV$^2$. We have
chosen $\mu=Q$.}
\label{cratio}
\end{figure}

Let us continue our discussion of DIS, describing
what is known about the $N$-dependence of the coefficient
functions $C$, as a step toward understanding
the role of higher orders.  Specializing again for simplicity to
nonsinglet or valence, the resummed coefficient function may
be written as \cite{oldDY,Cea1PI}
\beq
\tilde C^{\rm res}(N,Q,\mu) = \tilde C^{\rm NLO}_{sub}(N,Q,\mu)+
C_{\delta}^{\rm DIS} \, {\rm e}^{E_{\rm DIS}(N,Q,\mu)},
\label{Cresum}
\eeq
where ``{\it sub}" implies a subtraction on
$\tilde{C}^{\rm NLO}$ to keep $\tilde{C}^{\rm res}$ exact at order
$\alpha_s$, and where $C_{\delta}^{\rm DIS}$ corresponds to the NLO 
$N$-independent (``hard virtual'') terms.  
The exponent resums logarithms of $N$:
\beqa
E_{\rm DIS}(N,Q,\mu) &=& \\
&\ & \hspace{-30mm}
\int_{Q^2/\bar{N}}^{\mu^2}{d\mu'{}^2\over\mu'{}^2}
\Big[ A(\alpha_s(\mu'{}^2))\ln( \bar{N}\mu'{}^2/Q^2)
  + B(\alpha_s(\mu'{}^2))\Big], \nonumber
\label{Edisdef}
\eeqa
with $\bar{N} \equiv N {\rm e}^{\gamma_E}$, and with 
\beqa
A (\alpha_s) &=& \nonumber\\
&\ & \hspace{-10mm}{\alpha_s\over \pi}\, C_F\, 
\left[ 1+  {\alpha_s\over 2\pi} \left( C_A\left({67\over 18}-
{\pi^2\over 6} \right)-{10\over 9}T_F\right)\right] 
\nonumber\\
B (\alpha_s) &=& {3\over 2}C_F{\alpha_s\over 2\pi}\, \, .
\label{ABdef}
\eeqa
Eq.\ (\ref{Edisdef}) is accurate to leading (LL) and next-to-leading 
logarithms (NLL)
in $N$ in the exponent: $\alpha_s^m\ln^{m+1}N$ and $\alpha_s^m\ln^mN$,
respectively.   The $N$ dependence of the ratio
$\tilde{C}_2^{\rm res}(N,Q,Q) / \tilde{C}_2^{\rm NLO}(N,Q,Q)$
is shown in Fig.\ \ref{cratio},
with $Q^2$ = 1, 5, 10, 100 GeV$^2$.
At $N=1$ the ratio is unity.  It is less than unity for
moderate $N$, but then begins to rise, with a slope that
increases strongly for small $Q$.   At low $Q^2$
and large $N$, higher orders can be quite important.
What does this  mean for PDFs?   We can certainly refit
PDFs with resummed coefficient functions, and 
we see that the high moments of such PDFs are likely
to be quite different from those from NLO fits.  

To get a sense of how such an NLL/NLO-$\overline{\rm MS}$ scheme
might differ from a classic NLO-$\overline{\rm MS}$
scheme, we resort to a model set of resummed
distributions, determined as follows.
We define valence PDFs in the resummed scheme
by demanding that their contributions to $F_2$ match those of the 
corresponding NLO
valence PDFs at a fixed $Q=Q_0$, which is ensured by
\beq
\tilde{\phi}^{\rm res}(N,Q_0^2)=\tilde{\phi}^{\rm NLO}(N,Q_0^2)\;
{ \tilde{C}_2^{\rm NLO}(N,Q_0,Q_0) \over 
\tilde{C}_2^{\rm res}(N,Q_0,Q_0)}\, .
\label{fresdef}
\eeq
Using the resummed parton densities from Eq.\ (\ref{fresdef}),
we can generate the ratios 
$F_2^{\rm res}(x,Q)/F_2^{\rm NLO}(x,Q)$.

\begin{figure}[t]
\epsfig{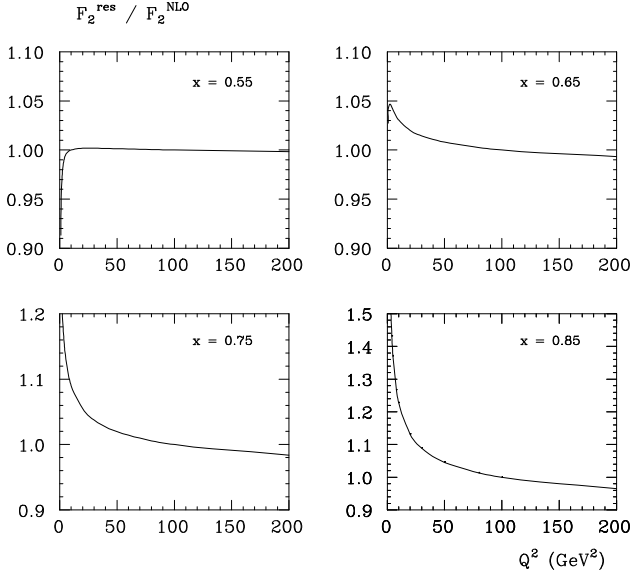}
\vspace{-0.8cm}
\caption{Ratio of the valence parts of the resummed and NLO proton 
structure function $F_2 (x,Q^2)$, as a function of $Q^2$ for various
values of Bjorken-$x$. For $F_2^{\rm res}$, the `resummed' parton densities 
have been determined through Eq.~(\ref{fresdef}).}
\label{fratio}
\end{figure}

The result of this test, picking $Q_0^2=100$ GeV$^2$
is shown in Fig.\ \ref{fratio}, for the valence $F_2(x,Q)$ of
the proton, with $x$ = 0.55, 0.65, 0.75 and 0.85.
The NLO distributions were those
of \cite{grv98}, and the inversion of moments
was performed as in \cite{minimal}.  The effect of
resummation is moderate for most $Q$.
At small values of $Q$, and large $x$, the resummed
structure function shows a rather sharp upturn. 
One also finds a gentle decrease toward very large $Q$ \cite{CM}.
We could interpret this difference as the
{\it uncertainty} in the purely NLO valence PDFs 
implied by resummation.

 From this simplified example, we can already see that the
use of resummed coefficient functions is not likely to make
drastic differences in global fits to PDFs based on DIS data, at least
so long as the region of small $Q^2$, of 10 GeV$^2$ or below, is avoided
at very large $x$.  At the same time, it is clear that a resummed
fit will make some difference at larger $x$, where PDFs are
not so well known. We stress that a full global fit
will be necessary for complete confidence.

\section{Resummed hadronic scattering}

Processes other than DIS play an important role
in global fits, and in any case are of paramount phenomenological interest.
Potential sources of large corrections can be identified
quite readily in Eq.\ (\ref{momentact}).  At higher orders, 
factors such as $\alpha_s\ln^2N$, can
be as large as unity over the physically relevant range of 
$z$ in some processes.
In this case, they, and their scale dependence can be competitive with
NLO contributions.
Since they make up well-defined parts of the correction at each higher
order, however, it is possible to resum them.  
To better determine PDFs in regions of phase space where
such corrections are important, we may
incorporate resummation in the hard-scattering functions that determine
PDFs.

The Drell-Yan cross section is the benchmark for
the resummation of logs of $1-z$, or equivalently, logarithms of the
moment variable $N$ \cite{oldDY},
\beqa
\hat{\sigma}_{q\bar q}^{\rm DY}(N,Q,\mu) &=& \sigma_{\rm Born}(Q)\;
C_{\delta}^{\rm DY} \, {\rm e}^{E_{\rm DY}(N,Q,\mu)}
\nonumber \\
&\ & \hspace{10mm} + {\cal O}(1/N)\, .
\label{dyexp}
\eeqa
The exponent is given in the $\overline{\rm MS}$ scheme by
\beqa
E_{\rm DY}(N,Q,\mu) 
&=&  2\int^{\mu^2}_{Q^2/\bar{N}^2}{d\mu'{}^2\over\mu'{}^2}
A(\alpha_s(\mu'{}^2))\ln\bar{N}
\nonumber\\ 
&\ & \hspace{-15mm}
+ 2\int^{Q^2}_{Q^2/\bar{N}^2}{d\mu'{}^2\over\mu'{}^2}
A(\alpha_s(\mu'{}^2))\ln\left({\mu'\over Q}\right),
\label{Edydef}
\eeqa
with $A$ as in Eq.\ (\ref{ABdef}), and where we have
exhibited the dependence on the factorization scale,
setting the renormalization scale to $Q$.  Just as in Eq.\ (\ref{Edisdef})
for DIS,
Eq.\ (\ref{Edydef}) resums all leading and next-to-leading 
logarithms of $N$.

It has been noted
in several phenomenological applications that threshold resummation,
and even fixed-order expansions based upon it,
significantly reduce sensitivity to the factorization scale
\cite{scalered}.
To see why, we
rewrite the moments of the Drell-Yan cross section 
in resummed form as
\beqa
\sigma^{\rm DY}_{AB}(N,Q)
&& \nonumber\\
&& \hspace{-20mm}
= \sum_q\;
  \phi_{q/A}(N,\mu)\;
\hat{\sigma}_{q\bar q}^{\rm DY}(N,Q,\mu)\;
\phi_{\bar q/B}(N,\mu)
\nonumber\\
&& \nonumber\\
&& \hspace{-20mm} =
\sum_q\;
\phi_{q/A}(N,\mu)\; e^{E_{\rm DY}(N,Q,\mu)/2}
\sigma_{\rm Born}(Q) \, \, C_{\delta}^{\rm DY}
\nonumber\\
&& \hspace{-15mm} \times\;
\phi_{\bar q/B}(N,\mu)\; e^{E_{\rm DY}(N,Q,\mu)/2} + {\cal 
O}({1/ N})\, .
\eeqa
The exponentials compensate for the $\ln N$ part of the
evolution of the parton distributions, and the
  $\mu$-dependence of the resummed expression
is suppressed by a power of the moment variable,
\beq
\mu{d\over d\mu}\left[\, \phi_{q/A}(N,\mu)\; e^{E_{\rm 
DY}(N,Q,\mu)/2}\; \right]
=   {\cal O}({1/ N})\, .
\eeq
This surprising relation holds because the function
$A(\alpha_s)$ in Eq.\ (\ref{ABdef}) equals the
residue of the $1/(1-x)$ term in the splitting function $P_{qq}$.
Thus, the remaining $N$-dependence in a resummed cross section
still begins at order $\alpha_s^2$, but the part associated
with the $1/(1-x)$ term in the
splitting functions has been canceled to all orders.
Of course, the importance of the remaining sensitivity to $\mu$
depends on the kinematics and the process.   In addition,
although resummed cross sections can be made independent
of $\mu$ for all $\ln N$,
they are still uncertain at next-to-next-to leading logarithm
in $N$, simply because we do not know the function $A$ at three loops.
Notice that none of these results depends on using PDFs
from a resummed scheme, because
$\overline{\rm MS}$ PDFs, whether resummed or NLO, evolve the same way.
The remaining, uncanceled dependence on the scales leaves room for an educated
use of scale-setting arguments \cite{scales}.  The connection
between resummation and the elimination of scale
dependence has also been emphasized in \cite{maxwell}.

Scale dependence aside, 
can we in good conscience combine resummed hard scattering
functions in Eq.\ (\ref{basicfact}) with PDFs from an NLO scheme?
This wouldn't make much sense if resummation
significantly changed the coefficient functions
with which the PDFs were originally fit.
As Fig.\ \ref{fratio}
shows, however, this is unlikely to be the case for DIS at moderate $x$.
Thus, it makes sense to apply threshold resummation with NLO PDFs
to processes and regions of phase space where there is reason
to believe that logs are more important at higher orders than
for the input data to the NLO fits.

At the same time, a set of
fits that includes threshold resummation in their hard-scattering
functions can be made \cite{Cea1PI}, and their comparison to
strict NLO fits would be quite interesting.  Indeed, such a comparison
would be a new measure of the influence of higher orders.
A particularly interesting example might be to compare resummed and NLO
fits using high-$p_T$ jet data \cite{cteq5}.

\section{Power-suppressed corrections}

In addition to higher orders in $\alpha_s(\mu^2)$,
Eq.\ (\ref{basicfact}) 
has corrections that fall
off as powers of the hard-scattering scale $Q$.
In contrast to higher orders, these
corrections require a generalization of the {\it form}
of the factorized cross section.  
Often power corrections are parameterized
as $h(x)/[(1-x)Q^2]$ in inclusive DIS, where
they begin at twist four.
In DIS, this higher twist term influences PDFs when
included in joint fits
with the NLO and NNLO models, and vice-versa 
\cite{bodekyang,kateevetal,vanneervenvogt}.
As in the case with higher orders, such ``power-improved" fits
should be treated as new schemes.

\section{Conclusions}

The success of NLO fits to DIS and
the studies of resummation above
suggest that over most of the range of $x$, 
theoretical uncertainties of the NLO model
are not severe.   At the
same time, to fit large $x$
with more confidence than is now possible may
require including the 
resummed coefficient functions.

Resummation is especially desirable for
global fits that employ a variety of processes, such
as DIS and high-$p_T$ jet production,
which differ in available phase space near partonic
threshold.
In a strictly NLO approach, uncalculated large
corrections are automatically incorporated in the
PDFs themselves.
As a result, the NLO model cannot
be expected to fit simultaneously the
large-$x$ regions of processes with differing logs
of $1-x$ in their hard-scattering functions, unless
these higher-order corrections are taken into account.

The results illustrated
in the figures suggest that these
considerations may be important in
DIS with $Q^2$ below a few GeV$^2$ and at large $x$, where
they may have substantial effects on estimates
of higher twist in DIS.
In hadronic scattering, large-$N$ ($x\rightarrow 1$) resummation, which
automatically reduces scale dependence, may play
an even more important role than in DIS.

\subsection*{Acknowledgments}
We thank Andreas Vogt and Stephane Keller for useful discussions.

\newpage
\setcounter{section}{0}
\begin{center}
\section*{PARTON DISTRIBUTION FUNCTIONS: EXPERIMENTAL DATA AND THEIR 
INTERPRETATION}
\end{center}

L.~de~Barbaro\\

Northwestern University, Evanston, IL, USA.\\

\section{Introduction}

The last few years have seen both new
and improved measurements of deep inelastic and related hard scattering 
processes and invigorated efforts to test the limits of our knowledge
of parton distributions (PDF) and assess their uncertainty.
Recent global analysis fits to the wealth of structure functions
and related data provide PDFs of
substantial sophistication compared to the previous parametrizations
~\cite{cteq5_b}\cite{mrst_b}.
The new PDF sets account for correlated uncertainty in strong coupling
constant, variation from normalization uncertainty of data sets,
theoretical assumptions regarding higher twists effects,
 initial parametrization form and
starting $Q_o$ value, etc. Range of potential variation in gluon density,
strange and charm quark densities or, recently, also in d quark 
distribution~\cite{kuhl} 
are also provided.
Participants of this Workshop in the PDF group primarily concentrated on 
finding new ways of inclusion of systematic uncertainties associated
with experimental data into the framework of global analyses.
Development in likelihood calculation by Giele, Keller, and Kossover, 
studies by CDF and D0 collaborators, and a parallel
work of CTEQ collaboration are presented in these proceedings.

New or improved results from several experiments have contributed to 
better knowledge of PDFs, however, there are still areas
where the interpretation of experimental data is not clear.
Few of these contentious issues will be discussed in this note.

\section{Issues in the Interpretation of Experimental Data}

\subsection{Gluon distribution at moderate to high x}

In principle, many processes are sensitive to the gluon
distribution, but its measurement is difficult beyond x $>$ 0.2
where it becomes very small.
Fermilab second generation--
direct photon experiment E706, although quite challenging 
experimentally, was designed to constrain gluon distribution at high x.
For proton-nucleon interactions in
 LO, direct photons are produced through Compton scattering off gluon
($gq \to \gamma q$) 90\% of the time 
in the E706 kinematic range.

\begin{figure}[htb]
\centering\leavevmode
\epsfxsize=3in\epsffile{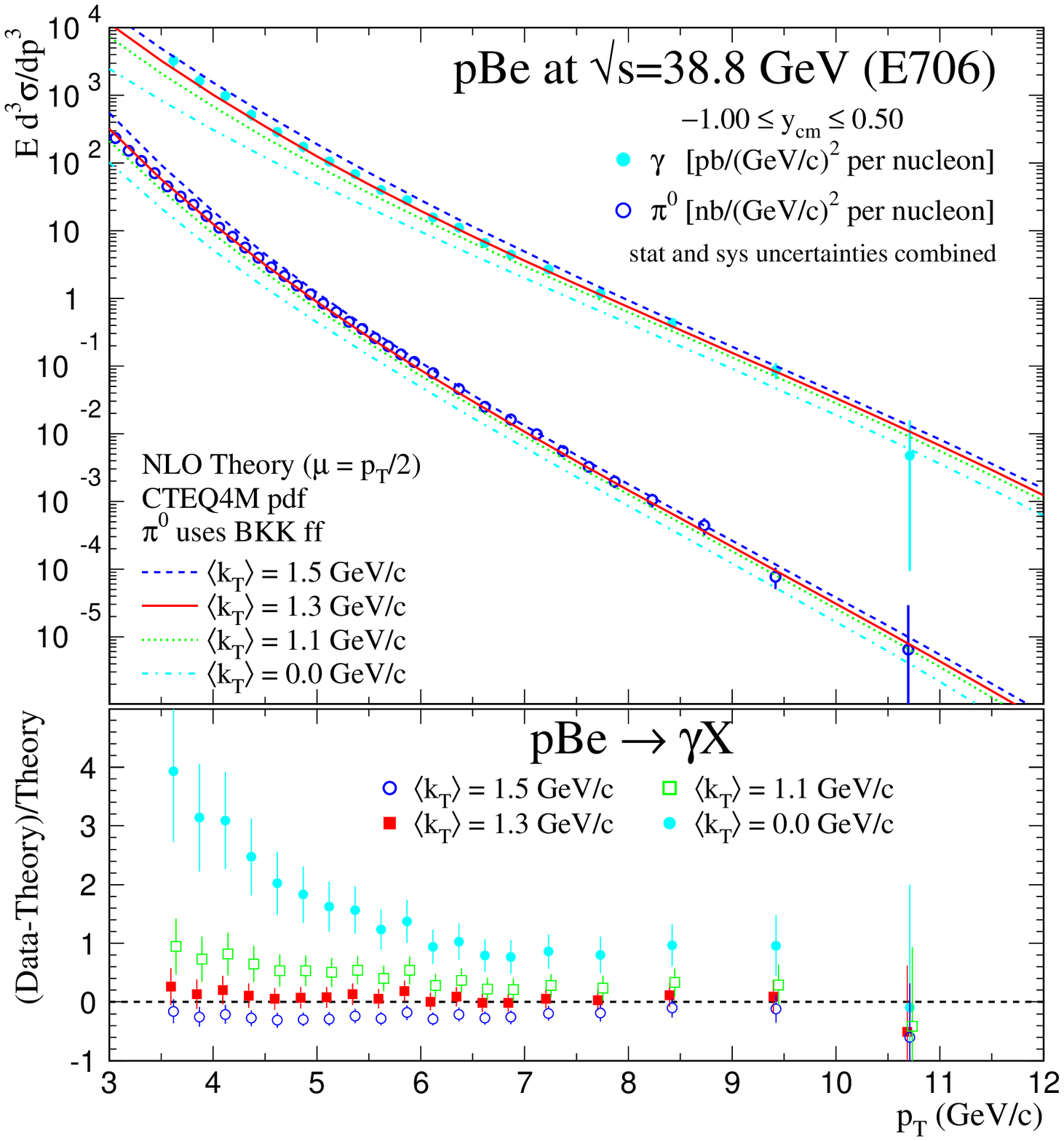}
\caption{Data and Theory agree after $k_T$ smearing for $\pi^0$
and $\gamma$ production in pBe interactions at 800 GeV. 
 Data-Theory/Theory comparison for various values of $k_T$ is shown
in lower insert.}
\label{fig:dp1}
\end{figure}

The first direct photon measurements, as well as WA70~\cite{bone}
were in agreement with the NLO theory and were used in several generations
of global analysis fits.
However, series of revisits of theoretical issues in 1990-ties (see, e.g.,
discussion in~\cite{vogt})
pointed to a large dependence of the NLO calculation
on renormalization and factorization scales and necessity
to include yet-unknown photon fragmentation function in the calculations.
Since the available $\sqrt(s)$ energy is low (20-40 GeV) for the fixed 
target experiments missing perturbation orders in the calculation 
are important.
Moreover, as shown by the E706 analysis, the transverse momentum of 
initial state  partons ($k_T$) dramatically affects
the differential cross sections measured versus transverse momentum
of the outgoing photon ($p_T$).
E706 measured the so-called $k_T$ smearing
by observing kinematic imbalance in production of $\pi^o$  pairs,
$\pi^o\gamma$, and double-direct photons and  found $k_T$ values $\approx$ 1 GeV
and increasing with $\sqrt(s)$~\cite{apana}. Similar results are
obtained in dijets and Drell-Yan data.
$K_T$ is believed to arise  
from both soft gluon emissions and non-perturbative phenomena.
NLO calculations smeared with $k_T$ estimated from these measurements
are increased by a factor of 2 to 4 (see Figure~\ref{fig:dp1}) and 
agree with the E706 direct photon and $\pi^0$ data on proton and Be targets, 
at $\sqrt(s)$ of 31 and 38 GeV.
A strong indication of $k_T$ effects and the need for soft gluon
resummation comes also from the analysis 
of double direct photon production. Both the NLO resummed theory
and $k_T$ smeared NLO theory describe the double direct photon
kinematics and cross section very well, in stark
contrast to the ``plain'' NLO prediction~\cite{begel}.

\begin{figure}[htb]
\centering\leavevmode
\epsfxsize=3.1in\epsffile{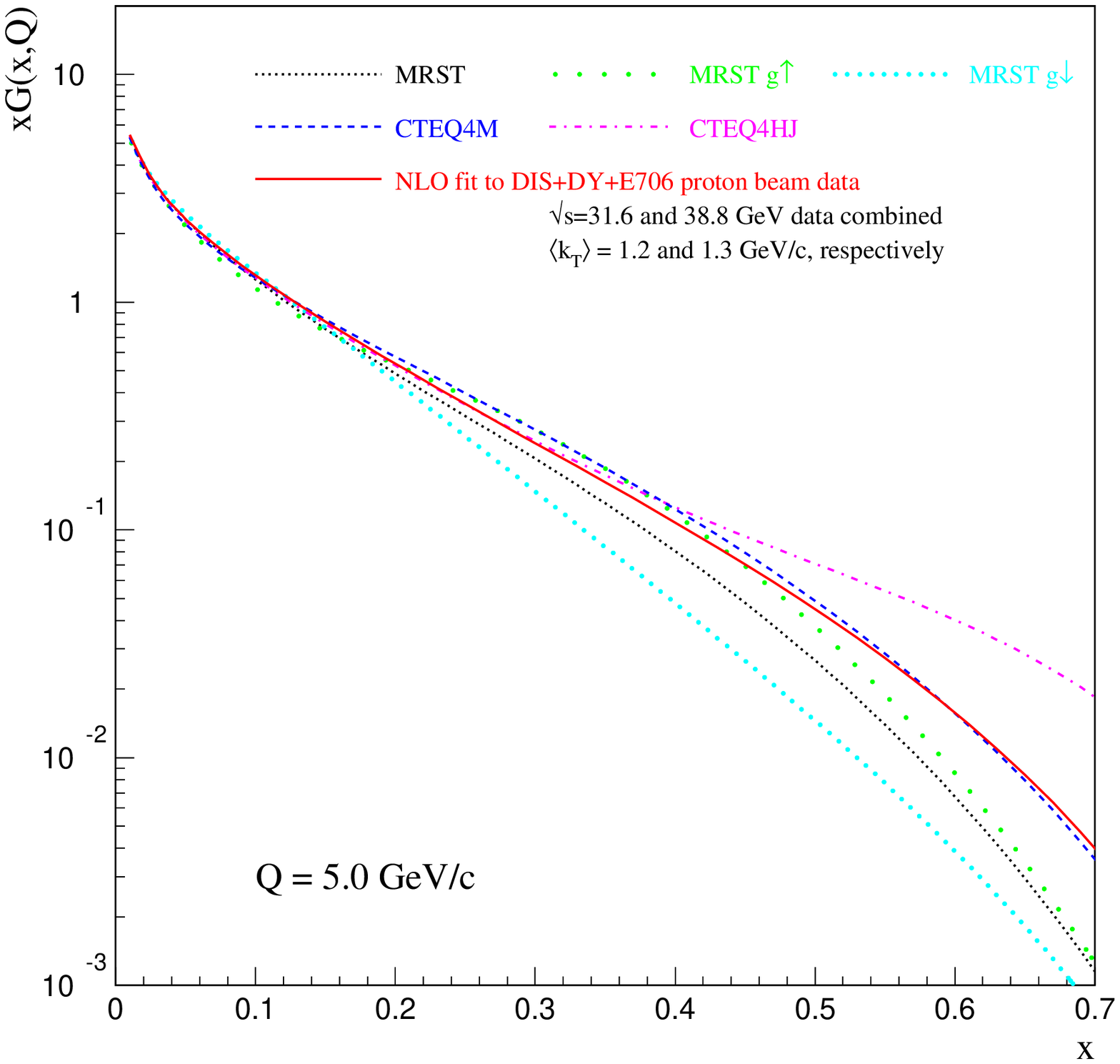}
\caption{Recent PDF sets indicate substantial disagreement 
about the shape and size of gluon distribution at moderate to high x.
CTEQ5 results closely follow CTEQ4M curve shown here.}
\label{fig:gluon}
\end{figure}

A comparison of current gluon distribution parametrizations indicates
our lack of knowledge of gluon in the moderate to high x range, (see
Figure~\ref{fig:gluon}).  The hardest gluon is the CTEQ4HJ
distribution. Here the gluon distribution is forced to follow the high
$E_T$ inclusive differential jet cross section measured at CDF.
Latest PDF sets by CTEQ match the WA70 direct photon data at
$\sqrt(s)$=23 GeV with no $k_T$, and require $k_T$=1.1 (1.3) GeV/c for
the E706 data at $\sqrt(s)$=31 (38) GeV.  Due to the difficulty in
reconciling this approach no direct photon data is used in the CTEQ5
global analysis.  The MRST group chose a different treatment: gluon
distributions are reduced at high x to accommodate some $k_T$ smearing
for both WA70 and E706 resulting in a moderately good description of
the data and three PDF sets spanning the extremes (shown in
Figure~\ref{fig:gluon}). The variety of predictions agree at low x,
but differ widely at high x.  The uncertainty in the $k_T$ modeling,
its unknown shape versus $p_T$, and potential discrepancy between WA70
and E706 measurements (see discussions in~\cite{apana}
and~\cite{fontaz}) require theoretical work to help resolve this
outstanding controversy.  Luckily, the interest in direct photon
physics and its importance for gluon determination has caught on, and
98 and 99 have seen a flurry of publications, notably: ``Soft-gluon
resummation and NNLO corrections for direct photon production'' by
N. Kidonakis, J. Owens (hep-ph/9912388), ``Results in
next-to-leading-log prompt-photon hadroproduction'' by S.Catani,
M.Mangano, C.Oleari (hep-ph/9912206), ``Unintegrated parton
distributions and prompt photon hadroproduction'' by M.Kimber,
A.Martin, M.Ryskin (DTP/99/100), ``Origin of $k_T$ smearing in direct
photon production'' by H.Lai, H.Li (hep-ph/9802414), ``Sudakov
resummation for prompt photon production in hadron collisions'' by
S.Catani, M.Mangano, P.Nason (hep-ph/9806484), etc.  New resummation
results are also expected from a group of G.Sterman and Vogelsang.

In addition to direct photons, the Tevatron jet and dijet measurements
are also sensitive to the gluon distribution (in the moderate x region). 
These measurements and comparisons to theory have their own set of
concerns, e.g. 
jet definition, which is never exactly the same 
in the data and in the NLO calculation or higher order correlations
in the underlying event (see discussion in, e.g.,~\cite{joey}).
The jet cross sections, strongly 
dominated by $q \bar q$ scattering,
are also sensitive to changes in high x valence distributions.
An unresolved issue in the jet cross section analysis
is also a lack of full 
 scaling between 630 and 1800 GeV data,
 predicted by QCD,  and a discrepancy
between the D0 and CDF measurements of this scaling ratio at lowest 
$x_T= E_T/\sqrt(s)$.

\subsection{Valence distributions at high x}

\begin{figure}[htb]
\centering\leavevmode
\epsfxsize=3.8in\epsffile{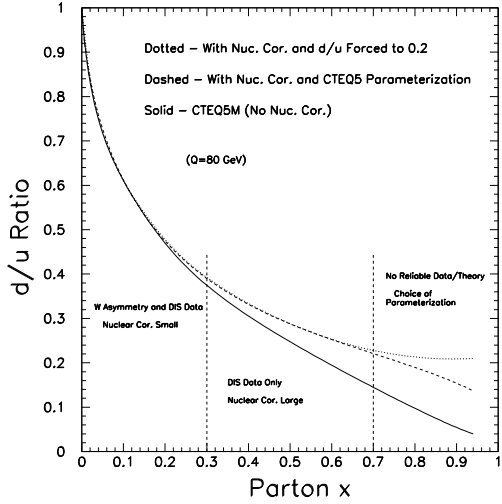}
\vspace{-1.in}
\caption{The d/u ratio for CTEQ5 and CTEQ5UD PDF sets, illustrating 
difference from nuclear correction for NMC $F_2$ on deuterium. The 
dotted and dashed lines correspond to two different assumptions regarding
value of d/u at x=1.} 
\label{fig:kulm}
\end{figure}

Apart from modifications to gluon and charm quark distributions,
the valence d quark has received the biggest boost in high x region
compared to previous PDF sets. 
The change is on the order of 30\% at x=0.6
and $Q^2=20$GeV$^2$
and comes from inclusion of a new observable in the global
analysis fits, namely W-lepton asymmetry measured at CDF.
Precise measurement of W-lepton asymmetry serves as an independent
check on the u and d quark distributions obtained from fits to deep inelastic
 data. The observable is directly correlated with the slope of the
d/u ratio in the x range of 0.1-0.3.
The consequence of this new constraint is that the predicted
$F_2^n/F_2^p$ ratio
is increased and the description of the NMC 
 measurement of  $F_2^d/F_2^p$ is improved relative to earlier PDF
sets.
There remain, however, two areas of uncertainty regarding valence 
distributions at high x:
the value of d/u ratio at x=1 and a  question
regarding a need for nuclear corrections to  $F_2^d/F_2^p$ NMC measurement.
Deuterium is a loosely bound nucleus, of low A, and traditionally no 
corrections for nuclear effects have been applied. 
However, an analysis of SLAC $F_2$ data on different targets under 
the assumption
that nuclear effects scale with the nuclear binding for all
nuclei predicts nuclear correction for deuterium of 4$\pm$1\% at
x=0.7~\cite{bodek}.
There is also a lack of clarity regarding d/u value at 1.
A non-perturbative QCD-motivated models of the 1970's argue that
the d/u ratio should approach 0.2 at highest x, whereas any standard 
form of the parametrization
used in global fits drive this ratio to zero.
The CTEQ collaboration has performed studies of
change in d/u ratio, depending on assumptions regarding 
nuclear effects in deuterium and the value of d/u ratio at x=1~\cite{kuhl}.
CTEQ5UD PDF set includes nuclear corrections for deuterium in 
$F_2^d/F_2^p$;
its change relative to CTEQ5 is a plausible range for d distribution 
uncertainty in light of this unresolved question, see Figure~\ref{fig:kulm}.

\subsection{Resolved discrepancies between PDF fits
and the data}

During the duration of this Workshop (March - Nov 1999), two
of the outstanding discrepancies between PDF fits or two sets of the 
experimental data 
have been resolved.

One of these was the near 20\% discrepancy at small x (0.007-0.1)
between structure function $F_2$ measured in muon (NMC) and neutrino
(CCFR) deep inelastic scattering~\cite{selig}.
For the purpose of comparison of these structure functions, 
NMC $F_2^{\mu p}$ was ``corrected'' for nuclear shadowing, measured
in muon scattering, to correspond to $F_2^{\mu Fe}$, and rescaled
by the 5/18 charge rule to convert from muon to neutrino $F_2$.
On the other hand, CCFR result was obtained in the framework of massless
charm quark to avoid kinematic differences between
muon and neutrino scattering off the strange quark ($\nu s \to \mu c$ 
versus $\mu s \to \mu s$) resulting from mass of the charm.
Any one of the above procedures could have had an unquantified
systematic uncertainty resulting in the observed disagreement.

\begin{figure}[hbt]
\centering\leavevmode
\epsfxsize=3.5in\epsffile{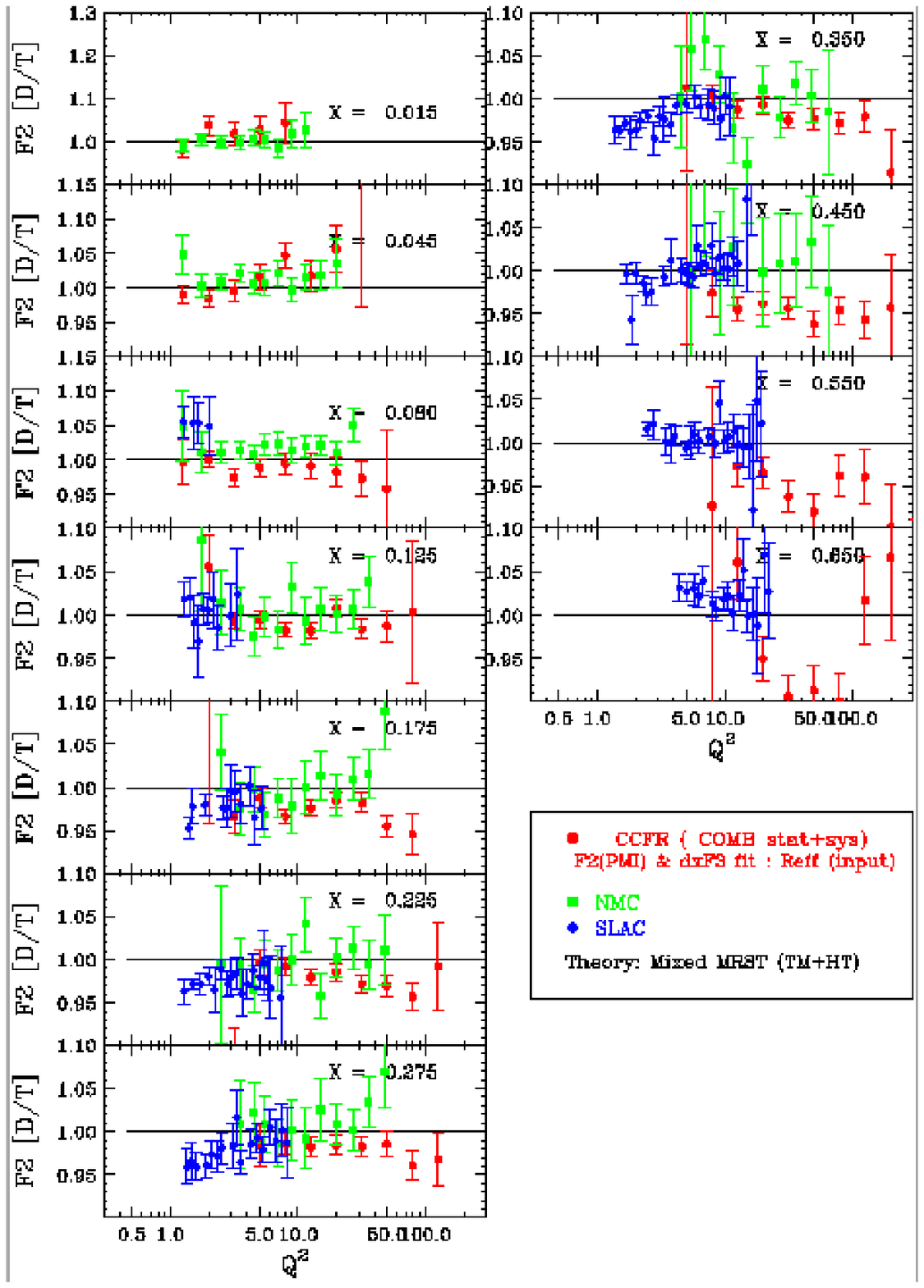}
\caption{The ratio of the massive $F_2^{\nu}$ measured at CCFR to the
prediction of MFS MRST prediction with target mass and higher twist
corrections applied. Also shown are the ratios of $F_2{\mu}$ (NMC)
and $F_2^e$ (SLAC) to the MFS MRST predictions.}
\label{fig:f2}
\end{figure}

New analysis from CCFR, presented at this Workshop~\cite{unki}, indicates
that the SF measured in CCFR is in agreement with the $F_2$ of NMC,
within experimental uncertainties.
The analysis used a new measurement of the difference between
neutrino and antineutrino structure functions $xF_3$,
 rather than the  $\Delta xF_3$=4(s-c) parametrization
used earlier. 
Comparison between calculations~\cite{charm_b} indicated that there were large
theoretical uncertainties in the charm production modeling for
both $\Delta xF_3$ and the ``slow rescaling'' correction that 
converts from massive
to massless charm quark framework.
Therefore, in the new analysis ``slow rescaling'' correction was not
applied and  $\Delta xF_3$ and $F_2$ were extracted from two parameter
fits to the data. 
The new measurement agrees well with the Mixed Flavor  Scheme  (MFS)  
for heavy quark production
as implemented by MRST group. To compare with charged lepton scattering data
each of the experimental results were divided by the theoretical 
predictions for $F_2$, using either light or heavy quark schemes implemented
by MRST.
The ratios of Data/Theory for $F_2^{\nu}$ (CCFR), $F_2^{\mu}$ (NMC), and 
$F_2^{e}$ (SLAC) are shown in Figure~\ref{fig:f2}. 
Systematic errors, except for
the overall normalization uncertainties, are included. The MFS MRST 
predictions have higher twist and target mass correction applied.
Apart from resolving the NMC-CCFR discrepancy, the new measurement had
also implication of ruling out one of the Variable Flavor
Scheme calculations available
on the market~\cite{unki2}.

\begin{figure}[htb]
\centering\leavevmode
\epsfxsize=3.in\epsffile{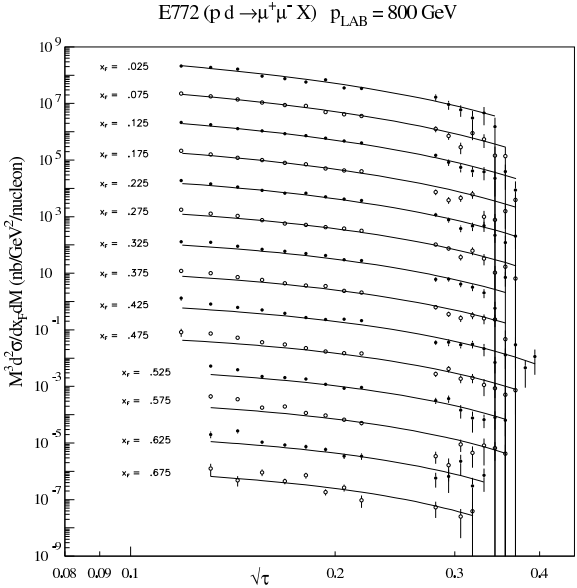}
\caption{Drell-Yan production from E772 compared to the MRST prediction.
The theory curves include K factor of 0.96 and the cross sections for different
values of $x_F$ are offset by a factor of 10. Corrected E772 data
reduce the discrepancy at high $x_F$ and low $\tau$.} 
\label{fig:e772}
\end{figure}

Another example is that of Drell-Yan production ($pd \to \mu^+\mu^-$)
as measured by Fermilab experiment E772, shown in Figure~\ref{fig:e772}. 
The MRST fits are
compared to the differential
cross section in $x_F=x_1 - x_2$ and in $\sqrt(\tau)=\sqrt(M^2/s)$,
where $x_1$ and $x_2$ are
 the target and 
projectile fractional momenta, and 
 $M$ - dimuon pair mass. The discrepancy, visible at high $x_F$ and
low $\sqrt(\tau)$
was hard to reconcile, since in this kinematic region the dominant
contribution to the cross section comes from $u(x_1)\times[\bar u(x_2)+
\bar d(x_2)]$ evaluated at $x_1\approx x_F$ and $x_2\approx$ 0.03,
well constrained by the deep inelastic scattering data.
Since then, the E772 experiment has reexamined their acceptance corrections
and released an erratum to their earlier measurement~\cite{erratum}.
The new values differ from the old ones only for large $x_F$ and
small values of mass $M$, and the new cross section is decreased
in this region by a factor up to two.

\section{Outlook for New Structure Function Measurements}

Measurements of neutral and charged current cross sections
in positron - proton collisions at large $Q^2$ 
from the 1994-97 data have just
been published by HERA experiments~\cite{h1}\cite{zeus}.
The data sample corresponds to an integrated luminosity of 35 pb$^{-1}$.
The $Q^2$ evolution of the parton densities of the proton is tested 
over  150-30000 GeV$^2$, Bjorken x between 0.0032-0.65,
 and yields no
significant deviation from the prediction of perturbative QCD.
These data samples are not yet sensitive eunough to pin down the d quark
distribution at high x, however, an expected 1000 pb$^{-1}$ 
in positron and electron running in years 2001-2005,
achievable after HERA luminosity upgrade, will have a lot to say about
20\% -like effects at high x in the ratio of valence 
distributions\footnote{Charge current $ep \to \nu X$ 
component, of the same order as the neutral current scattering at
very high $Q^2$, directly probes u ($e^-p$) and d ($e^+p$)
distributions at high x.}.

HERA's 1995-1999 data sets, not yet included in the global analysis
fits, were ploted against standard PDFs and showed a good
agreement over the new kinematic range that these data span
(extention to lower yet x and higher $Q^2$ compared to 1994 data)~\cite{mrst_b}.
HERA's very large statistics and improved precision will allow
further reduction of normalization uncertainty of PDF fits.
This is important for QCD prediction like W and Z total cross sections
at Tevatron - current 3\% normalization uncertainty in PDFs directly 
translates
to 3\% uncertainty for these cross sections.
Improvements in the measurements may need to go in hand with progress
in the perturbative calculations; it is likely that NNLO analysis
of deep inelastic scattering data
will change the level and/or x dependence of PDFs at the percentish-type
level.

One can expect continued progress in heavy quark treatment and 
in the theoretical understanding of soft gluon and
non-perturbative effects in the direct
photon production. In that case, the E706 data are sufficiently
precise to severely constrain the gluon distribution.

One of the few currently active structure function -- related experiments
is also NuTeV (Fermilab E815).
Better understanding of charm quark issues (see discussion in
preceeding section) and much improved calibration of NuTeV detector
relative to CCFR's (with a similar statistical power of the data set)
is expected to yield a more precise measurement
of structure functions and differential cross section for $\nu$ and $\bar \nu$
interaction in Fe.
Sign-selected beam and several advancements in the NLO theory of heavy
quark production 
 will allow NuTeV to improve systematic uncertainty
in the new measurement of the strange seas $s$ and $\bar s$.

Last but not least, Run II physics promises to be a good source of new
constraints on parton distributions. W-lepton asymmetry will be
measured with much improved  precision and in an expanded
rapidity range. New observables are proposed for further exploring
collider constrains
on PDFs, e.g., W and Z rapidity distributions~\cite{bodek3}.
And hopefully, many of the issues in jet 
measurements will be addressed and understood -- they are high on
J.Wormesley Christmas wish-list!~\cite{worm}

\newpage
\setcounter{section}{0}
\newcommand{\md}{\mbox{d}}
\def\simgt{\rlap{\lower 3.5 pt \hbox{$\mathchar \sim$}} \raise 1pt \hbox
{$>$}}
\def\simlt{\rlap{\lower 3.5 pt \hbox{$\mathchar \sim$}} \raise 1pt \hbox
{$<$}}
\def \MSbar {\vbox{\hrule\kern 1pt\hbox{\rm MS}}}
\def \GeV { {\ \rm GeV} }
\def\lsim{\simlt}
\def\gsim{\simgt}
\def\eq#1{Eq.~(\ref{#1})}
\def\eqs#1{Eqs.~(\ref{#1})}
\def\fig#1{Fig.~\ref{fig:#1}}
\def\tab#1{Table.~\ref{tab:#1}}
\def\junk#1{}


\def\tabgkl{
\begin{table}[hbtp]
\begin{tabular}{||l||c|c|c|c||} \hline \hline
 PDF Set           &Mass (GeV)    & LO    & $WQ\bar{Q} $  &  NLO \\   \hline \hline
  CTEQ1M        &$m_c$=1.7      & 96    &20             &161\\  \hline
  MRSD0'        &$m_c$=1.7      & 81    &20             &138\\  \hline
  CTEQ3M        &$m_c$=1.7      & 83    &20             &141\\  \hline
  CTEQ3M        & $m_b$=5.0     & 0.17   &9.09           &9.33 \\  \hline \hline
\end{tabular}
\vskip -0pt
\caption{
The $W$ + charm-tagged one-jet inclusive cross section in $pb$ for
LO, $W+Q\bar{Q}$, and NLO (including the $W+Q\bar{Q}$ contribution)
using  different sets of
parton distribution functions. Table is taken from 
Ref.~\protect\cite{gkl}.
\label{tab:gkl} 
}
\end{table}
}
\def\figband{
\begin{figure}[ht] 
\begin{center}
\leavevmode
 \epsfxsize=0.85\hsize \epsfbox{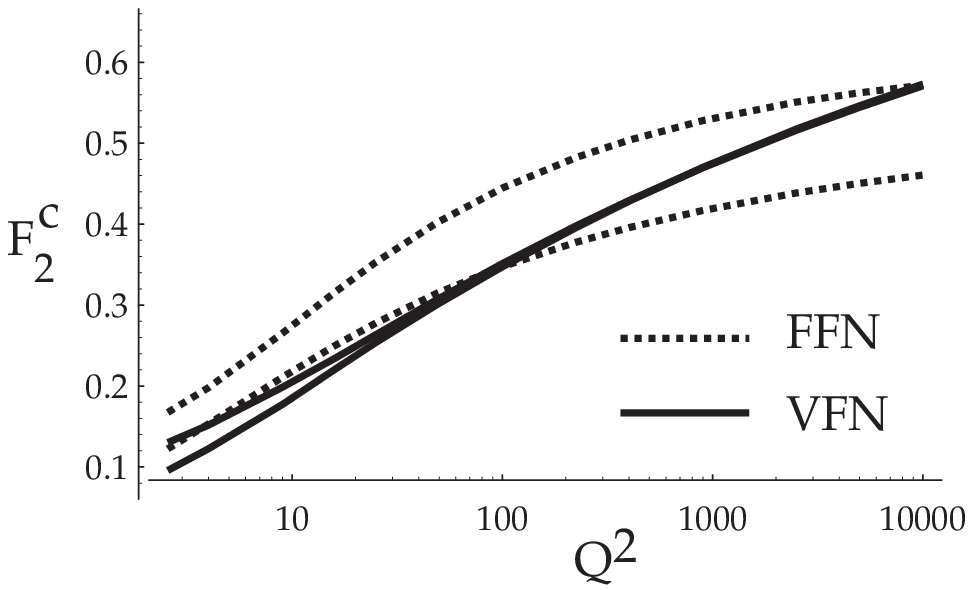} 
\vskip -20pt
\caption{
 $F_2^c$  for $x=0.01$ as a function of $Q^2$  in GeV 
 for two choices of $\mu$ 
 as obtained within the ${\cal O}(\alpha_s^1)$
 FFN and (ACOT) VFN schemes. 
 For details, see Ref.~\protect{\cite{schmidt}}.
\label{fig:band} 
}
\vskip -20pt
\end{center}
\end{figure}
}
\def\figsacot{
\begin{figure}[ht] 
\vbox{
 \hbox{
 \epsfxsize=0.85\hsize \epsfbox{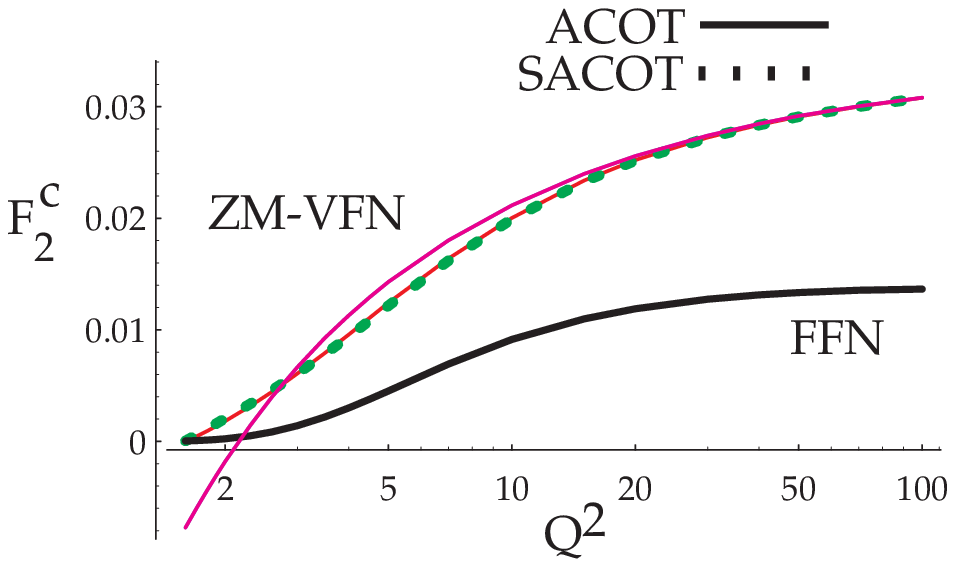} 
 }
 \hbox{
 \epsfxsize=0.85\hsize \epsfbox{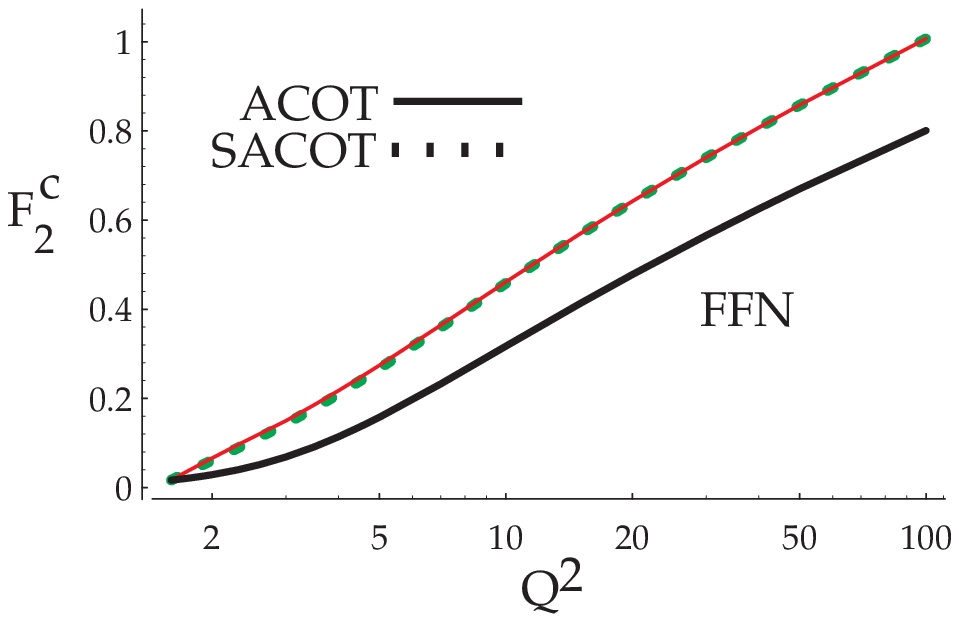}
 }
}
\vskip -20pt
\caption{
 $F_2^c$  as a function of $Q^2$ in GeV 
 computed to ${\cal O}(\alpha_s^1)$ in the 
 ZM-VFN, FFN, ACOT, and SACOT schemes using CTEQ4M PDF's. 
 Fig.~a)  $x=0.1$ , and 
 Fig.~b)  $x=0.001$.
 Figures taken from Ref.~\protect{\cite{sacot}}.
\label{fig:sacot} 
}
\vskip -20pt
\end{figure}
}
\def\figcgn{
\begin{figure}[ht] 
\begin{center}
\leavevmode
 \epsfxsize=0.95\hsize 
\epsfbox{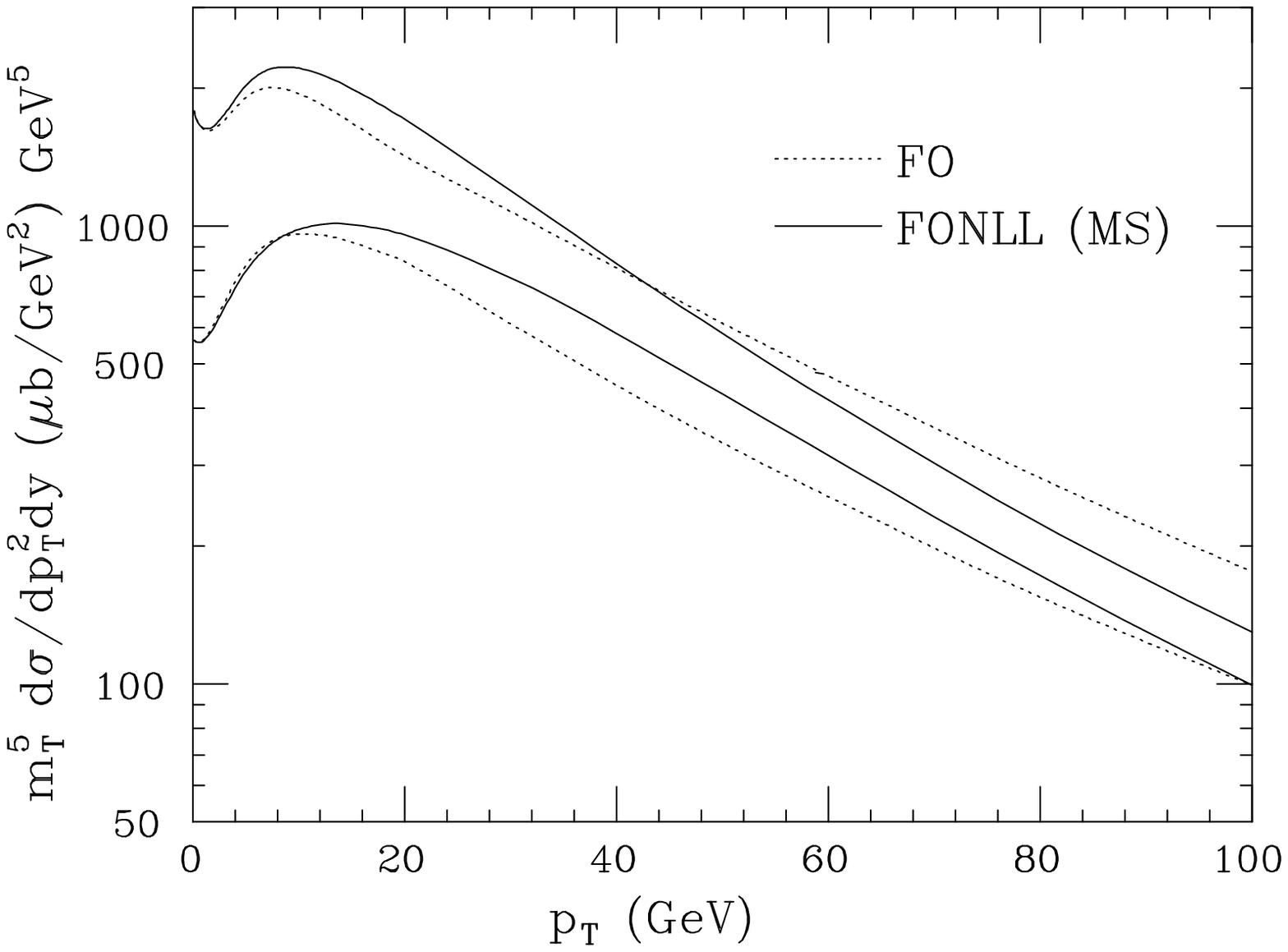} 
\vskip -20pt
\caption{
Differential cross section for b-production {\it vs.} $p_T$
comparing the Fixed-Order (FO) and the Fixed-Order Next-to-Leading-Log
(FONLL) result in the $\overline{MS}$ scheme. 
The bands are obtained by varying independently the renormalization 
and factorization scales. 
 The cross section is scaled by $m_T^5$ with $m_T = \sqrt{m_b^2+p_T^2}$,
and  $\sqrt{s}=1800\, GeV$, $m_b=5\, GeV$, $y=0$, with 
CTEQ3M PDF's. 
 Figure taken from Cacciari,  Greco, and Nason, Ref.~\protect{\cite{cgn}}.  
\label{fig:cgn} 
}
\vskip -20pt
\end{center}
\end{figure}
}
\def\figstk{
\begin{figure}[ht] 
\begin{center}
\leavevmode
 \epsfxsize=0.95\hsize \epsfbox{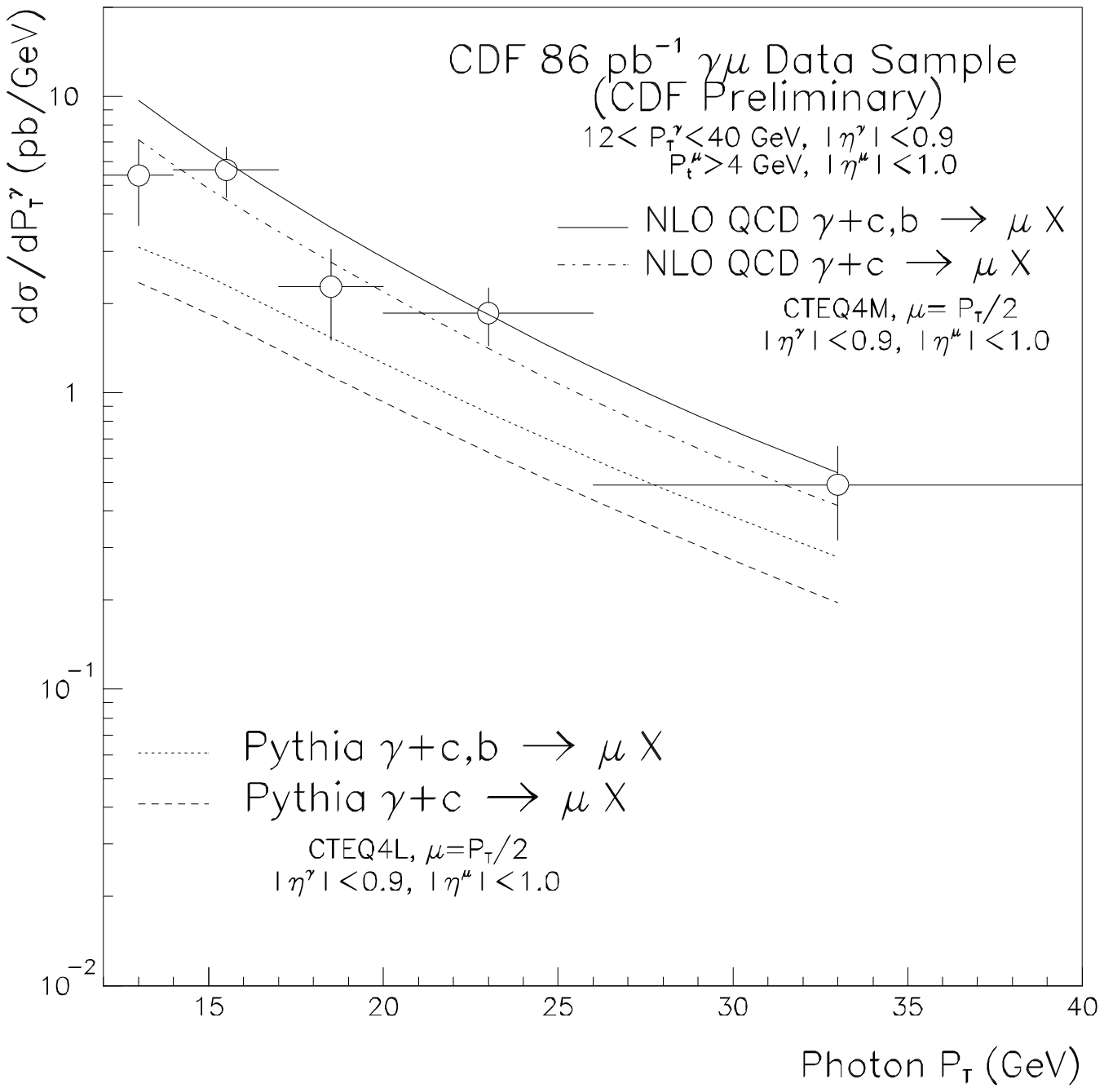} 
\vskip -20pt
\caption{
 Differential $d\sigma/dp_T^{\gamma}$ 
for $\gamma$ plus tagged heavy quark production 
as compared with Pythia and the NLO QCD results. 
 Figure taken from Ref.~\protect{\cite{stk}}.  
 NLO QCD calculations  from Ref.~\protect{\cite{berger}}.  
\label{fig:stk} 
}
\vskip -20pt
\end{center}
\end{figure}
}
\def\figxsc{
\begin{figure}[ht] 
\begin{center}
\leavevmode
 \epsfxsize=0.95\hsize \epsfbox{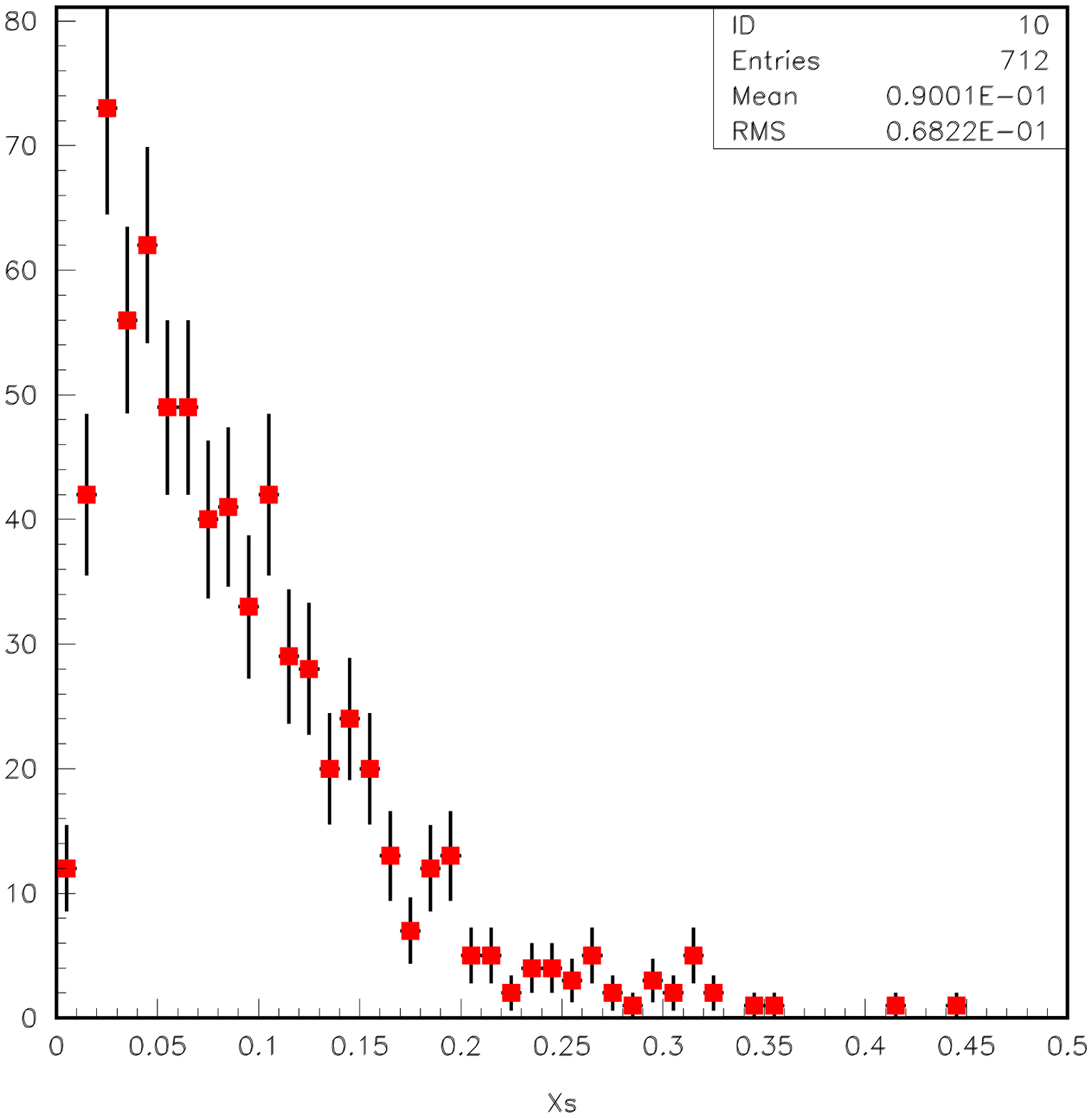} 
\vskip -20pt
\caption{
 Distribution of $Events/0.01$
{\it vs.} $x$ of the s-quarks which contribute to 
the $s+W \to c$ process. 
 Figure taken from Ref.~\protect{\cite{regina}}.  
\label{fig:xsc} 
}
\vskip -20pt
\end{center}
\end{figure}
}
\def\figdff{
\begin{figure}[ht] 
\begin{center}
\leavevmode
\vbox{
 \hbox{
 \epsfxsize=0.85\hsize \epsfbox{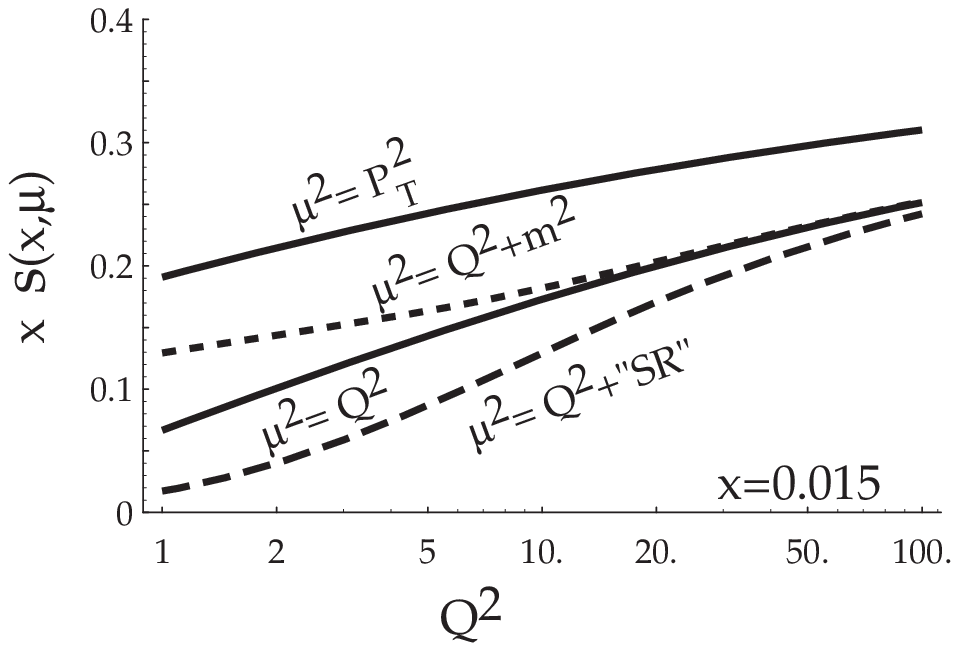} 
 }
 \hbox{
 \epsfxsize=0.85\hsize \epsfbox{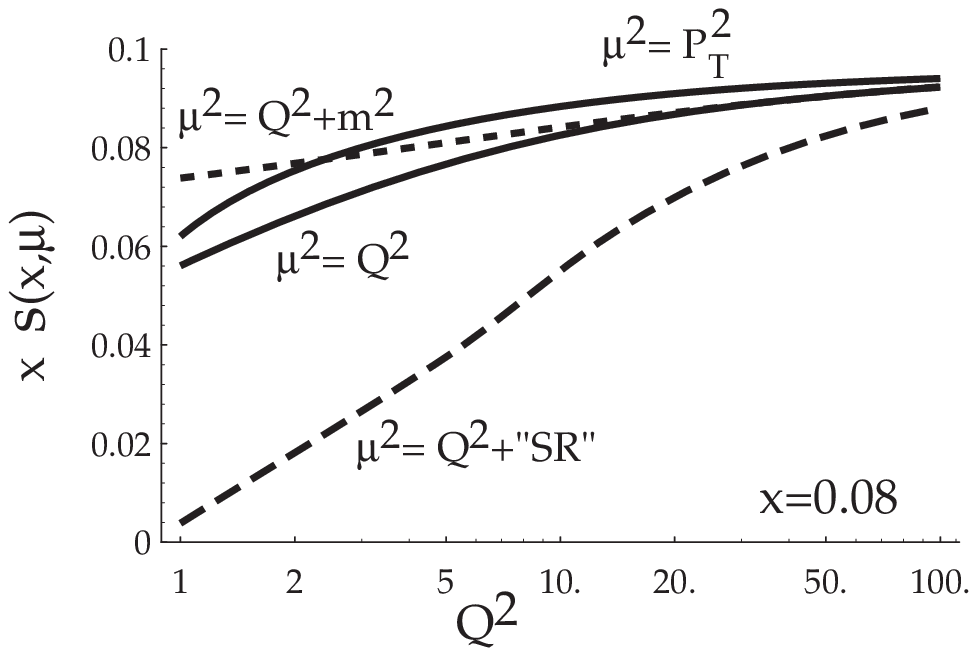}
 }
}
\vskip -20pt
\caption{
 Variation of $x \, s(x,\mu)$ for three choices of $\mu$, and also with
a  ``SR" (slow-rescaling) type correction: $x \to x(1+m_c^2/Q^2)$. 
\label{fig:dff} 
}
\vskip -20pt
\end{center}
\end{figure}
}
\def\figdf{
\begin{figure}[ht] 
\begin{center}
\leavevmode
 \epsfxsize=0.99\hsize \epsfbox{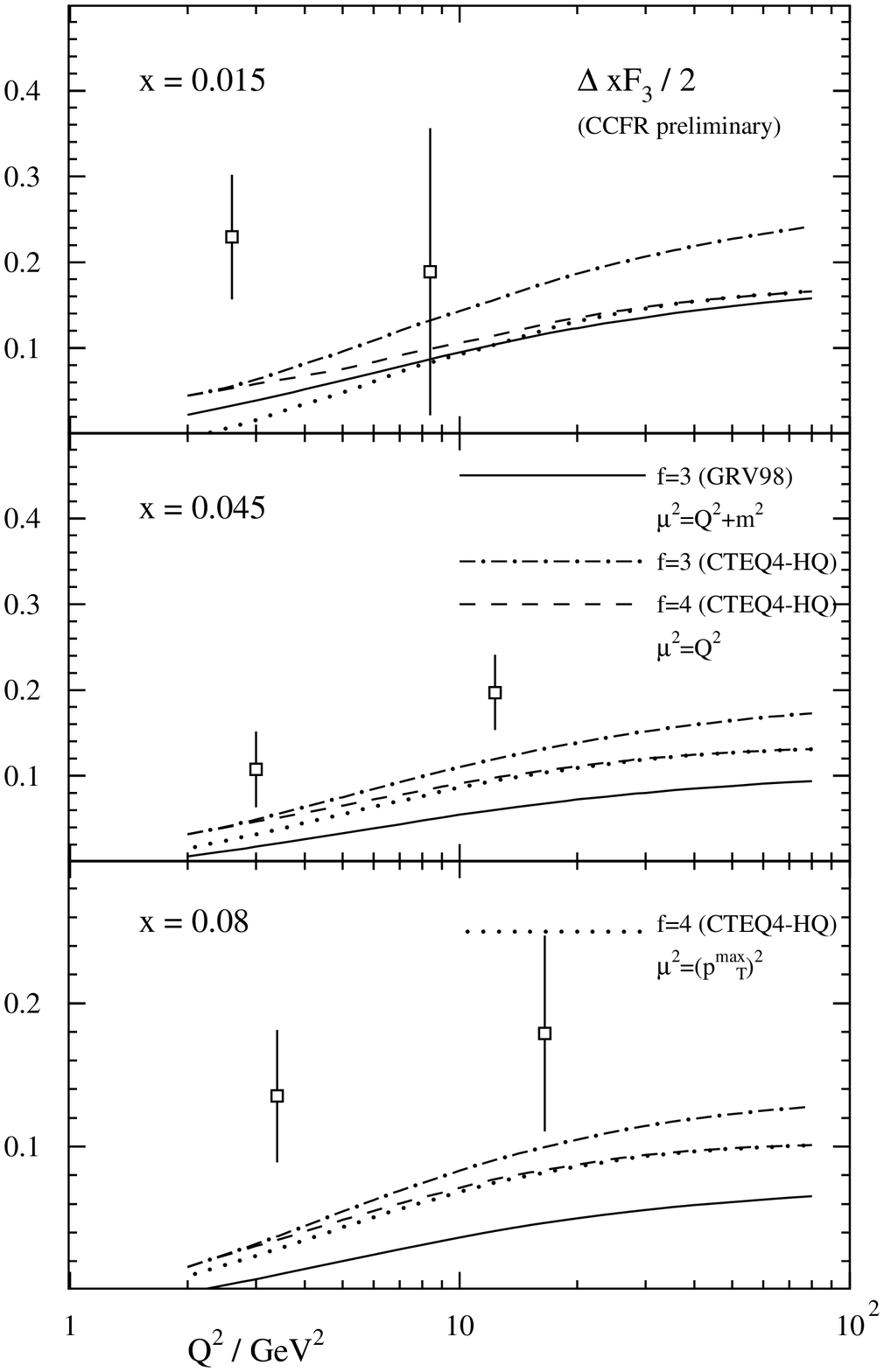} 
\vskip -20pt
\caption{
 $\Delta x F_3/2$ {\it vs.} $Q^2$ for three choices of $x$. 
 Calculations provided by S.~Kretzer. 
\label{fig:df} 
}
\vskip -20pt
\end{center}
\end{figure}
}

\begin{center}
\section*{HEAVY QUARK PRODUCTION}
\end{center}

\def\addr#1{\address{#1 \vskip -8pt}}

R.~Demina$^a$, 
S.~Keller$^b$, 
M.~Kr\"amer$^c$, 
S. Kretzer$^d$, 
R.~Martin$^e$, 
F.I.~Olness$^f$, 
R.J.~Scalise$^f$, 
D.E.~Soper$^g$, 
W.-K.~Tung$^{h,i}$, 
N.~Varelas$^e$, 
and  U.K. Yang$^j$.\\

a) {Kansas State University, Physics Department, 116 Cardwell Hall,
  Manhattan, KS 66506}; 
b) {Theoretical Physics Division, CERN, CH-1211 Geneva 23, 
      	Switzerland};
c) {Department of Physics, University of Edinburgh, 
       Edinburgh EH9 3JZ, Scotland};
d) {Univ. Dortmund, Dept. of  Physics, 
      D-44227 Dortmund, Germany};
e) {Univ. of Illinois at Chicago, Dept. of Physics, 
      Chicago, IL 60607};
f) {Southern Methodist University, Department of Physics, 
      Dallas, TX 75275-0175};
g) {Institute of Theoretical Science, University of Oregon, 
      	Eugene OR 97403, USA};
h) {Michigan State University, Department of Physics and Astronomy, 
      East Lansing, Michigan 48824-1116};
i) {Fermi National Accelerator Laboratory,  
  Batavia, IL 60510};
j) {University of Chicago, Enrico Fermi Institute, 
      Chicago, IL 60637-1434}.\\


\begin{center}
		Abstract
\end{center}

We present a status report of a variety of projects related to 
heavy quark production and parton distributions for the 
Tevatron Run~II. 

\def\subsection#1{\section{#1}\null \vskip -15pt}
\subsection{Introduction} 

\addtocounter{footnote}{-1}

\addtocounter{footnote}{-1}

The production of heavy quarks, both hadroproduction and leptoproduction, 
 has become an important theoretical and phenomenological
issue. 
 While the hadroproduction mode is of direct interest to this
workshop,\cite{tev} we shall find that the simpler 
leptoproduction process
can provide important insights into the fundamental production 
mechanisms.\cite{nutev,charm,hera,kramer}
 Therefore, in preparation for the Tevatron Run~II, we must consider 
information from a variety of sources including charm and bottom 
production at fixed-target and collider lepton and hadron facilities. 

For example, the charm contribution to the total structure function $F_2$
at
 HERA, is sizeable, up to $\sim 25\%$ in the small $x$ region.\cite{hera}
 Therefore a proper description of charm-quark
production is required for a global analysis of structure
function data,  and hence a precise extraction of the parton densities in
the
proton.
 These elements are important for addressing a variety of issues 
at the Tevatron.

In addition to the studies investigated at the Run~II workshop 
series,\footnote{%
In particular,  in the Run~II B-Physics workshop, 
the studies of {\it Working Group 4: Production, Fragmentation, Spectroscopy}, 
organized by  
Eric Braaten, Keith Ellis, Eric Laenen, William Trischuk, Rick Van Kooten, and Scott Menary,
addressed many issues of direct interest to this subgroup. 
The report is in progress, and the web page is located at: 
\hbox{http://www-theory.fnal.gov/people/ligeti/Brun2/}
}
we want to call attention to the extensive work done in the 
 {\it Standard Model Physics
(and more) at the LHC Workshop} organized by Guido Altarelli, Daniel
Denegri, Daniel Froidevaux, Michelangelo Mangano, Tatsuya Nakada which
was held at CERN during the same period.\footnote{%
The main web page is located at: 
 http://home.cern.ch/$\sim$mlm/lhc99/lhcworkshop.html}
 In particular, the investigations of the LHC {\it b-production group} 
(convenors: Paolo Nason, Giovanni Ridolfi, Olivier Schneider, Giuseppe
Tartarelli, Vikas Pratibha) and the {\it QCD group} (convenors: Stefano
Catani, Davison Soper, W. James Stirling, Stefan Tapprogge, Michael
Dittmar) are directly relevant to the material discussed here.
  Furthermore, our report limits its scope to the issues discussed 
within the Run~II workshop; for a recent comprehensive review,
see Ref.~\cite{Frixione:1997ma}.

\subsection{Schemes for Heavy Quark Production\label{sec:schemes}} 

Heavy quark production also provides an important theoretical challenge as 
the presence of the heavy quark mass, $M$, 
introduces a new scale into the problem. 
 The heavy quark mass scale, $M$, 
in addition to the 
characteristic energy scale of the process 
(which we will label here generically as $E$),
will require a different organization of the perturbation series 
depending on the relative magnitudes of $M$ and $E$. 
  We find there are essentially two cases to 
consider.\footnote{%
We emphasize that the choice of a prescription for dealing with
quark masses in the hard scattering coefficients for deeply inelastic
scattering is a separate issue from the choice of definition of the
parton distribution functions. For all of the prescriptions discussed
here, one uses the standard \MSbar\ definition of parton distributions.}

\begin{enumerate}

\item 
 For the case of $E \sim  M$,  heavy-quark production  is
calculated in the so-called fixed flavor number  (FFN) scheme
from hard processes initiated by light quarks ($u,d,s$) and gluons,
where all effects of the charm quark are contained in the perturbative
coefficient functions. The FFN scheme incorporates the correct
threshold behavior, but for large scales, $E\gg M$, the coefficient
functions in the FFN scheme at higher orders in $\alpha_s$ contain
potentially large logarithms $\ln^n(E^2/M^2)$, which may need to be
resummed.\cite{LRSN,buza,csn,or} 

\item 
 For the case of $E \gg  M$, it is necessary to include the heavy
quark as an active parton in the proton. 
 This serves to resum the potentially large logarithms $\ln^n(E^2/M^2)$
discussed above. 
 The simplest approach
incorporating this idea is the so-called zero mass variable flavor
number  (ZM-VFN) scheme, where heavy quarks are treated as
infinitely massive below some scale $E\sim M$ and massless above this
threshold. This prescription has been used in global fits for many
years, but it has an error of ${\cal O}(M^2/E^2)$ and is not suited
for quantitative analyses unless $E\gg M$.

\end{enumerate}  

While the extreme limits $E \gg  M$ and $E \sim  M$
are straightforward, much of the experimental data lie in the 
intermediate region 
 As such, the correct PQCD formulation of heavy quark production, 
capable of spanning the full energy range,  must incorporate
the physics of both the FFN scheme and the ZM-VFN scheme. 
 Considerable effort has
been made to devise a prescription for heavy-flavor production that
interpolates between the FFN scheme close to threshold and the ZM-VFN
scheme at large $E$. 

The generalized VFN scheme includes the heavy quark as an
active parton flavor and involves matching between the FFN scheme with
three
active flavors and a four-flavor prescription with non-zero
heavy-quark mass. It employs the fact that the mass singularities
associated with the heavy-quark mass can be resummed into the parton
distributions without taking the limit $M\to 0$ in the short-distance
coefficient functions, as done in the ZM-VFN scheme. 
 This is precisely the underlying idea of the
Aivazis--Collins--Olness--Tung
(ACOT) ACOT scheme\cite{acot}
which is based on the renormalization method of 
Collins--Wilczek--Zee (CWZ).\cite{cwz}
 The order-by-order procedure to implement this approach has now
been systematically established to all orders in 
PQCD by Collins.\cite{collins98}

Recently, additional implementations of 
VFN schemes have been
defined in the literature. 
 While these schemes all agree in principle on the
result summed to all orders of perturbation theory,
the way of
ordering the perturbative expansion is not unique and the results
differ at finite order in perturbation theory.
 The Thorne--Roberts (TR) \cite{TR} prescription
has been used in the MRST recent global analyses of parton
distributions.\cite{MRST} 
 The BMSN and CNS prescriptions have made use of the 
${\cal O}(\alpha_s^2)$ calculations by Smith, van~Neerven, and 
collaborators\cite{buza,csn}
to carry these ideas to higher order. 
 The boundary conditions on the PDF's at the
flavor threshold become more complicated at this order; 
in particular, the PDF's are no longer continuous  across the 
N to N+1 flavor threshold. 
Buza {\it et al.},\cite{buza} have computed the matching conditions,
and this has been implemented in an evolution program by CSN.\cite{csn}
 More recently, a Simplified-ACOT (SACOT) 
scheme inspired by the prescription advocated by
Collins~\cite{collins98} was introduced;\cite{sacot}
we describe this new scheme in Sec.~\ref{sec:acot}.

\subsection{From Low To High Energy  Scale}
\figband

To compare the features of the FFN scheme with the ACOT VFN scheme\footnote{%
In this section we shall use the ACOT VFN scheme for this illustration. 
The conclusions extracted in comparison to the FFN scheme are largely independent
of which VFN scheme are used.}
 concretely,
we will take  the example of heavy quark production in DIS; the features 
we extract from this example are directly applicable to the
hadroproduction case
relevant for the Tevatron Run~II.
 One measure we have of estimating the uncertainty of a calculated
quantity
is to examine the variation of the renormalization and factorization 
scale dependence. While this method can only provide a lower bound on the
uncertainty, it is a useful tool.  

In \fig{band}, we display the component of $F_2^{c}$ for the $s+W\to c$
sub-process 
at $x=0.01$ plotted {\it vs.} $Q^2$. We gauge the scale uncertainty by
varying 
$\mu$ from $1/2\, \mu_0$to $2.0\, \mu_0$ with $\mu_0= \sqrt{Q^2+m_c^2}$. 
In this figure, both schemes are applied to ${\cal O}(\alpha_s^1)$.
 We observe  that the FFN scheme is narrower at low $Q$, and increases
slightly
at larger $Q$. This behavior is reasonable given that we expect this 
scheme to work best in the threshold region, but to decrease in accuracy
as the unresummed logs of $\ln^n(Q^2/m_c^2)$ increase. 
 
Conversely, the ACOT VFN scheme has quite the opposite behavior. 
At low $Q$, this calculation displays mild scale uncertainty,
but at large $Q$ this uncertainty is significantly reduced. 
This is an indication that the resummation of the  $\ln^n(Q^2/m_c^2)$ 
terms via the heavy quark PDF serves to decrease the scale uncertainty
at a given order of perturbation theory. 
 While these general results were to be expected, what is surprising
is the magnitude of the scale variation. Even in the threshold region
where
$Q\sim m_c$ we find that the VFN scheme is comparable or better than 
the FFN scheme. 

 At present, the FFN scheme has been calculated to one further order
in perturbation theory, ${\cal O}(\alpha_s^2)$. While the higher order
terms
do serve to reduce the scale uncertainty, it is only at the lowest 
values of $Q$ that the ${\cal O}(\alpha_s^2)$ FFN band is smaller than the
${\cal O}(\alpha_s^1)$ VFN band.
 Recently, ${\cal O}(\alpha_s^2)$ calculations in the VFN scheme 
have been performed;\cite{csn} it would be interesting to extend such
comparisons
to these new calculations.  

Let us also take this opportunity to clarify a misconception that
has occasionally appeared in the literature. The VFN scheme is {\it not} 
required to reduce to the FFN scheme at $Q=m_c$. 
While it is true that the VFN scheme does have the FFN scheme as a limit, 
this matching depends on the definitions of the PDF's, and the 
choice of the $\mu$ scale.\footnote{%
The general renormalization scheme is laid out in the CWZ paper\cite{cwz}.
The matching of the PDF's at ${\cal O}(\alpha_s^1)$ was computed in 
Ref.~\cite{collinstung} and Ref.~\cite{qian}.
The  ${\cal O}(\alpha_s^2)$ boundary conditions were computed in 
Ref.~\cite{buza}.
}
In this particular example, even at $Q=m_c$, 
the resummed logs 
in the heavy quark PDF can yield a non-zero contribution which help
to stabilize the scale dependence of the VFN scheme result.\footnote{%
{\it Cf.}, Ref.~\cite{schmidt} for a detailed discussion.}

The upshot is that even in the threshold region, the resummation of the
logarithms via the heavy quark PDF's can help the stability of the theory.

\subsection{Simplified ACOT (SACOT) prescription\label{sec:acot}}
 \figsacot

We  investigate a modification of the ACOT
scheme inspired by the prescription advocated by
Collins.\cite{collins98} This prescription has the advantage of being
easy to state, and allowing relatively simple calculations. Such
simplicity could be crucial for going beyond one loop order in
calculations.\footnote{%
See Ref.~\cite{sacot} for a detailed definition, discussion, and
comparisons.}

\begin{quote}
{\it Simplified ACOT (SACOT) prescription}.
Set $M_H$ to zero in the calculation of the hard
scattering partonic functions $\widehat \sigma$ for incoming heavy quarks.
\end{quote}

For example, this scheme tremendously simplifies the calculation of 
the neutral current structure function 
$F_2^{charm}$ even at ${\cal O}(\alpha_s^1)$. 
 In  other prescriptions, 
the tree process $\gamma +c \to c+g$ 
and the one loop process $\gamma +c \to c$ must 
be computed with non-zero charm mass, and this results
in a complicated expression.\cite{kretzer}
In the SACOT scheme, the charm mass can be set to zero
so that the final result for these sub-processes reduces 
to the very simple massless result.

While the SACOT scheme allows us to simplify the calculation,
the obvious question is: does this simplified version contain the
full dynamics of the process. 
 To answer this quantitatively,  
we compare prediction for $F_2^{charm}$ obtained with
 1) the SACOT scheme at order $\alpha_s^1$ with 
 2) the predictions obtained with
the original ACOT scheme, 
 3) the ZM-VFN procedure in which the charm quark
can appear as a parton but has zero mass, and 
 4) the FFN procedure in which
the charm quark has its proper mass but does not appear as a parton.
For simplicity, we take $\mu=Q$.

In \fig{sacot} we show $F_2^c(x,Q)$ as a function of $Q$ for $x = 0.1$
and $x = 0.001$ using the CTEQ4M parton distributions.\cite{cteqo,LaiTun97a}
 We observe that the ACOT and SACOT schemes are effectively identical
throughout the kinematic range. There is a slight  difference observed in
the threshold region, but this is small in comparison to 
the  renormalization/factorization 
$\mu$-variation (not shown). Hence the difference between the
ACOT and SACOT results is of no physical consequence. The fact that  the
ACOT and SACOT match extremely  well throughout the full kinematic range
provides  explicit numerical verification that the SACOT scheme fully
contains the physics.

Although we have used the example of heavy quark leptoproduction, let
us comment briefly on the implications of this scheme for 
the more complex case of hadroproduction.\cite{tev,nde,Beenakker,ost} 
 At present, we have calculations for the all the ${\cal O}(\alpha_s^2)$ 
hadroproduction sub-processes such as $gg \to Q \bar{Q}$ and $gQ \to gQ$. 
 At ${\cal O}(\alpha_s^3)$ we have the result for the  $gg \to g Q
\bar{Q}$
sub-processes, but not the general result for  $gQ \to g g Q$ with
non-zero heavy quark mass. With the SACOT scheme, 
we can set the heavy quark mass to zero in the  $gQ \to g g Q$ sub-process
and thus make use of the simple result already in the literature.\footnote{%
For a related idea, see the fragmentation function formalism 
of Cacciari and Greco\cite{Cacciari} in the following section.}
 This is just one example of how the SACOT has the practical advantage of 
allowing us to extend our calculations to higher orders in the
perturbation theory. 
 We now turn to the case of heavy quark production for hadron colliders.

\subsection{Heavy Quark Hadroproduction}

\figcgn

There has been notable  progress in the area of 
hadroproduction of heavy quarks. 
The original NLO calculations of the 
$g g \to b \bar{b}$ subprocess were performed by 
 Nason, Dawson, and Ellis \cite{nde},
and by 
Beenakker, Kuijf, van Neerven, Meng, Schuler, and Smith\cite{Beenakker}.
Recently, Cacciari and Greco\cite{Cacciari} 
have used a NLO fragmentation 
formalism to resum the heavy quark contributions in the limit of large $p_T$; 
the result is a decreased renormalization/factorization scale variation in the large $p_T$ region. 
The ACOT scheme was applied to the hadroproduction case by Olness, Scalise, and Tung.\cite{ost}
More recently, the NLO fragmentation 
formalism of Cacciari and Greco has been merged with the massive FFN calculation
of  Nason, Dawson, and Ellis
by Cacciari,  Greco, and Nason,\cite{cgn}; 
the result is a calculation which matches the FFN calculation 
at low $p_T$, and  takes advantage of the NLO fragmentation formalism 
in the high $p_T$ region, thus yielding good behavior 
throughout the full $p_T$ range.
 This is displayed in \fig{cgn} where we see that this 
 Fixed-Order Next-to-Leading-Log (FONLL) calculation 
 displays reduced  scale variation in the large $p_T$ region,
 and matches on the the massive NLO calculation in the small $p_T$ region.
Further details can be found in the report of the LHC Workshop
{\it b-production group}.\footnote{
 The LHC Workshop {\it b-production group} is organized by
 Paolo Nason, Giovanni Ridolfi, Olivier Schneider, Giuseppe
Tartarelli, Vikas Pratibha, and the report is currently in preparation. 
 The webpage for the {\it b-production group} is located at
 http://home.cern.ch/n/nason/www/lhc99/
 }

\subsection{$W$ + Heavy Quark Production}
\tabgkl

\figstk

\figxsc

The precise measurement of $W$ plus heavy quark ($W$+$Q$) events provides
an important 
information on a variety of issues. 
 Measurement of $W$+$Q$ allows us to 
test  NLO  QCD theory  at high scales and  investigate  questions about 
 resummation  and heavy quark PDF's. 
 For example, if sufficient statistics are available,  $W$+$charm$ final
states
can be used to extract information about the strange quark distribution. 
 In an analogous manner, the  $W$+$bottom$ final states 
are sensitive to the charm PDF; 
furthermore, $W$+$bottom$ can fake Higgs events, and are  
also an important background for sbottom ($\widetilde{b}$) searches.

The cross sections for $W$ plus tagged heavy quark jet were computed
in Ref.~\cite{gkl}, and are shown in \tab{gkl}.  
 Note that this process has a large $K$-factor, and hence comparison
between
data and theory will provide discerning test of the NLO QCD theory. 
 While the small cross sections of these channels hindered analysis in
Run~I,
the increased luminosity in Run~II can make this a discriminating tool. 
 For example,  Run~I provided minimal statistics on $W$+$Q$, 
but there was data in the analogous 
neutral current channel $\gamma$+$Q$. 
The NLO QCD cross sections for $\gamma$ plus  heavy quark  were computed
in Ref.~\cite{berger}.
 \fig{stk} displays preliminary Tevatron data  from Run~I 
and the comparison with both the PYTHIA Monte Carlo and the NLO QCD
calculations;
again, note the large $K$-factor.  If similar results are attainable
in the charged current channel at Run~II, this would be revealing. 

Extensive analysis the $W$+$Q$ production channels were performed in 
Working Group~I:  ``QCD tools for heavy flavors and new physics searches,"
and we can make use of these results
to estimate the precision to which the strange quark distribution can be
extracted. 
We display \fig{xsc} (taken from the WGI
report\cite{regina}) which shows the distribution in $x$ of the s-quarks 
which contribute to the $W$+$c$ process.\footnote{%
For a detailed analysis of this work including 
selection criteria, see the report of 
Working Group~I:  
``QCD Tools For Heavy Flavors And New Physics Searches," 
as well as Ref.~\cite{regina}.}
  This figure indicates that there will good statistics
in an $x$-range comparable to that investigated by 
neutrino DIS experiments;\cite{nutev,charm}
hence, comparison with this data should provide an important test of 
the strange quark sea and the underlying mechanisms for computing 
such processes.

\subsection{The Strange Quark Distribution}
 \figdff

 \figdf

A primary uncertainty for $W$+$charm$ production 
discussed above  comes from the strange sea PDF, $s(x)$,
which has been the subject of controversy for sometime now. 
  One possibility is that new analysis of present data will resolve this
situation
prior to Run~II, and provide precise distributions as an input the the
Tevatron data analysis. 
The converse would be that this situation remains unresolved, in which
case new data
from Run~II may help to finally solve this puzzle.

The strange distribution is directly measured by dimuon production in 
neutrino-nucleon 
scattering.\footnote{%
Presently, there are a number of LO analyses, 
and one NLO  analysis.\cite{nutev,charm}}
 The basic sub-process is $\nu N \to \mu^- c X$ with a subsequent charm
decay $c \to \mu^+ X'$. 

The strange distribution can also be extracted indirectly using a
combination 
of charged ($W^\pm$) and neutral ($\gamma$) current data; however, the systematic uncertainties
involved
in this procedure make an accurate determination
difficult.\cite{yang}
The basic idea is to use the relation
 \begin{equation}
\frac{F_2^{NC}}{F_2^{CC}} =
\frac{5}{18}
\left\{ 
1 - \frac{3}{5} \, \frac{(s+\bar{s})-(c+\bar{c}) + ...}{q+\bar{q}}
\right\}  
\end{equation}
to extract the strange distribution. This method is complicated
by a number of issues including the $xF_3$ component which can 
play a crucial role in the small-$x$ region---precisely
the region where there has been a long-standing discrepancy.

The structure functions are defined in 
terms of the neutrino-nucleon 
cross section via: 
\[
\frac{d^2 \sigma^{\nu,\bar{\nu}}}{dx \, dy}
=
\frac{\scriptstyle G_F^2 M E}{\pi}
\left[ 
F_2 (1-y) + x F_1 y^2 \pm x F_3 y (1-\frac{y}{2}) 
\right]
\nonumber  
\]
It is instructive to recall the simple leading-order correspondence 
between the $F$'s and the PDF's:\footnote{%
To exhibit the basic structure, 
the above is taken the limit of 4 quarks, a symmetric sea, and a 
vanishing Cabibbo angle.
 Of course, the actual analysis takes into account the full
structure.\cite{yang}
}
\begin{eqnarray}
F_2^{(\nu,\bar{\nu}) N} &=&
  x 
\left\{ 
u + \bar{u} + d + \bar{d} + 2s + 2c
\right\} 
\nonumber \\
x F_3^{(\nu,\bar{\nu}) N} &=&
 x 
\left\{ 
u - \bar{u} + d - \bar{d} \pm 2s \mp 2c
\right\} 
\end{eqnarray}
 Therefore, the combination $\Delta xF_3$:
\begin{equation}
\Delta xF_3 = x F_3^{\nu N} - x F_3^{\bar{\nu} N} =
4 x \{ s - c \}
\end{equation}
 can be used to
probe the strange sea distribution, 
and to understand heavy quark (charm) production. 
 This information, together with the exclusive dimuon events,
may provide a more precise determination of the strange quark sea. 

To gauge the dependence of $\Delta xF_3$ upon various factors, we first 
consider $x s(x,\mu)$  in \fig{dff}, and then the full NLO $\Delta xF_3$
in \fig{df}; this allows us to see the connection between $\Delta xF_3$ and
$x s(x,\mu)$ beyond leading order. 
 In \fig{dff} we have plotted the quantity $x s(x,\mu)$  
{\it vs.} $Q^2$ for two choices of $x$ in a range relevant to the
the dimuon measurements. 
 We use three choices of the $\mu^2$ scale: $\{Q^2, Q^2+m_c^2, P_{T_{max}}^2  \}$.
The choices $Q^2$ and $Q^2+m_c^2$ differ only at lower values of $Q^2$; 
the choice $P_{T_{max}}^2$ is comparable to $Q^2$ and $Q^2+m_c^2$ at $x=0.08$
but lies above for $x=0.015$. 
The fourth curve labeled $Q^2+``SR"$ uses $\mu^2=Q^2$ with a ``slow-rescaling" 
type of correction which (crudely) includes mass effects by 
shifting $x$ to $x(1+m_c^2/Q^2)$; note, the result of this 
correction is significant at large $x$ and low $Q^2$. 

In \fig{df} we have plotted the quantity $\Delta xF_3/2$ 
for an isoscalar target computed to 
order $\alpha_s^1$.
 We display three calculations for three different $x$-bins relevant to 
strange sea measurement.
 1)~A 3-flavor calculation using the GRV98\cite{grv98o} distributions,\footnote{%
 The scale choice $\mu=\sqrt{Q^2+m^2}$ for the 3-flavor GRV calculation 
precisely cancels the
collinear strange quark mass logarithm in the coefficient function
thereby making the coefficient function an exact scaling 
function, {\it i.e.} independent of $\mu^2$.}
 and $\mu=\sqrt{Q^2+m^2}$. 
 2)~A 3-flavor calculation using the CTEQ4HQ distributions, and $\mu=Q$. 
 3)~A 4-flavor calculation using the CTEQ4HQ distributions, and $\mu=Q$. 

The two CTEQ curves show the effect of the charm distribution, and
the GRV curve shows the effect of using a different PDF set.
Recall that the GRV calculation corresponds to a FFN scheme.

The pair of curves using the CTEQ4HQ distributions nicely illustrates
how the charm distribution $c(x,\mu^2)$ evolves as $\ln(Q^2/m_c^2)$ for
increasing $Q^2$; note, $c(x,\mu^2)$ enters with a negative sign 
so that the 4-flavor result is below the 3-flavor curve. 
 The choice $\mu=Q$ ensures the 3- and 4-flavor calculation coincide
at $\mu=Q=m_c$; while this choice is useful for instructive purposes, 
a more practical choice might be $\mu \sim \sqrt{Q^2+m^2}$, 
{\it cf.}, Sec.~\ref{sec:schemes}, and Ref.~\cite{schmidt}.
 
For comparison, we also display preliminary data from the 
CCFR analysis.\cite{yang} While there is much freedom in the
theoretical calculation, the difference between these calculations
and the data at low $Q$ values warrants further investigation.

\subsection{Conclusions and Outlook}

A detailed understanding of heavy quark production and heavy quark PDF's
at the Tevatron Run~II will require analysis of fixed-target and HERA
data as well as Run~I results.  
 Comprehensive analysis of the combined data set can provide incisive tests
of the theoretical methods in an unexplored regime, and enable precise
predictions that will facilitate new particle searches in a variety of channels. 
 This document serves as a progress report, and work on these 
topics will continue in preparation for   the Tevatron Run~II.


 This work is supported
by the U.S. Department of Energy, 
the National Science Foundation, 
and the Lightner-Sams Foundation.



\newpage
\setcounter{section}{0}
\section*{PARTON DENSITIES FOR HEAVY QUARKS}

J. Smith\footnote{Work supported in part by the NSF grant
PHY-9722101}
\\ 

C.N. Yang Institute for Theoretical Physics, SUNY at Stony Brook ,
Stony Brook, NY 11794-3840) \\

\begin{center}
			Abstract
\end{center}
We compare parton densities for heavy quarks.

\vskip 1cm
Reactions with incoming heavy (c,b) quarks are often
calculated with heavy quark densities
just like those with incoming light mass (u,d,s) quarks
are calculated with light quark densities.
The heavy quark densities are derived 
within the framework of the so-called zero-mass 
variable flavor number scheme (ZM-VFNS). In this scheme these quarks 
are described by massless densities which are zero below 
a specific mass scale $\mu$. The latter depends on $m_c$ or $m_b$.
Let us call this scale the matching point.
Below it there are $n_f$ 
massless quarks described by $n_f$ massless densities.
Above it there are
$n_f + 1$ massless quarks described by $n_f + 1$ massless densities.
The latter densities are used to  
calculate processes with a hard scale $M \gg m_c, m_b$.
For example in the production of single top quarks via the weak process
$q_i + b \rightarrow  q_j + t$, where $q_i$, $q_j$ are light mass 
quarks in the proton/antiproton, one can argue that $M = m_t$
should be chosen as the large scale and $m_b$ can be neglected. 
Hence the incoming bottom quark
can be described by a massless bottom quark density.

\begin{figure}[ht]
   \epsfig{file=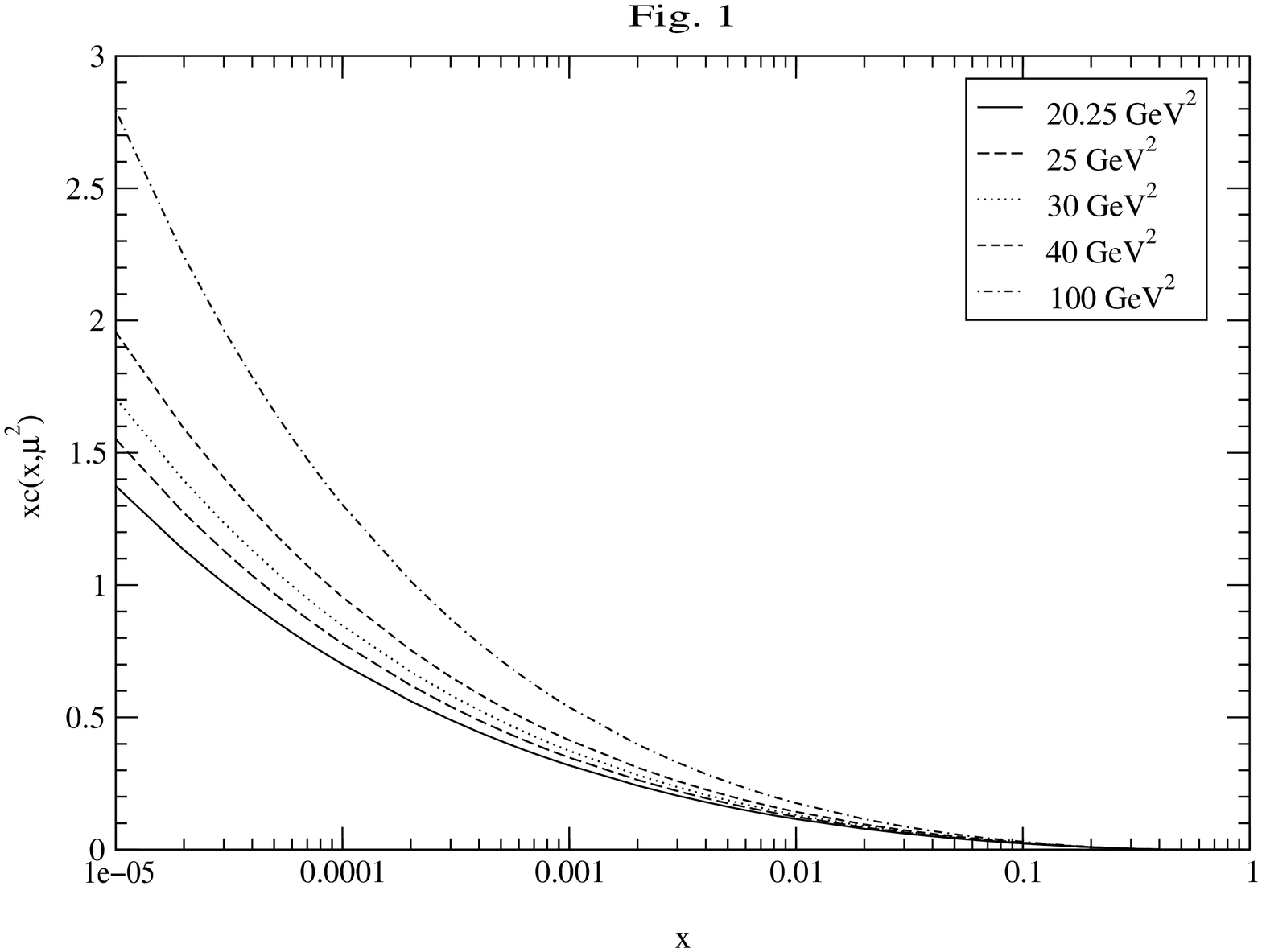,width=6cm,angle=270 }
\vspace{-0.8cm}
\caption{The charm quark density $xc_{\rm NNLO}(5,x,\mu^2)$ in the range
$10^{-5} < x < 1$ for $\mu^2 =$ 20.25, 25, 30, 40 and 100 in units
of $({\rm GeV/c}^2)^2$.}
\vspace{-0.4cm}
\end{figure}                                                              

The generation of these densities starts from the solution of
the evolution equations for $n_f$ massless quarks below the matching point.
At and above this point one solves the evolution equations for $n_f+1$
massless quarks. However in contrast to the parameterization
of the $x$-dependences of the light quarks and gluon at the initial
starting scale,
the $x$ dependence of the heavy quark 
density at the matching point is fixed. In perturbative QCD it is
defined by convolutions of the densities for the $n_f$ quarks 
and the gluon with specific operator matrix elements (OME's), 
which are now know up to $O(\alpha_s^2)$ \cite{bmsn1}. 
These matching conditions determine both the ZM-VFNS density
and the other light-mass quark and gluon densities at the matching points.
Then the evolution equations determine the new densities at larger
scales. The momentum sum rule is satisfied for the $n_f + 1$ quark
densities together with the corresponding gluon density.
\begin{figure}[ht]
   \epsfig{file=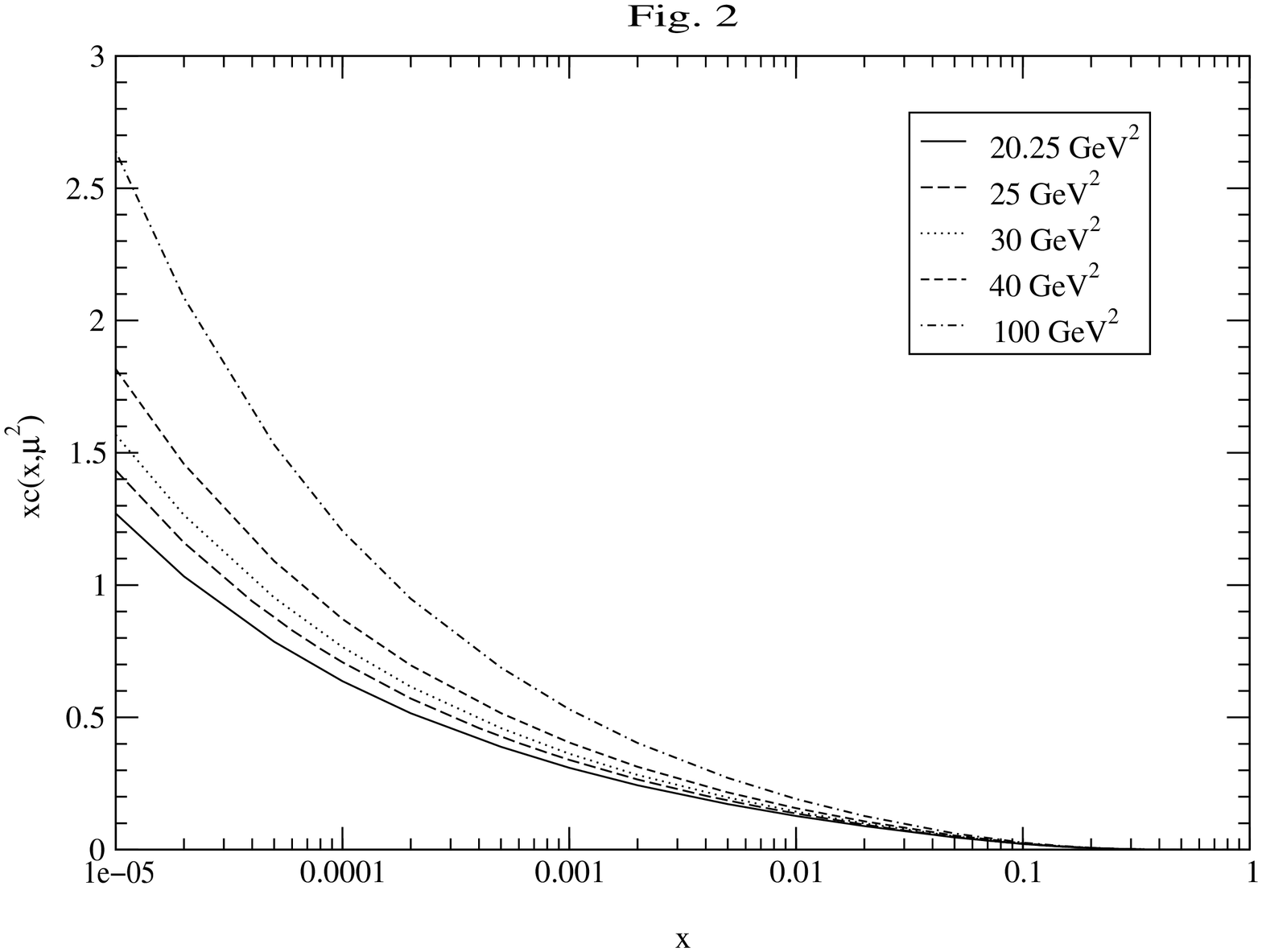,width=6cm,angle=270 }
\vspace{-0.5cm}
\caption{Same as Fig.1 for the NLO results from MRST98 set 1.}
\vspace{-0.4cm}
\end{figure}                                                              
\begin{figure}[ht]
   \epsfig{file=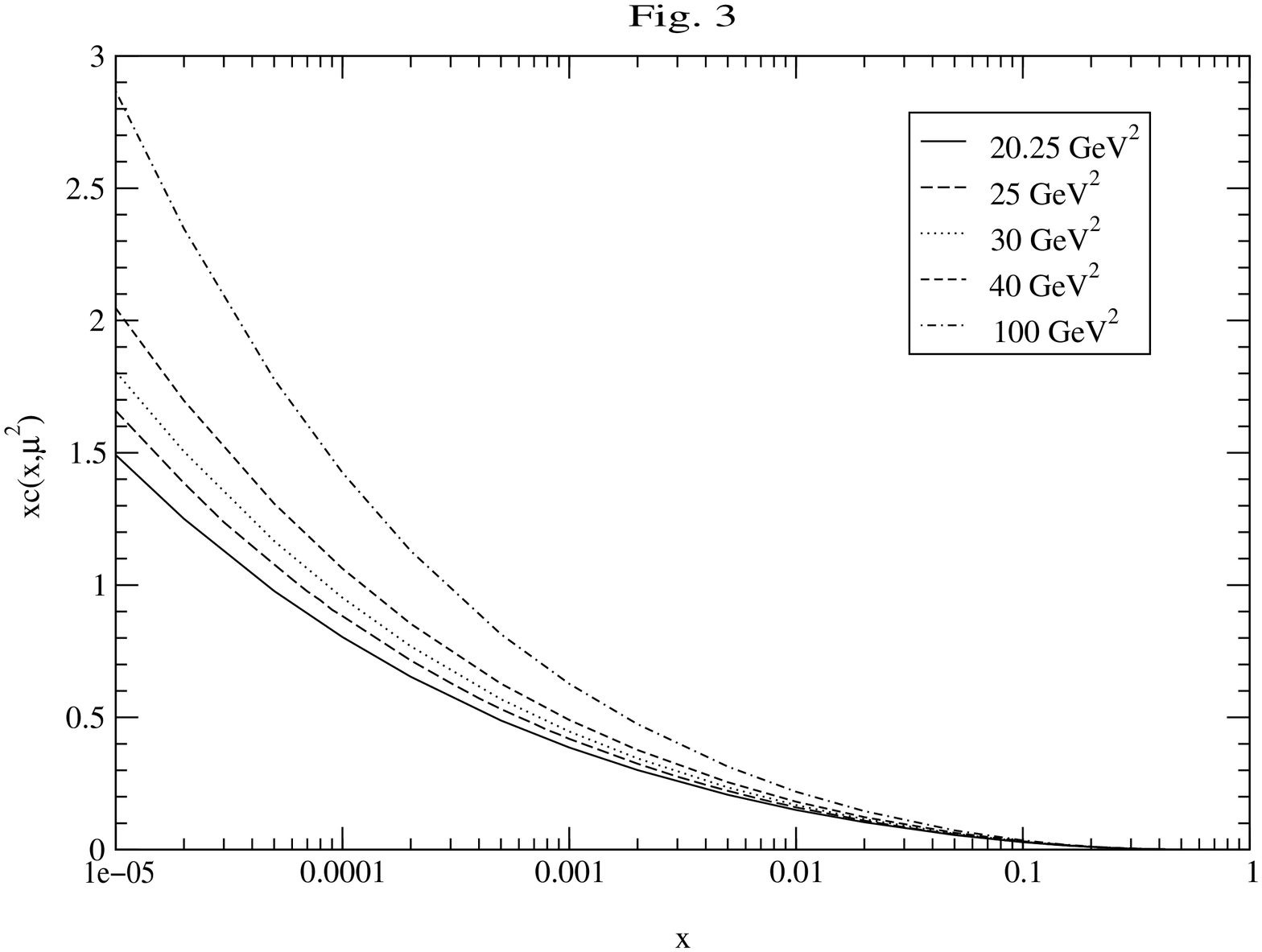,width=6cm,angle=270 }
\vspace{-0.5cm}
\caption{Same as Fig.1 for the NLO results from CTEQ5HQ.} 
\vspace{-0.4cm}
\end{figure}                                                              

Parton density sets contain densities for charm and bottom quarks, which
generally directly follow this approach or some modification of it. 
The latest CTEQ densities \cite{cteq5_s} 
use $O(\alpha_s)$ matching conditions. The $x$ 
dependencies of the heavy c and b-quark densities 
are zero at the matching points.
The MRST densities \cite{mrst98} have more complicated matching conditions 
designed so that the derivatives of the deep inelastic structure 
functions $F_2$ and $F_L$ with 
regard to $Q^2$ are continuous at the matching points.
Recently we have provided another set of ZM-VFNS densities \cite{cs1}, 
which are based on extending the GRV98 three-flavor densities in \cite{grv98_s}
to four and five-flavor sets. GRV give the formulae for their 
LO and NLO three flavor densities at very small scales. 
They never produced a c-quark density but advocated that charm quarks
should only exist in the final state of production reactions, which
should be calculated from NLO QCD with massive quarks as in \cite{lrsn}.
We have evolved their LO and NLO densities across the 
matching point $\mu = m_c$ with $O(\alpha_s^2)$
matching conditions to provide LO and NLO four-flavor densities containing 
massless c-quark densities. Then these LO and NLO densities were evolved 
between $\mu = m_c$ and $\mu = m_b$ with four-flavor LO and
NLO splitting functions. 
At this new matching point the LO and NLO four-flavor densities were then 
convoluted with the $O(\alpha_s^2)$ OME's to form five-flavor
sets containing massless b-quarks. These LO and NLO
densities were then evolved 
to higher scales with five-flavor LO and NLO splitting functions. 
Note that the $O(\alpha_s^2)$ matching conditions should
really be used with NNLO splitting functions to produce NNLO
density sets. However the latter splitting functions are not yet available, 
so we make the approximation of replacing the NNLO splitting functions
with NLO ones. 
\begin{figure}[ht]
   \epsfig{file=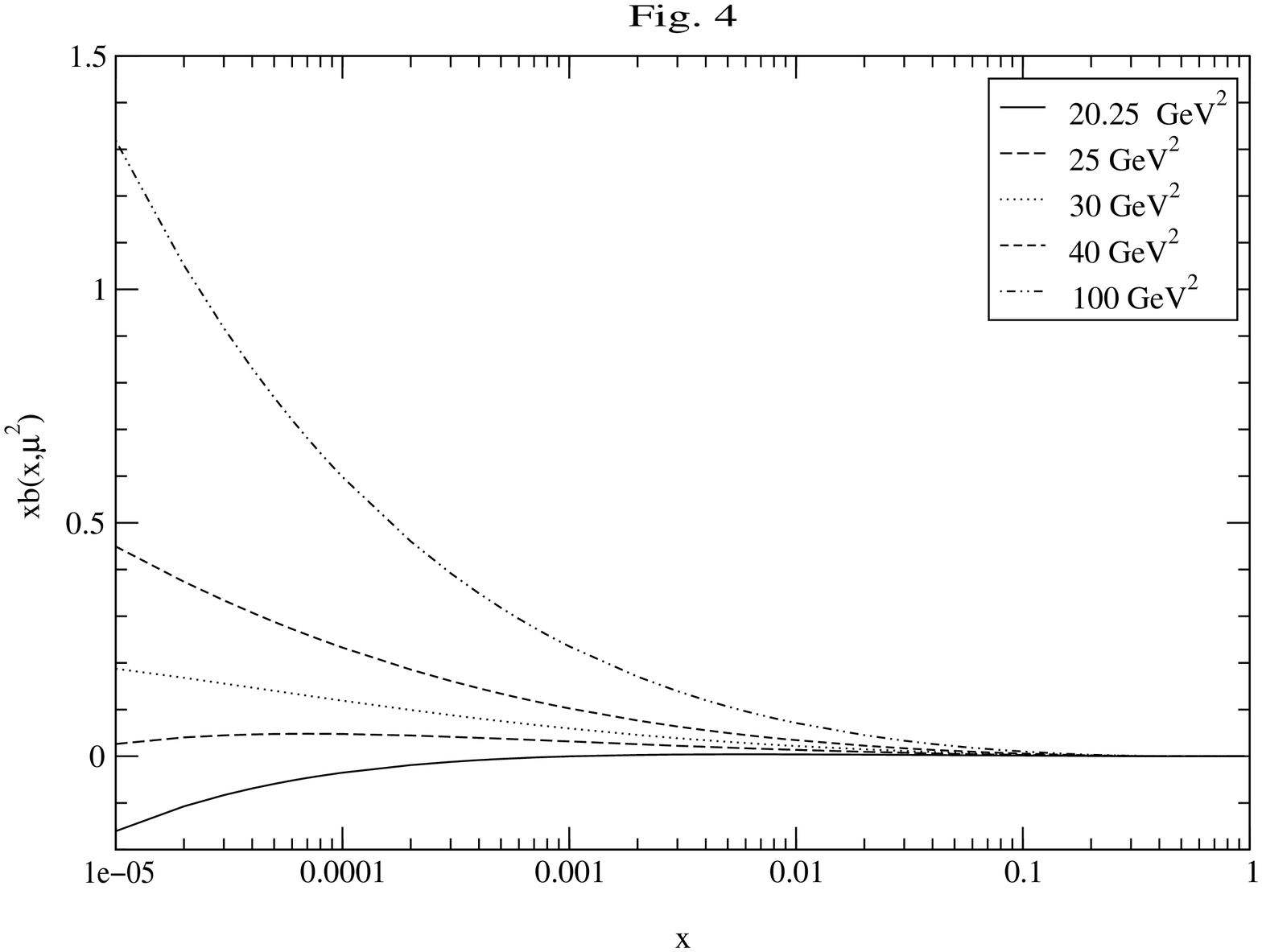,width=6cm,angle=270 }
\vspace{-0.8cm}
\caption{The bottom quark density $xb_{\rm NNLO}(5,x,\mu^2)$ in the range
$10^{-5} < x < 1$ for $\mu^2 =$ 20.25, 25, 30, 40 and 100 in units
of $({\rm GeV/c}^2)^2$.}
\vspace{-0.4cm}
\end{figure}                                                              

In this short report we would like to compare the charm and bottom 
quark densities in the CS, MRS and CTEQ sets.
We concentrate on the five-flavor densities, which are more important
for Tevatron physics. In the CS set they start at $\mu^2 = m_b^2 = 20.25$ 
${\rm GeV}^2$. At this scale the charm densities in the
CS, MRST98 (set 1) and CTEQ5HQ sets are shown in Figs.1,2,3
respectively. Since the CS charm density starts off negative for small $x$ at
$\mu^2 = m_c^2 = 1.96$ ${\rm GeV}^2$ it evolves less than the corresponding
CTEQ5HQ density. At larger $\mu^2$ all the CS curves in Fig.1 are below 
those for CTEQ5HQ in Fig.3 although the differences are small.
In general the CS c-quark densities
are more equal to those in the MRST (set 1) in Fig.2.

\begin{figure}[ht]
   \epsfig{file=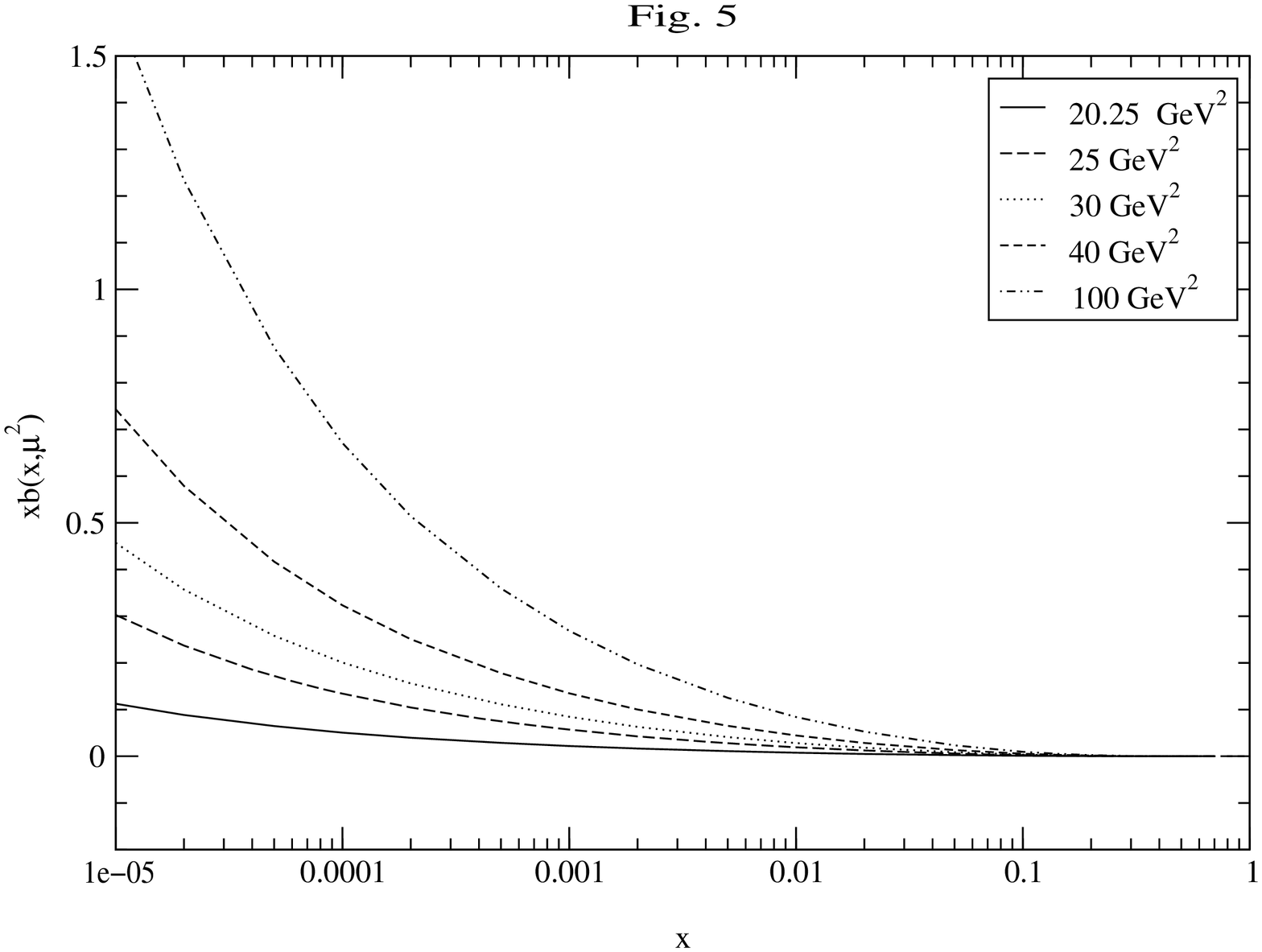,width=6cm,angle=270 }
\vspace{-0.5cm}
\caption{Same as Fig.4 for the NLO results from MRST98 set 1.}
\vspace{-0.4cm}
\end{figure}                                                              

At the matching point $\mu^2 = 20.25$ ${\rm GeV}^2$
the b-quark density also starts off negative at small $x$ 
as can be seen in Fig.4, which is a consequence of the explicit
form of the OME's in \cite{bmsn1}. At $O(\alpha_s^2)$ the OME's
have nonlogarithmic terms which do not vanish at the matching point
and yield a finite function in $x$, which is the boundary value
for the evolution of the b-quark density.
This negative start slows down the evolution of the b-quark density
at small $x$ as the scale $\mu^2$ increases. Hence the CS densities 
at small $x$ in Fig.4
are smaller than the MRST98 (set 1) densities in Fig.5 and the CTEQ5HQ 
densities in Fig.6 at the same values of $\mu^2$.
The differences between the sets are still small, of the order of five percent
at small $x$ and large $\mu^2$. Hence it should not really matter
which set is used to calculate cross sections for processes
involving incoming b-quarks at the Tevatron.

\begin{figure}[ht]
   \epsfig{file=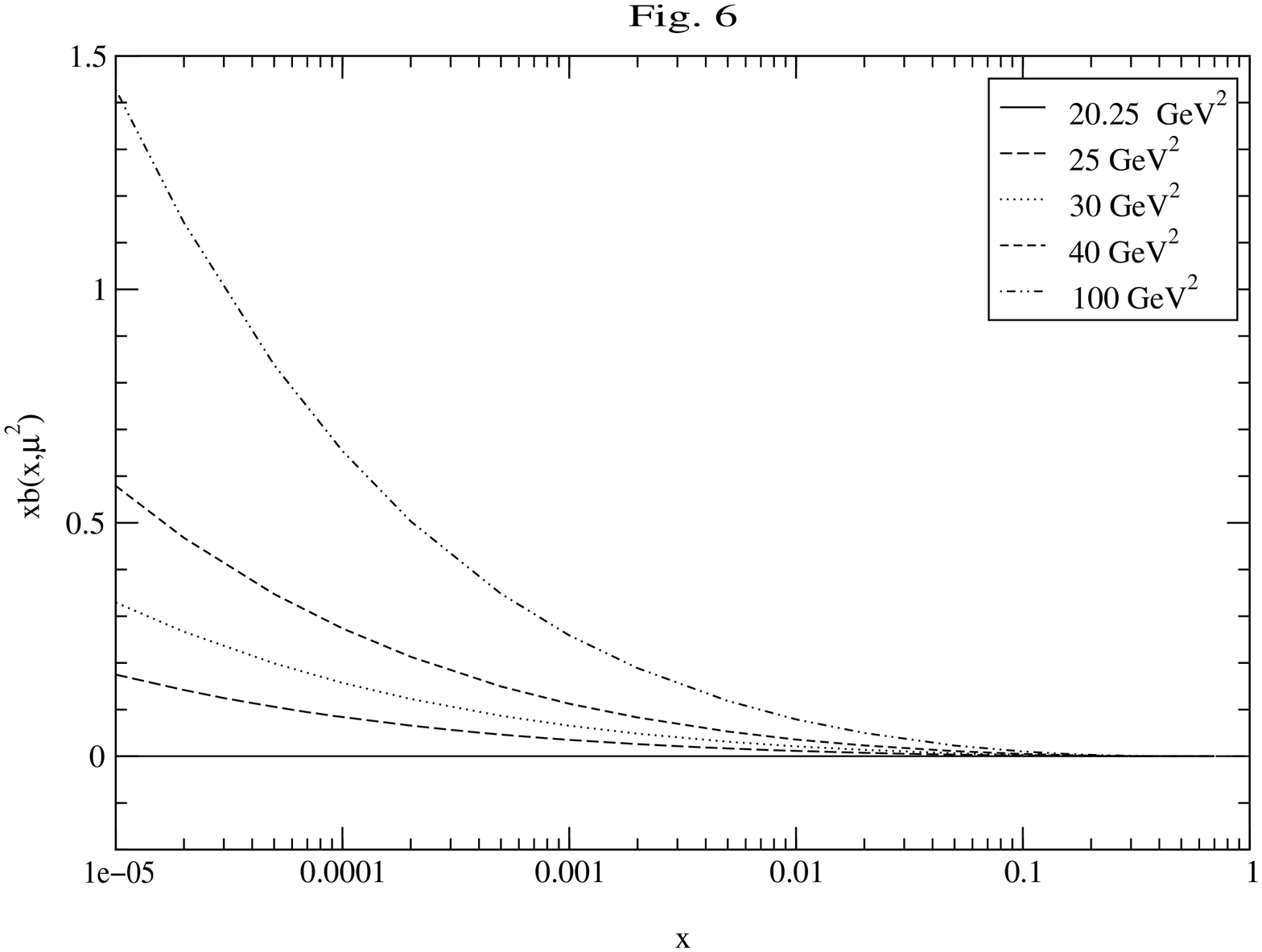,width=6cm,angle=270 }
\vspace{-0.5cm}
\caption{Same as Fig.4 for the NLO results from CTEQ5HQ.}
\vspace{-0.4cm}
\end{figure}                                                              

We suspect that the differences between these results 
for the heavy c and b-quark densities are primarily due to the
different gluon densities in the three sets rather to than the effects
of the different boundary conditions.
This could be checked theoretically if both LO and
NLO three-flavor sets were provided by MRST and CTEQ
at small scales. Then we could rerun our programs to generate
sets with $O(\alpha_s^2)$ boundary conditions. However these inputs are
not available. We note that CS uses the GRV98 LO and NLO gluon densities, 
which are rather steep in $x$ and generally
larger than the latter sets at the same values of $\mu^2$.
Since the discontinuous boundary conditions
suppress the charm and bottom densities at small $x$, they enhance the
gluon densities in this same region (in order that the
momentum sum rules are satisfied).
Hence the GRV98 three flavour gluon densities and the 
CS four and five flavor gluon densities are generally significantly  
larger than those in MRST98 (set 1) and CTEQ5HQ. Unfortunately 
experimental data are not yet precise enough to decide which set is
the best one. We end by noting that all these densities are 
given in the $\overline{\rm MS}$ scheme.

\newpage
\setcounter{section}{0}
\section*{CONSTRAINTS ON THE GLUON DENSITY FROM LEPTON PAIR PRODUCTION}

E.~L.~Berger$^{a,}$\footnote{Supported by the U.S.\ Department of
Energy, Division of High Energy Physics, under Contract
W-31-109-ENG-38.} and M.~Klasen$^{b,}$\footnote{Supported by
Bundesministerium f\"ur Bildung und Forschung under Contract 05 HT9GUA
3, by Deutsche Forschungsgemeinschaft under Contract KL 1266/1-1, and
by the European Commission under Contract ERBFMRXCT980194.}.\\

a) HEP Theory Group, Argonne National Laboratory, 9700 South Cass
Avenue, Argonne, IL 60439,USA; b) II.\ Institut f\"ur Theoretische
Physik, Universit\"at Hamburg, Luruper Chaussee 149, D-22761 Hamburg,
Germany\\ 
       
\begin{center}
		Abstract
\end{center}

The hadroproduction of lepton pairs with mass $Q$ and finite transverse
momentum $Q_T$ is described in perturbative QCD by the same partonic
subprocesses as prompt photon production. We demonstrate that, like prompt
photon production, lepton pair production is dominated by quark-gluon
scattering in the region $Q_T>Q/2$. This feature leads to sensitivity to the 
gluon density in kinematical regimes accessible in collider and 
fixed target experiments, and it provides a new independent method for 
constraining the gluon density.

\section{Introduction}
The production of lepton pairs in hadron collisions $h_1h_2\rightarrow\gamma^*
X;\gamma^*\rightarrow l\bar{l}$ proceeds through an intermediate
virtual photon via $q {\bar q} \rightarrow \gamma^*$, and the subsequent 
leptonic decay of the virtual photon. Traditionally, interest in
this Drell-Yan process has concentrated on lepton pairs with
large mass $Q$ which justifies the application of perturbative QCD and allows 
for the extraction of the antiquark density in hadrons \cite{Drell:1970wh}.

Prompt photon production $h_1h_2\rightarrow\gamma X$ can be calculated in
perturbative QCD if the transverse momentum $Q_T$ of the photon is
sufficiently large. Because the quark-gluon Compton subprocess is dominant, 
$g q \rightarrow \gamma X$, this reaction provides essential information on the
gluon density in the proton at large $x$ \cite{Martin:1998sq}. Unfortunately,
the analysis suffers from fragmentation, isolation, and intrinsic transverse
momentum uncertainties. Alternatively, the gluon density can be constrained
from the production of jets with large transverse momentum at hadron colliders
\cite{Lai:1999wy}, but the information from different experiments and colliders 
is ambiguous.

In this paper we demonstrate that, like prompt photon production, lepton pair
production is dominated by quark-gluon scattering in the region $Q_T>Q/2$.
This realization means that new independent constraints on the gluon density 
may be derived from Drell-Yan data in kinematical regimes that are accessible 
in collider and fixed target experiments but without the theoretical and 
experimental uncertainties present in the prompt photon case.

In Sec.~\ref{sec:2}, we review the relationship between virtual and
real photon production in hadron collisions in next-to-leading order QCD.
In Sec.~\ref{sec:3} we present our numerical results, and Sec.~\ref{sec:4}
is a summary.

\section{Next-to-leading order qcd formalism}
\label{sec:2}

In leading order (LO) QCD, two partonic subprocesses contribute to the
production of virtual and real photons with non-zero transverse momentum:
$q\bar{q}\rightarrow\gamma^{(*)}g$ and $qg\rightarrow\gamma^{(*)}q$.
The cross section for lepton pair production is related to the cross section
for virtual photon production through the leptonic branching ratio of the
virtual photon $\alpha/(3\pi Q^2)$. The virtual photon cross section reduces
to the real photon cross section in the limit $Q^2\rightarrow 0$.

The next-to-leading order (NLO) QCD corrections arise from virtual one-loop
diagrams interfering with the LO diagrams and from real emission diagrams. At
this order $2 \rightarrow 3$ partonic processes with incident gluon pairs $(gg)$, 
quark pairs $(qq)$, and non-factorizable quark-antiquark $(q\bar{q}_2)$ processes 
contribute also.  Singular contributions are regulated in $n$=4-2$\epsilon$ 
dimensions and removed through $\overline{\rm MS}$ renormalization, factorization, 
or cancellation between virtual and real contributions. An important difference
between virtual and real photon production arises when a quark emits a
collinear photon. Whereas the collinear emission of a real photon leads to a
$1/\epsilon$ singularity that has to be factored into a fragmentation
function, the collinear emission of a virtual photon yields a finite
logarithmic contribution since it is regulated naturally by the photon
virtuality $Q$. In the limit $Q^2\rightarrow 0$ the NLO virtual photon
cross section reduces to the real photon cross section if this logarithm is
replaced by a $1/\epsilon$ pole. A more detailed discussion can be found
in \cite{Berger:1998ev}.

The situation is completely analogous to hard
photoproduction where the photon participates in the scattering in the initial
state instead of the final state. For real photons, one encounters an
initial-state singularity that is factored into a photon structure function.
For virtual photons, this singularity is replaced by a logarithmic dependence
on the photon virtuality $Q$ \cite{Klasen:1998jm}.

A remark is in order concerning the interval in $Q_T$ in which our analysis is 
appropriate.  In general, in two-scale situations, a series of logarithmic 
contributions will arise with terms of the type $\alpha_s^n \ln^n (Q/Q_T)$.  Thus, 
if either $Q_T >> Q$ or $Q_T << Q$, resummations of this series must be considered. 
For practical reasons, such as event rate, we do not venture into the domain 
$Q_T >> Q$, and our fixed-order calculation should be adequate.  On the 
other hand, the cross section is large in the region $Q_T << Q$.  In previous 
papers~\cite{Berger:1998ev}, we compared our cross sections with available 
fixed-target and collider data on massive lepton-pair production, and we were able
to establish that fixed-order perturbative calculations, without resummation, 
should be reliable for $Q_T > Q/2$.  At smaller values of $Q_T$, non-perturbative 
and matching complications introduce some level of phenomenological ambiguity.  For 
the goal we have in mind, viz., contraints on the gluon density, it would appear 
best to restrict attention to the region $Q_T \geq Q/2$, but below $Q_T >> Q$.

\section{Predicted cross sections}
\label{sec:3}

In this section we present numerical results for the production of lepton pairs
in $p\bar{p}$ collisions at the Tevatron with center-of mass energy
$\sqrt{S}=1.8$ and 2.0 TeV.
We analyze the invariant cross section $Ed^3\sigma/
dp^3$ averaged over the rapidity interval -1.0 $<y<$ 1.0.
We integrate the cross section over various intervals of pair-mass 
$Q$ and plot it as a function of the transverse momentum $Q_T$.
Our predictions are based on a NLO QCD calculation \cite{Arnold:1991yk} and
are evaluated in the $\overline{\rm MS}$ renormalization scheme. The
renormalization and factorization scales are set to $\mu=\mu_f=
\sqrt{Q^2+Q_T^2}$. If not stated otherwise, we use the CTEQ4M
parton distributions \cite{Lai:1997mg} and the corresponding value of
$\Lambda$ in the two-loop expression of $\alpha_s$ with four flavors (five if
$\mu>m_b$). The Drell-Yan factor $\alpha/(3\pi Q^2)$ for the decay of the
virtual photon into a lepton pair is included in all numerical results.

\begin{figure}[htb]
 \begin{center}
  {\unitlength1cm
  \begin{picture}(7.6,10.5)
   \epsfig{file=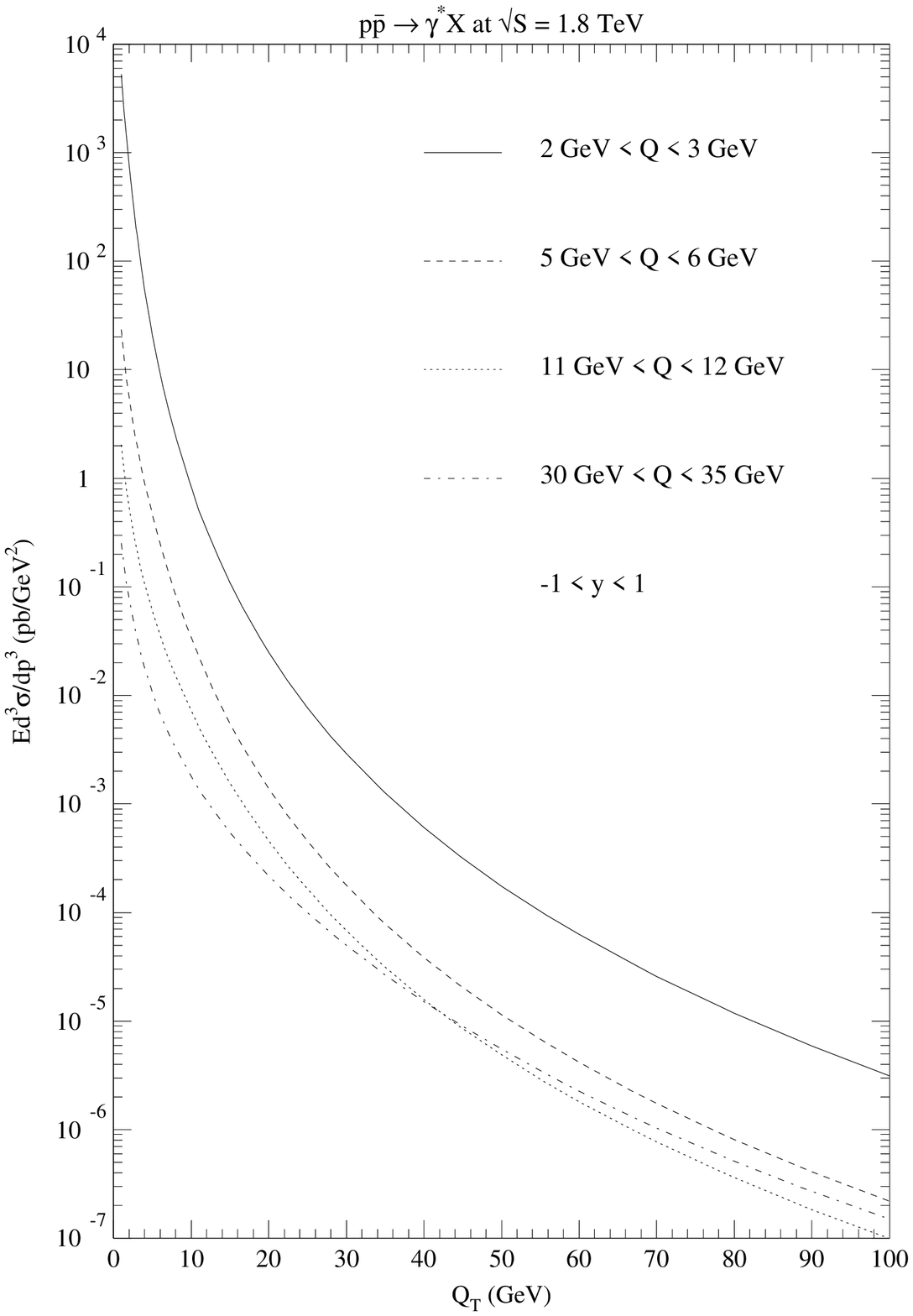,bbllx=60pt,bblly=100pt,bburx=495pt,bbury=725pt,%
           height=10.5cm}
  \end{picture}}
 \end{center}
\vspace*{-1cm}
\caption{Invariant cross section $Ed^3\sigma/dp^3$ as a function of $Q_T$
for $p\bar{p} \rightarrow \gamma^* X$ at $\sqrt{S}=1.8$ TeV in
non-resonance regions of $Q$. The cross section falls with the mass of the
lepton pair $Q$ and, more steeply, with its transverse momentum $Q_T$.}
\label{fig:1}
\end{figure}

In Fig.~\ref{fig:1} we display the NLO QCD cross section for lepton pair
production at the Tevatron at $\sqrt{S}=1.8$ TeV as a function of $Q_T$ for
four regions of $Q$. The regions of $Q$ have been chosen to avoid
resonances, {\it i.e.\ } between $2$ GeV and the $J/\psi$ resonance,
between the $J/\psi$ and the $\Upsilon$ resonances, above the $\Upsilon$'s,
and a high mass region. The cross section falls both with the mass of the
lepton pair $Q$ and, more steeply, with its transverse momentum $Q_T$.
No data are available yet from the CDF and D0 experiments.  However, prompt 
photon production data exist to $Q_T\simeq 100$ GeV, where the cross section
is about $10^{-3}$ pb/GeV$^2$. It should be possible to analyze Run I
data for lepton pair production to at least $Q_T\simeq 30$ GeV where one
can probe the parton densities in the proton up to $x_T = 2Q_T/\sqrt{S}\simeq
0.03$. The UA1
collaboration measured the transverse momentum distribution of lepton
pairs at $\sqrt{S}=630$ GeV up to $x_T=0.13$ \cite{Albajar:1988iq}, and their
data agree well with our theoretical results \cite{Berger:1998ev}.

\begin{figure}[htb]
 \begin{center}
  {\unitlength1cm
  \begin{picture}(7.6,10.5)
   \epsfig{file=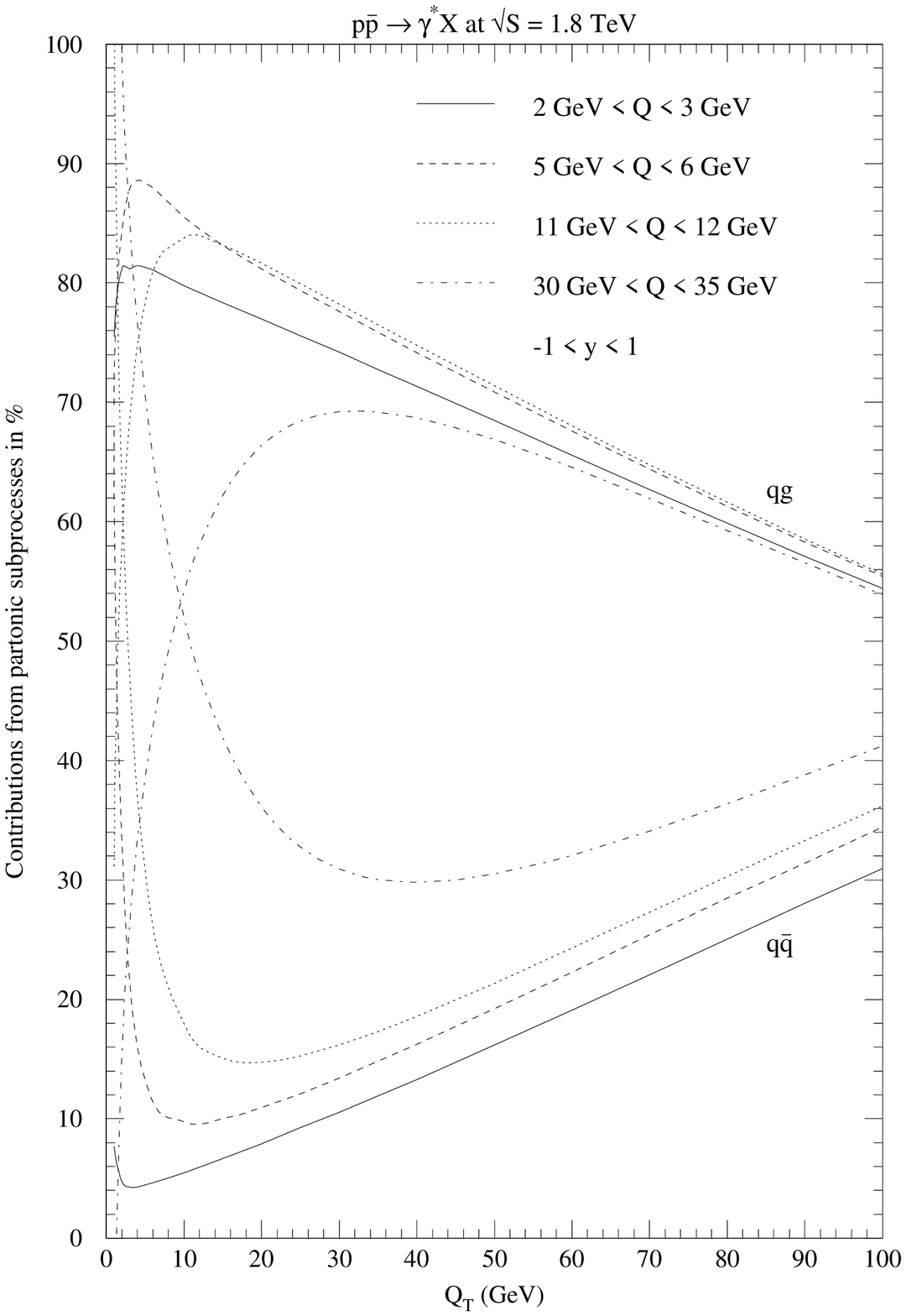,bbllx=60pt,bblly=100pt,bburx=495pt,bbury=725pt,%
           height=10.5cm}
  \end{picture}}
 \end{center}
\vspace*{-1cm}
\caption{Contributions from the partonic subprocesses $qg$ and $q\bar{q}$ to
the invariant cross section $Ed^3\sigma/dp^3$ as a function of $Q_T$
for $p\bar{p}\rightarrow \gamma^* X$ at $\sqrt{S}$ = 1.8 TeV. The
$qg$ channel dominates in the region $Q_T > Q/2$.}
\label{fig:2}
\end{figure}

The fractional contributions from the $qg$ and $q\bar{q}$ subprocesses 
through NLO are shown in Fig.~\ref{fig:2}. It is evident 
that the $qg$ subprocess is the most important subprocess as long as
$Q_T > Q/2$. The dominance of the $qg$ subprocess diminishes somewhat with $Q$,
dropping from over 80 \% for the lowest values of $Q$ to about 70 \%
at its maximum for $Q \simeq$ 30 GeV. In addition, for very large $Q_T$, the
significant luminosity associated with the valence dominated $\bar{q}$
density in $p\bar{p}$ reactions begins to raise the fraction of the cross
section attributed to the $q\bar{q}$ subprocesses.
Subprocesses other than those initiated by the $q\bar{q}$ and
$q g$ initial channels are of negligible import.

\begin{figure}[htb]
 \begin{center}
  {\unitlength1cm
  \begin{picture}(7.6,10.5)
   \epsfig{file=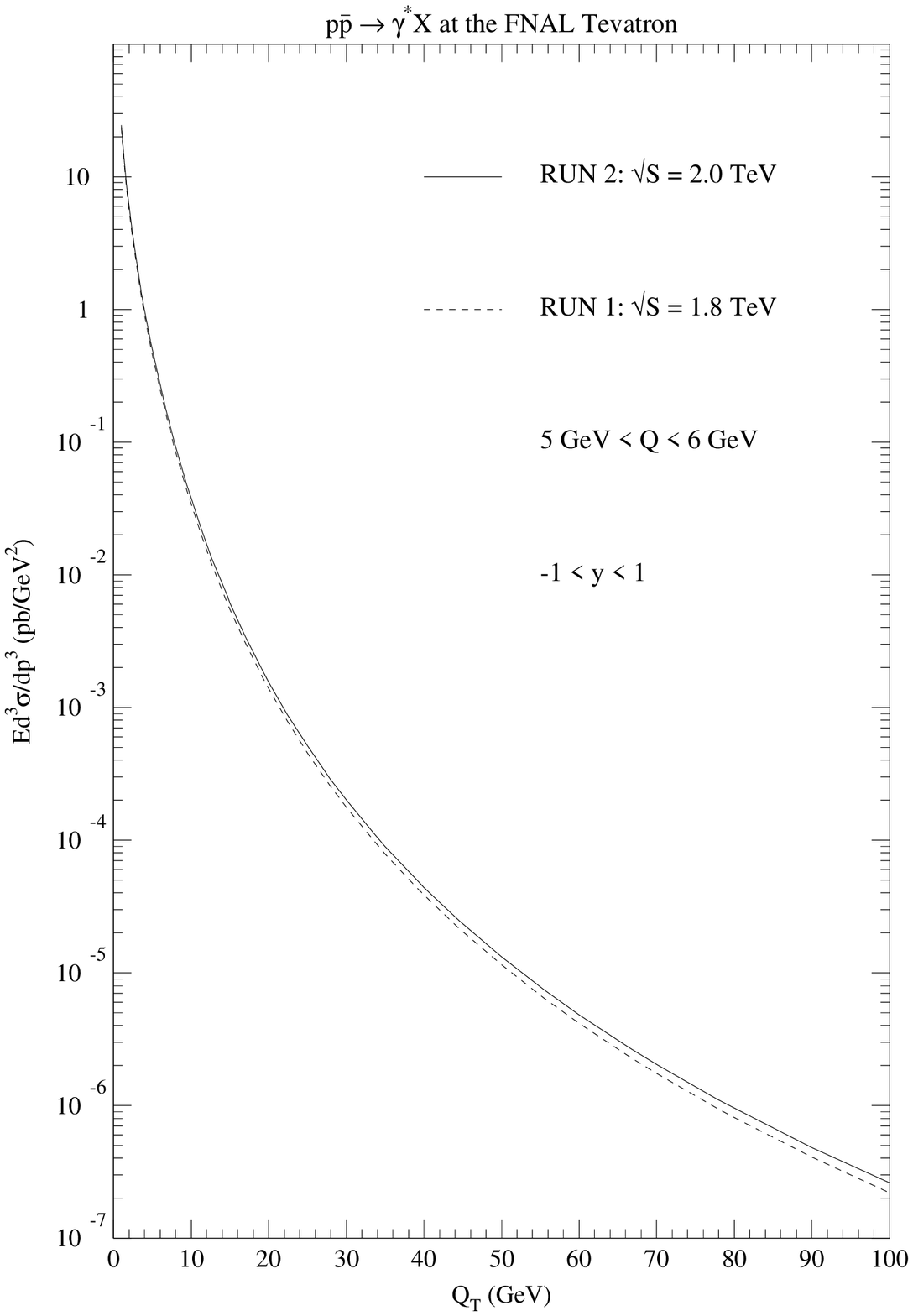,bbllx=60pt,bblly=100pt,bburx=495pt,bbury=725pt,%
           height=10.5cm}
  \end{picture}}
 \end{center}
\vspace*{-1cm}
\caption{Invariant cross section $Ed^3\sigma/dp^3$ as a function of
$Q_T$ for $p\bar{p} \rightarrow \gamma^* X$ and two different
center-of-mass energies of the Tevatron (Run 1: $\sqrt{S}=1.8$ TeV,
Run 2: $\sqrt{S}=2.0$ TeV). The cross section for Run 2 is
5 to 20 \% larger, depending on $Q_T$.}
\label{fig:7}
\end{figure}

We update the Tevatron center-of-mass energy to Run II conditions 
($\sqrt{S}= 2.0$ TeV) and use the latest global fit by the CTEQ 
collaboration (5M). Figure~\ref{fig:7}
demonstrates that the larger center-of-mass energy increases the
invariant cross section
for the production of lepton pairs with mass 5 GeV $<Q<$ 6 GeV by
5 \% at low $Q_T \simeq 1$ GeV and 20 \% at high $Q_T \simeq 100$ GeV. 
In addition, the expected luminosity for Run II of 2 fb$^{-1}$ should
make the cross section accessible to $Q_T\simeq 100$ GeV or
$x_T\simeq 0.1$. This extension would constrain the gluon density in the 
same regions as prompt photon production in Run I.

\setcounter{footnote}{0}

Next we present a study
of the sensitivity of collider and fixed target experiments to the gluon
density in the proton. The full uncertainty in the gluon density is not known.
Here we estimate this uncertainty from
the variation of different recent parametrizations. We choose the latest
global fit by the CTEQ collaboration (5M) as our point of reference
\cite{Lai:1999wy} and compare results to those based on their preceding analysis 
(4M\cite{Lai:1997mg}) and on a fit with a higher gluon density (5HJ) intended to
describe the CDF and D0 jet data at large transverse momentum.  We also compare 
to results based on global fits by MRST \cite{Martin:1998sq}, who provide three 
different sets with a central, higher, and lower gluon density, and to GRV98
\cite{Gluck:1998xa}\footnote{In this set a purely perturbative generation of
heavy flavors (charm and bottom) is assumed. Since we are working in a massless
approach, we resort to the GRV92 parametrization for the charm contribution
\cite{Gluck:1992ng} and assume the bottom contribution to be negligible.}.

\begin{figure}[htb]
 \begin{center}
  {\unitlength1cm
  \begin{picture}(7.6,10.5)
   \epsfig{file=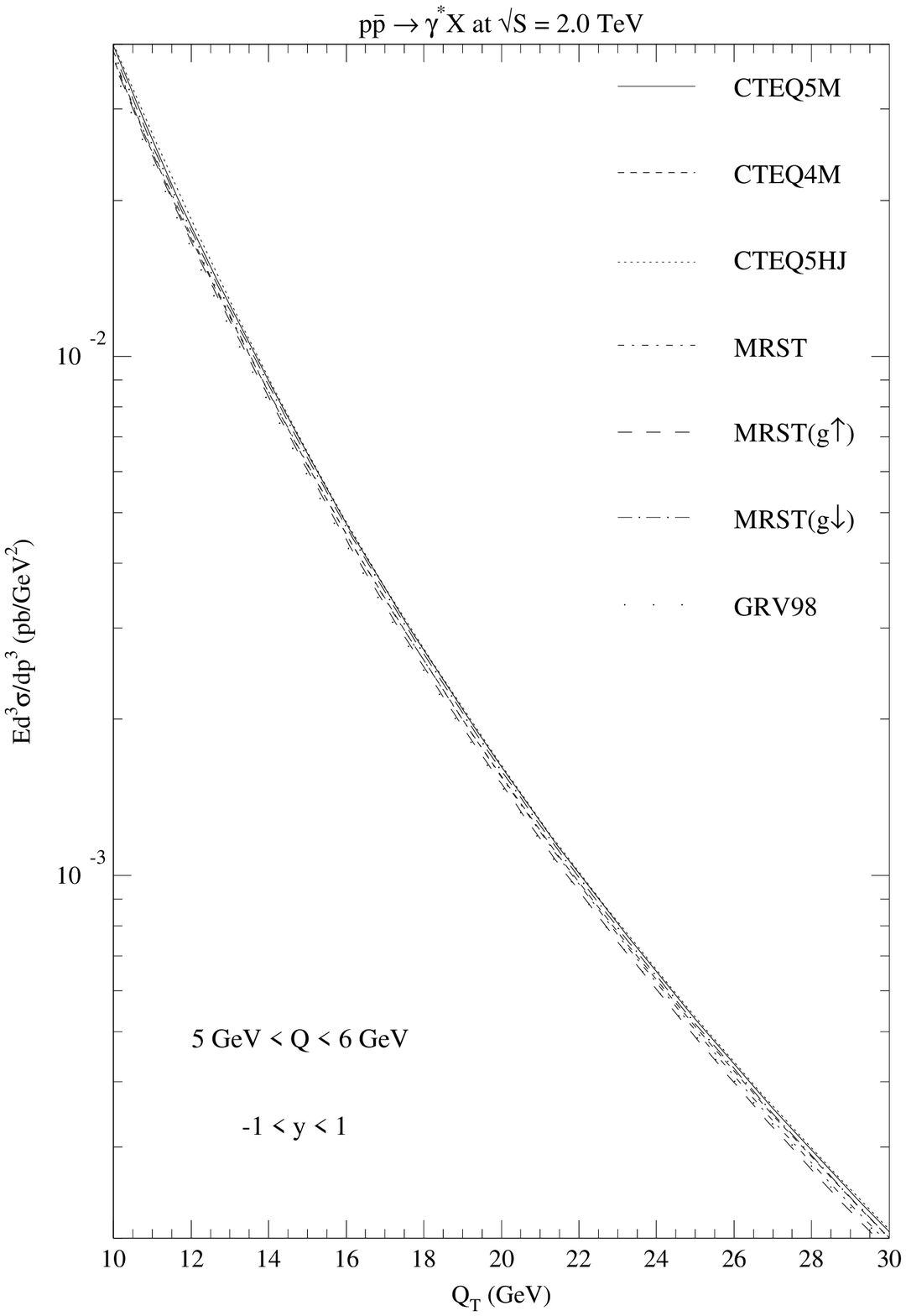,bbllx=60pt,bblly=100pt,bburx=495pt,bbury=725pt,%
           height=10.5cm}
  \end{picture}}
 \end{center}
\vspace*{-1cm}
\caption{Invariant cross section $Ed^3\sigma/dp^3$ as a function of $Q_T$
for $p\bar{p} \rightarrow \gamma^* X$ at $\sqrt{S}=2.0$ TeV in the
region between the $J/\psi$ and $\Upsilon$ resonances. The largest differences
from CTEQ5M are obtained with GRV98 at low $Q_T$ (minus 10 \%) and with
MRST(g$\uparrow$) at large $Q_T$ (minus 7 \%).}
\label{fig:5}
\end{figure}

In Fig.~\ref{fig:5} we plot the cross section for lepton pairs with mass 
between the
$J/\psi$ and $\Upsilon$ resonances at Run II of the Tevatron in the region
between $Q_T=10$ and 30 GeV ($x_T = 0.01 \dots 0.03$). For the CTEQ
parametrizations we find that the cross section increases from 4M to 5M by 2.5
\% ($Q_T=30$ GeV) to 5 \% ($Q_T=10$ GeV) and from 5M to 5HJ by 1 \% in the
whole $Q_T$-range. The largest differences from CTEQ5M are obtained with GRV98 at
low $Q_T$ (minus 10 \%) and with MRST(g$\uparrow$) at large $Q_T$ (minus 7\%).

\begin{figure}[htb]
 \begin{center}
  {\unitlength1cm
  \begin{picture}(9,8.08)
   \epsfig{file=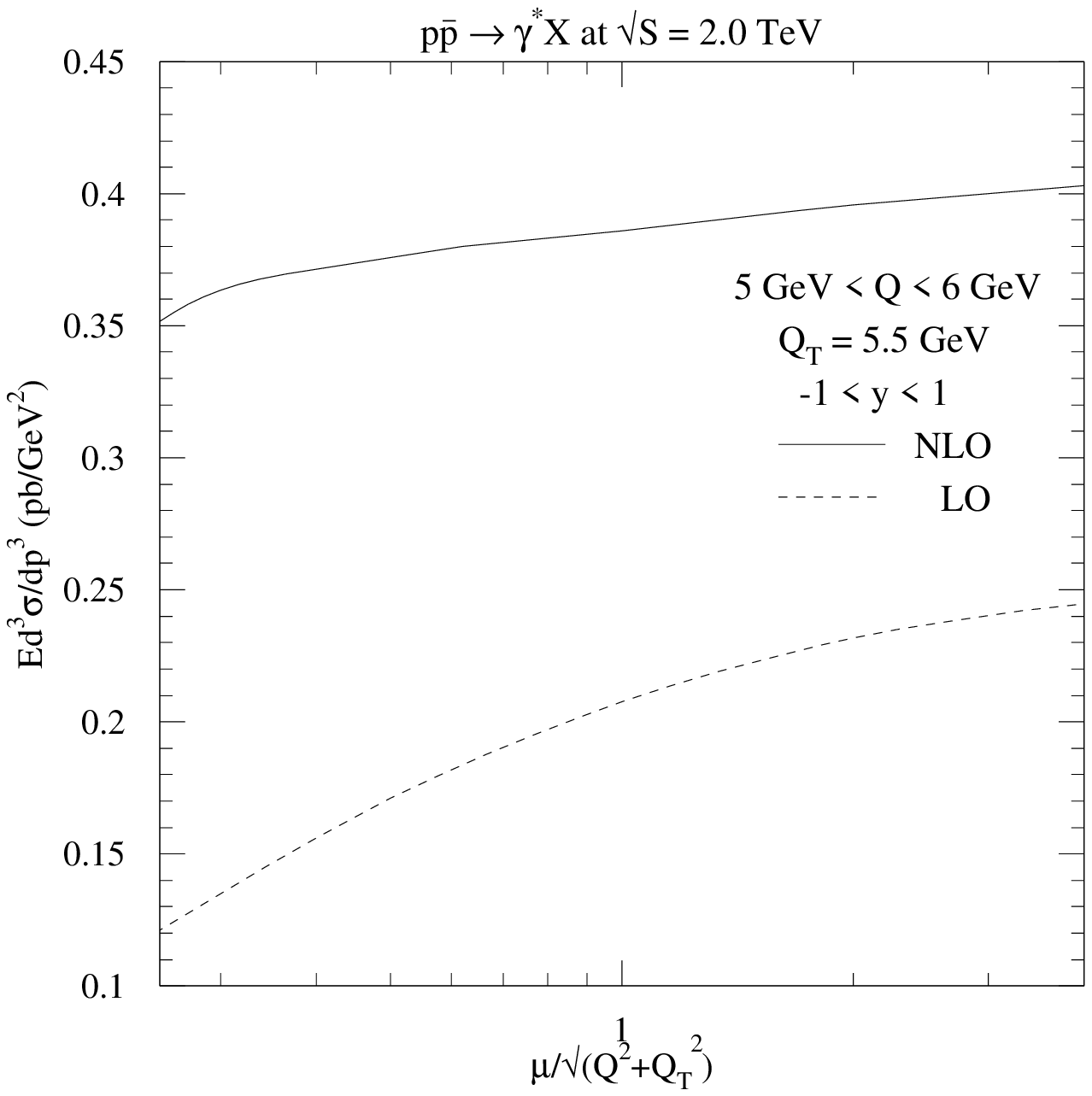,height=9cm}
  \end{picture}}
 \end{center}
\vspace*{-1cm}
\caption{Invariant cross section $Ed^3\sigma/dp^3$ as a function of
the renormalization and factorization scale $\mu=\mu_f$ for $p\bar{p}
\rightarrow \gamma^* X$ at $\sqrt{S}=2.0$ TeV in the region between
the $J/\psi$ and $\Upsilon$ resonances and $Q_T=5.5$ GeV.
In the interval $0.5 <
\mu/\sqrt{Q^2+Q_T^2} < 2$ the dependence of the cross section on the
scale $\mu=\mu_f$ drops from $\pm 15\%$ (LO) to $\pm 2.5\%$ (NLO).
The $K$-Factor (NLO/LO) is approximately 2.}
\label{fig:8}
\end{figure}

The theoretical uncertainty in the cross section can be estimated by varying
the renormalization and factorization scale $\mu=\mu_f$ around the central
value $\sqrt{Q^2+Q_T^2}$. Figure~\ref{fig:8}
shows this variation for $p\bar{p}
\rightarrow \gamma^* X$ at $\sqrt{S}=2.0$ TeV in the region between
the $J/\psi$ and $\Upsilon$ resonances. In the interval $0.5 <
\mu/\sqrt{Q^2+Q_T^2} < 2$ the dependence of the cross section on the
scale $\mu=\mu_f$ drops from $\pm 15\%$ (LO) to the small value 
$\pm 2.5\%$ (NLO). The $K$-factor ratio (NLO/LO) is approximately 2, 
as one might expect naively.  

\begin{figure}[htb]
 \begin{center}
  {\unitlength1cm
  \begin{picture}(7.6,10.5)
   \epsfig{file=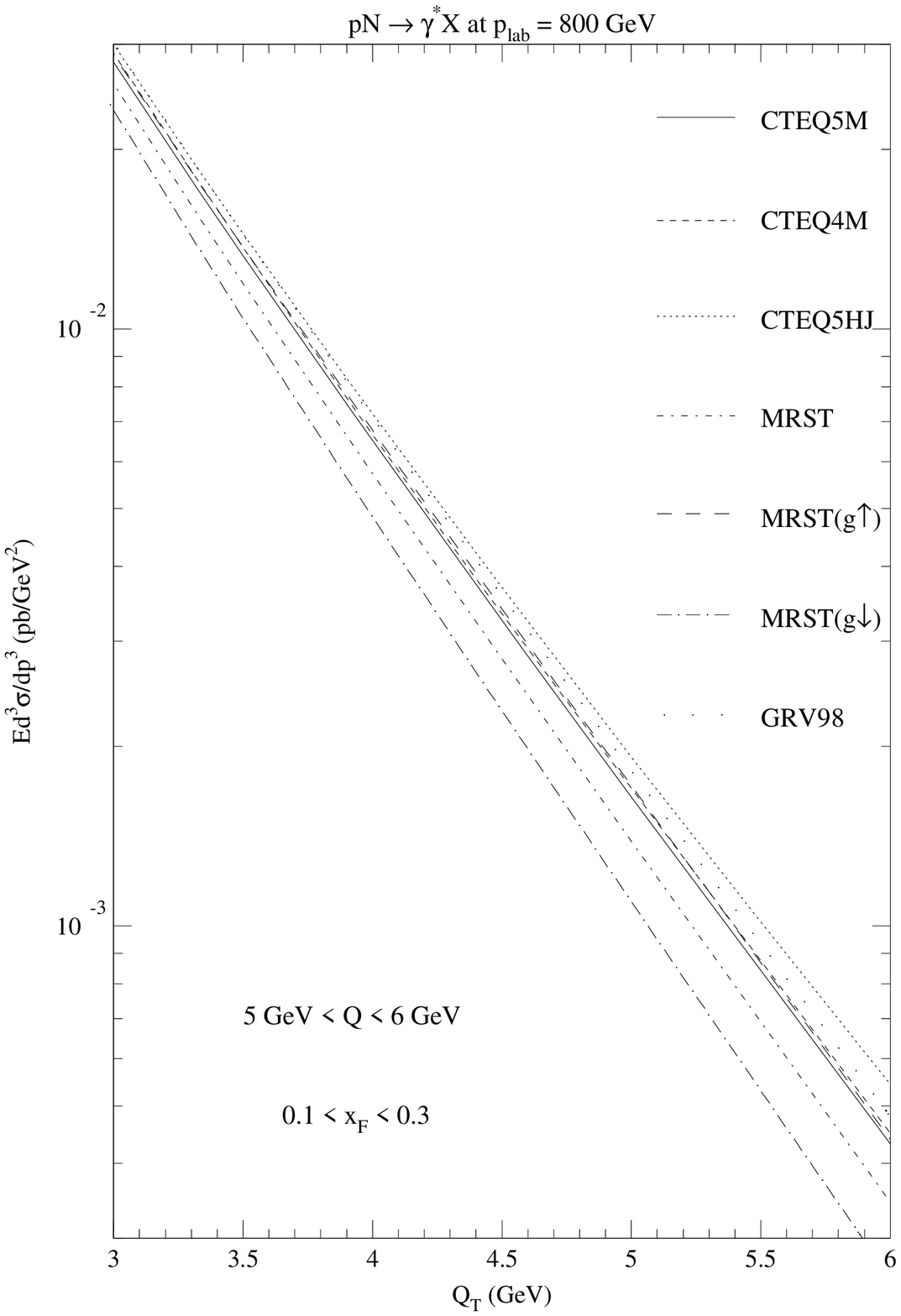,bbllx=60pt,bblly=100pt,bburx=495pt,bbury=725pt,%
           height=10.5cm}
  \end{picture}}
 \end{center}
\vspace*{-1cm}
\caption{Invariant cross section $Ed^3\sigma/d p^3$ as a function of 
$Q_T$ for $p N \rightarrow \gamma^* X$ at $p_{\rm lab}=$ 800 GeV. The cross
section is highly sensitive to the gluon distribution in the proton in regions
of $x_T$ where it is poorly constrained in current analyses.}
\label{fig:6}
\end{figure}

A similar analysis for Fermilab's fixed target experiment E772
\cite{McGaughey:1994dx} is shown in Fig.~\ref{fig:6}.
In this experiment, a deuterium target is bombarded with a proton beam of
momentum $p_{\rm lab}=$ 800 GeV, {\it i.e.} $\sqrt{S}=38.8$ GeV. The cross
section
is averaged over the scaled longitudinal momentum interval 0.1 $< x_F <$ 0.3.
In fixed target experiments one probes substantially larger regions of $x_T$
than in collider experiments.
Therefore one expects greater sensitivity to the gluon distribution in
the proton. We find that use of CTEQ5HJ increases the cross section by 7 \%
(26 \%) w.r.t.\ CTEQ5M at $Q_T=3$ GeV ($Q_T=6$ GeV) and by 134 \% at
$Q_T=10$ GeV. With MRST(g$\downarrow$) the cross section drops relative to 
the CTEQ5M-based values  
by 17 \%, 40 \%, and 59 \% for these three choices of $Q_T$. 

\begin{figure}[htb]
 \begin{center}
  {\unitlength1cm
  \begin{picture}(9,7.68)
   \epsfig{file=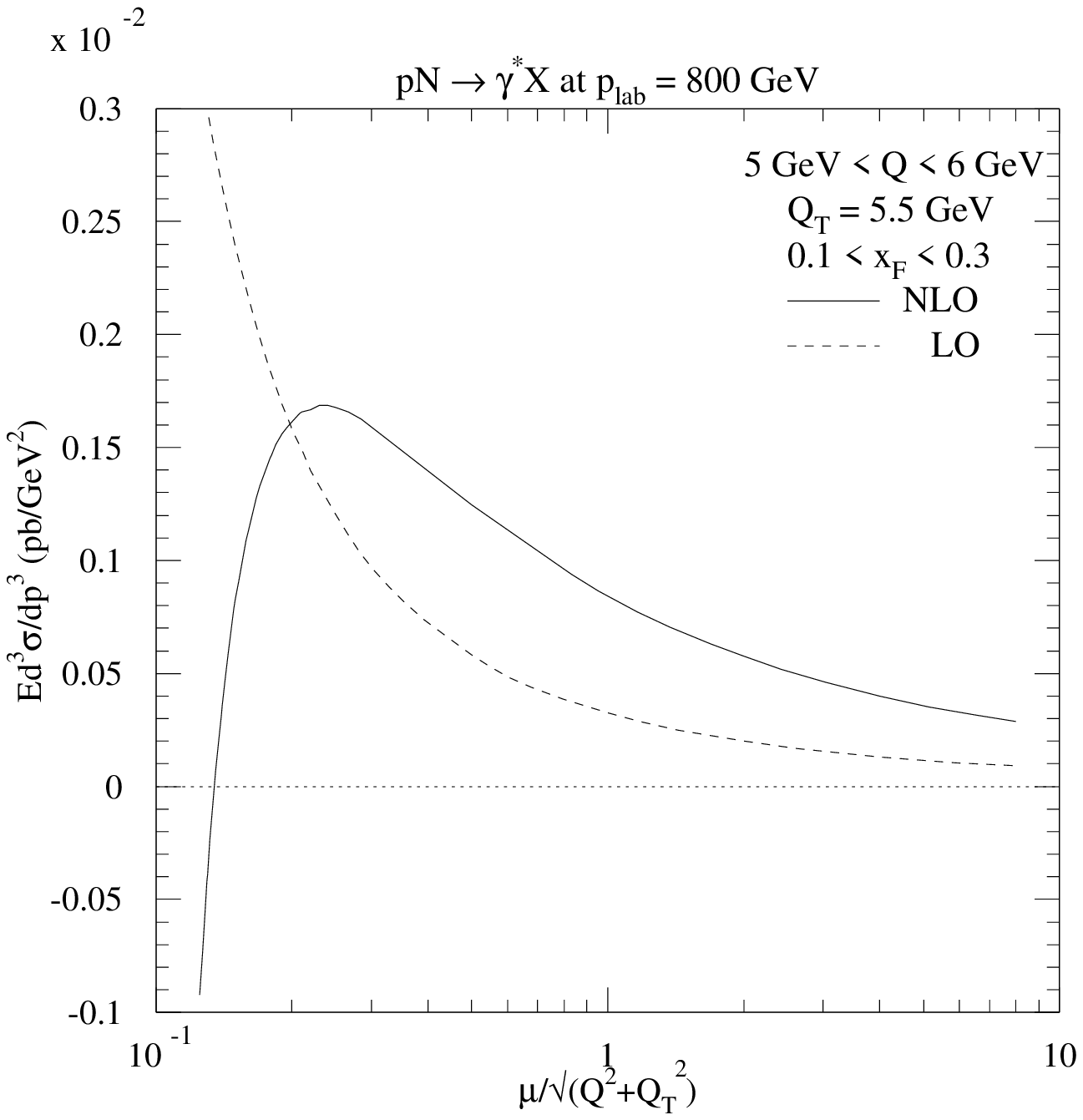,height=8.6cm}
  \end{picture}}
 \end{center}
\vspace*{-1cm}
\caption{Invariant cross section $Ed^3\sigma/d p^3$ as a function of 
the renormalization and factorization scale $\mu=\mu_f$ for $p N \rightarrow
\gamma^* X$ at $p_{\rm lab}=$ 800 GeV. In the interval
$0.5 < \mu/\sqrt{Q^2+Q_T^2} < 2$ the dependence of the cross section
on the scale $\mu$ drops from $\pm 49\%$ (LO) to $\pm 37\%$ (NLO).}
\label{fig:9}
\end{figure}

Figure~\ref{fig:9}
shows the variation of the fixed target cross section on the
renormalization and factorization scale $\mu=\mu_f$. In the interval
$0.5 < \mu/\sqrt{Q^2+Q_T^2} < 2$ the dependence decreases from 
$\pm 49\%$ (LO) to $\pm 37\%$ (NLO).
An optimal scale choice might be $\mu = \mu_f = \sqrt{Q^2+Q_T^2}/4$, where 
the points of Minimal Sensitivity (maximum of NLO) and of Fastest Apparent
Convergence (LO=NLO) nearly coincide.
At $\mu=\mu_f=\sqrt{Q^2+Q_T^2}$, the $K$-factor ratio is 2.6.
The NLO cross section turns negative at the lowest scale shown
$\mu=\mu_f=\sqrt{Q^2+Q_T^2}/8 \simeq 1$ GeV, a value too low to guarantee
perturbative stability.

\section{Summary}
\label{sec:4}
The production of Drell-Yan pairs with
low mass and large transverse momentum is dominated by gluon initiated
subprocesses. In contrast to prompt photon production, uncertainties 
from fragmentation, isolation, and intrinsic transverse momentum are absent.
The hadroproduction of low mass lepton pairs is therefore an advantageous
source of information on the parametrization and size of the gluon density.
The increase in luminosity of Run II increases the
accessible region of $x_T$ from 0.03 to 0.1. The theoretical uncertainty
has been estimated from the scale dependence of the cross sections and
found to be very small for collider experiments. 

\section*{Acknowledgment}
It is a pleasure to thank L.~E.~Gordon for his collaboration.

\newpage
\section*{CONCLUSION: MANIFESTO}

Our goal in this conclusion is not to summarize each of the individual
contributions, but to introduce simple guidelines, a ``Manifesto'', for RunII
analysis~\footnote{clearly this manifesto could be applied to any
experiment}:

\begin{itemize}
\item Each analysis should provide a way to calculate the Likelihood for 
their data, the probability of the data given a theory prediction.
\item The likelihood information should be stored permanently and 
made available.
\end{itemize}

The current practice is generally to take experimental data, correct
for acceptance and smearing and compare the result to the theoretical
predictions.  In many cases, the acceptance and smearing corrections
depend on the theoretical prediction and thus the practice may lead to
uncontrolled uncertainties.  Data are generally presented as tables of
central values with one-sigma standard deviation.  That information is
clearly not enough to reconstruct the Likelihood when the
uncertainties are not Gaussian distributed.  Hence the first guideline
of our Manifesto to provide a way to calculate the likelihood, the
probability of the data given a theory.  The likelihood contains all
the information about the experiment and is the basis for any
analysis.  It should consist of a code and necessary input tables of
``data''.  The code can be as simple as a $\chi^2$ calculation when
all the appropriate conditions are met, but will be significantly more
involved in the general case, see~\cite{GKK1}.  The likelihood
function should be stored in a format which remains valid for several
decades.  This means ASCII format for data and simplicity in the code.
This is important if we want the experimental data to remain useful
even as theoretical calculations evolve.  If the experimental results
are not tied to theory as it stands in the year 2001, they we will be
able to continue to use them, even as the theory evolves from NLO to
NNLO to resummed calculation.

The likelihood functions should be stored in a central repository and
treated in the same fashion as papers\footnote{Auxiliary files
in the FNAL preprints database may be one location or Web pages}.
This is important because Collaboration evolve over time and 
eventually disappear.

Note that the burden is of course not just on the experimental side.
Theoreticians need to provide predictions with understood theoretical
uncertainties over a defined kinematic range.  Numerical calculations
should be made more efficient.  Codes are usually written with the
anticipation that they will be run a few times with a few different
PDFs.  One can anticipate that if the goal to extract uncertainties
for the PDFs from data is to be reached that these codes will have to
be run many orders of magnitude more.  Event generators are preferable
as they allow a better match to experimental cuts and the possibility
of comparison of smeared theory to raw data.  A central repository for
the theoretical code would also be very helpful.

In this series of workshops several groups reported significant
progress towards extracting PDFs from data with 
uncertainties~\cite{GKK1,MSUfit}.
Note also that other groups, not connected to this workshop~\cite{others},  
have reported results on PDF uncertainties since this workshop started.
We are therefore optimistic that realistic PDF uncertainties will
be available from several groups by the start of Run II at the Tevatron.

Progress has also been made on the study of the best way to present
data~\cite{Hirosky} for Run II.  Clearly, the use of the Run II
Tevatron data to their full potential will require planning and care
through a collaborative effort between phenomenologists and
experimentalists.

\end{document}